\documentclass{aa}

\usepackage{graphicx}
\usepackage[dvipsnames]{xcolor}
\usepackage{txfonts}
\usepackage{libertine}
\usepackage{subcaption}
\usepackage{gensymb}
\usepackage[scaled=0.8]{FiraMono}
\usepackage{hyperref}
\begin{document}

   \title{The discovery of a radio galaxy of at least 5 Mpc}


   \author{Martijn S.S.L. Oei
          \inst{1}\thanks{\emph{In dear memory of Pallas. If your name hadn't been this popular with asteroid discoverers, you'd now be the giants' giant --- once again looking down at the sprawling ants below.}}
          \and
          Reinout J. van Weeren
          \inst{1}
          \and
          Martin J. Hardcastle
          \inst{2}
          \and
          Andrea Botteon
          \inst{1}
          \and
          Tim W. Shimwell
          \inst{1}
          \and
          Pratik Dabhade
          \inst{3}
          \and
          Aivin R.D.J.G.I.B. Gast
          \inst{4}
          \and
          Huub J.A. R\"ottgering
          \inst{1}
          \and
          Marcus Br\"uggen
          \inst{5}
          \and
          Cyril Tasse
          \inst{6, 7}
          \and
          Wendy L. Williams
          \inst{1}
          \and
          Aleksandar Shulevski
          \inst{1}
          }

   \institute{
              Leiden Observatory, Leiden University, Niels Bohrweg 2, NL-2300 RA Leiden, The Netherlands\\
              \email{oei@strw.leidenuniv.nl}
              \and
              Centre for Astrophysics Research, University of Hertfordshire, College Lane, Hatfield AL10 9AB, United Kingdom
              \and
              Observatoire de Paris, LERMA, Coll\`ege de France, CNRS, PSL University, Sorbonne University, 75014 Paris, France
              \and
              Somerville College, University of Oxford, Woodstock Road, Oxford OX2 6HD, United Kingdom
              \and
              Hamburger Sternwarte, University of Hamburg, Gojenbergsweg 112, 21029 Hamburg, Germany
              \and
              GEPI \& USN, Observatoire de Paris, Universit\'e PSL, CNRS, 5 Place Jules Janssen, 92190 Meudon, France
              \and
              Department of Physics \& Electronics, Rhodes University, PO Box 94, Grahamstown, 6140, South Africa
             }

   \date{\today}

 
  \abstract
   {
   Giant radio galaxies (GRGs, or colloquially \textit{`giants'}) are the Universe's largest structures generated by individual galaxies.
   They comprise synchrotron-radiating AGN ejecta and attain cosmological (Mpc-scale) lengths.
   However, the main mechanisms that drive their exceptional growth remain poorly understood.
   }
  {
  To deduce the main mechanisms that drive a phenomenon, it is usually instructive to study extreme examples.
  If there exist host galaxy characteristics that are an important cause for GRG growth, then the hosts of the \emph{largest} GRGs are likely to possess them.
  Similarly, if there exist particular large-scale environments that are highly conducive to GRG growth, then the \emph{largest} GRGs are likely to reside in them.
  For these reasons, we aim to perform a case study of the largest GRG available.
  }
   {
   We reprocessed the LOFAR Two-metre Sky Survey (LoTSS) DR2 by subtracting compact sources and performing multi-scale CLEAN deconvolution at $60''$ and $90''$ resolution.
   The resulting images constitute the most sensitive survey yet for radio galaxy lobes, whose diffuse nature and steep synchrotron spectra have allowed them to evade previous detection attempts at higher resolution and shorter wavelengths.
   We visually searched these images for GRGs.
   }
   {
   We discover \textit{Alcyoneus}, a low-excitation radio galaxy with a projected proper length $l_\mathrm{p} = 4.99 \pm 0.04\ \mathrm{Mpc}$.
   Its jets and lobes are all four detected at very high significance, and the SDSS-based identification of the host, at spectroscopic redshift $z_\mathrm{spec} = 0.24674\ \pm\ 6 \cdot 10^{-5}$, is unambiguous.
   The total luminosity density at $\nu = 144\ \mathrm{MHz}$ is $L_\nu = 8 \pm 1 \cdot 10^{25}\ \mathrm{W\ Hz^{-1}}$, which is below-average, though near-median (percentile $45 \pm 3\%$), for GRGs. 
   The host is an elliptical galaxy with a stellar mass $M_\star = 2.4 \pm 0.4 \cdot 10^{11}\ M_\odot$ and a supermassive black hole mass $M_\bullet = 4 \pm 2 \cdot 10^8\ M_\odot$, both of which tend towards the lower end of their respective GRG distributions (percentiles $25 \pm 9 \%$ and $23 \pm 11 \%$).
   The host resides in a filament of the Cosmic Web.
    Through a new Bayesian model for radio galaxy lobes in three dimensions, we estimate the pressures in the $\mathrm{Mpc}^3$-scale northern and southern lobe to be $P_\mathrm{min,1} = 4.8 \pm 0.3 \cdot 10^{-16}\ \mathrm{Pa}$ and $P_\mathrm{min,2} = 4.9 \pm 0.6 \cdot 10^{-16}\ \mathrm{Pa}$, respectively.
    The corresponding magnetic field strengths are $B_\mathrm{min,1} = 46 \pm 1\ \mathrm{pT}$ and $B_\mathrm{min,2} = 46 \pm 3\ \mathrm{pT}$.
    }
   {
   We have discovered what is in projection the largest known structure made by a single galaxy --- a GRG with a projected proper length $l_\mathrm{p} = 4.99 \pm 0.04\ \mathrm{Mpc}$.
   The true proper length is at least $l_\mathrm{min} = 5.04 \pm 0.05\ \mathrm{Mpc}.$
   Beyond geometry, Alcyoneus and its host are suspiciously ordinary: the total low-frequency luminosity density, stellar mass and supermassive black hole mass are all lower than, though similar to, those of the medial GRG.
   Thus, very massive galaxies or central black holes are not necessary to grow large giants, and, if the observed state is representative of the source over its lifetime, neither is high radio power.
   A low-density environment remains a possible explanation.
   The source resides in a filament of the Cosmic Web, with which it might have significant thermodynamic interaction.
   The pressures in the lobes are the lowest hitherto found, and Alcyoneus therefore represents the most promising radio galaxy yet to probe the warm--hot intergalactic medium.
   }

   \maketitle

\begin{figure*}
\centering
\includegraphics[width=.95\linewidth]{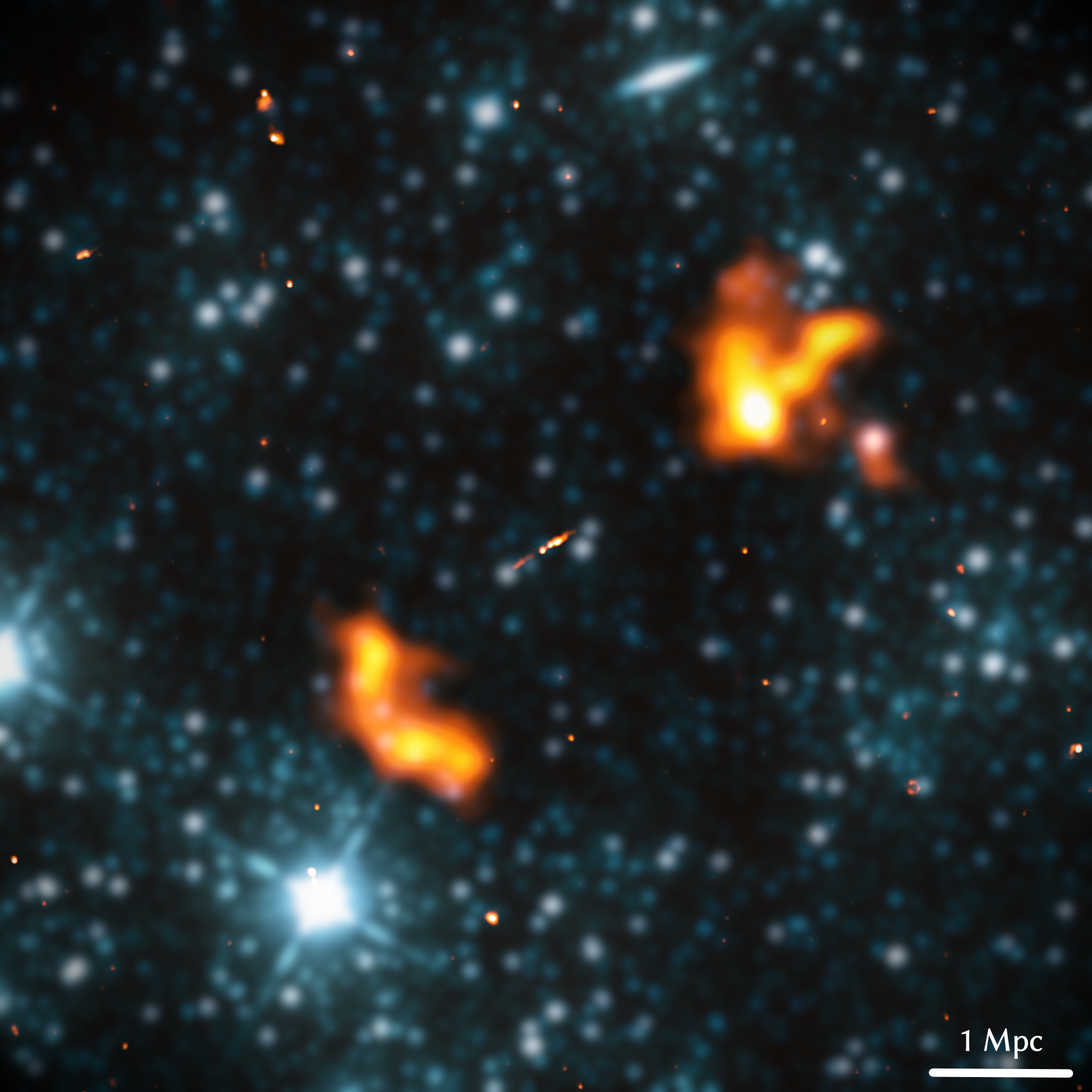}
\caption{
\textbf{Joint radio-infrared view of \textit{Alcyoneus,} a radio galaxy with a projected proper length of 5.0 Mpc.}
We show a $2048''\ \times\ 2048''$ solid angle centred around right ascension $123.590372\degree$ and declination $52.402795\degree$.
We superimpose LOFAR Two-metre Sky Survey (LoTSS) DR2 images at 144 MHz of two different resolutions ($6''$ for the core and jets, and $60''$ for the lobes) (orange), with the Wide-field Infrared Survey Explorer (WISE) image at 3.4 $\mu$m (blue).
To highlight the radio emission, the infrared emission has been blurred to $0.5'$ resolution.
}
\label{fig:Alcyoneus}
\end{figure*}\noindent
\section{Introduction}
Most galactic bulges hold a supermassive (i.e. $M_\bullet > 10^6\ M_\odot$) Kerr black hole \textcolor{blue}{\citep[e.g.][]{Soltan11982}} that grows by accreting gas, dust and stars from its surroundings \textcolor{blue}{\citep{Kormendy12013}}.
The black hole ejects a fraction of its accretion disk plasma from the host galaxy along two collimated, magnetised jets that are aligned with its rotation axis \textcolor{blue}{\citep[e.g.][]{Blandford11974}}.
The relativistic electrons contained herein experience Lorentz force and generate, through spiral motion, synchrotron radiation that is observed by radio telescopes.
The two jets either fade gradually or end in hotspots at the end of diffuse lobes, and ultimately enrich the intergalactic medium (IGM) with cosmic rays and magnetic fields.
The full luminous structure is referred to as a \textit{radio galaxy} (RG).
Members of a rare RG subpopulation attain megaparsec-scale proper (and thus comoving) lengths \textcolor{blue}{\citep[e.g.][]{Willis11974, Andernach11992, IshwaraChandra11999, Jamrozy12008, Machalski12011, Kuzmicz12018, Dabhade12020October}}.
The \emph{giant} radio galaxy (GRG, or colloquially \textit{`giant'}) definition accommodates our limited ability to infer an RG's true proper length from observations: an RG is called a GRG if and only if its proper length \emph{projected onto the plane of the sky} exceeds some threshold $l_\mathrm{p,GRG}$, usually chosen to be 0.7 or 1 Mpc.
Because the conversion between angular length and projected proper length depends on cosmological parameters, which remain uncertain, it is not always clear whether a given observed RG satisfies the GRG definition.\\
Currently, there are about a thousand GRGs known, the majority of which have been found in the Northern Sky.
About one hundred exceed 2 Mpc and ten exceed 3 Mpc; at 4.9 Mpc, the literature's projectively longest is J1420-0545 \textcolor{blue}{\citep{Machalski12008}}.
As such, GRGs --- and the rest of the megaparsec-scale RGs --- are the largest single-galaxy--induced phenomena in the Universe.
It is a key open question what physical mechanisms lead some RGs to extend for ${\sim}10^2$ times their host galaxy diameter.
To determine whether there exist particular host galaxy characteristics or large-scale environments that are essential for GRG growth, it is instructive to analyse the largest GRGs, since in these systems it is most likely that all major favourable growth factors are present.
We thus aim to perform a case study of the largest GRG available.\\
As demonstrated by \textcolor{blue}{\citet{Dabhade12020March}}'s record sample of 225 discoveries, the Low-frequency Array (LOFAR) \textcolor{blue}{\citep{vanHaarlem12013}} is among the most attractive contemporary instruments for finding new GRGs.
This Pan-European radio interferometer features a unique combination of short baselines to provide sensitivity to large-scale emission, and long baselines to mitigate source confusion.\footnote{
Source confusion is an instrumental limitation that arises when the resolution of an image is low compared to the sky density of statistically significant sources.
It causes angularly adjacent, but physically unrelated sources to blend together, making it hard or even impossible to distinguish them \textcolor{blue}{\citep[e.g.][]{Condon12012}}.}
These qualities are indispensable for observational studies of GRGs, which require identifying both extended lobes and compact cores and jets.
Additionally, the metre wavelengths at which the LOFAR operates allow it to detect steep-spectrum lobes far away from host galaxies.
Such lobes reveal the full extent of GRGs, but are missed by decimetre observatories.
\begin{figure}
    \centering
    \begin{subfigure}{\columnwidth}
    \includegraphics[width=\linewidth]{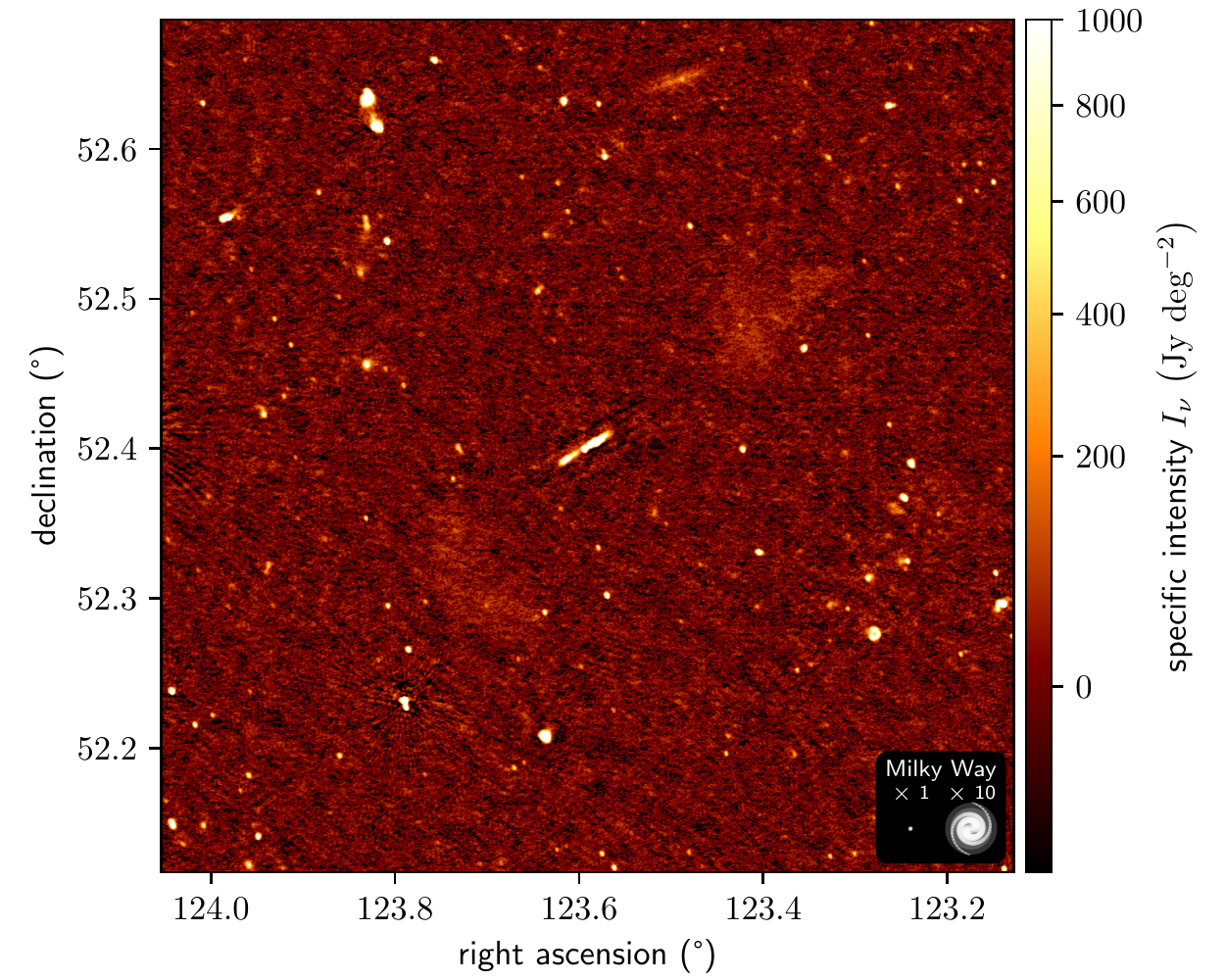}
    \end{subfigure}
    \begin{subfigure}{\columnwidth}
    \includegraphics[width=\linewidth]{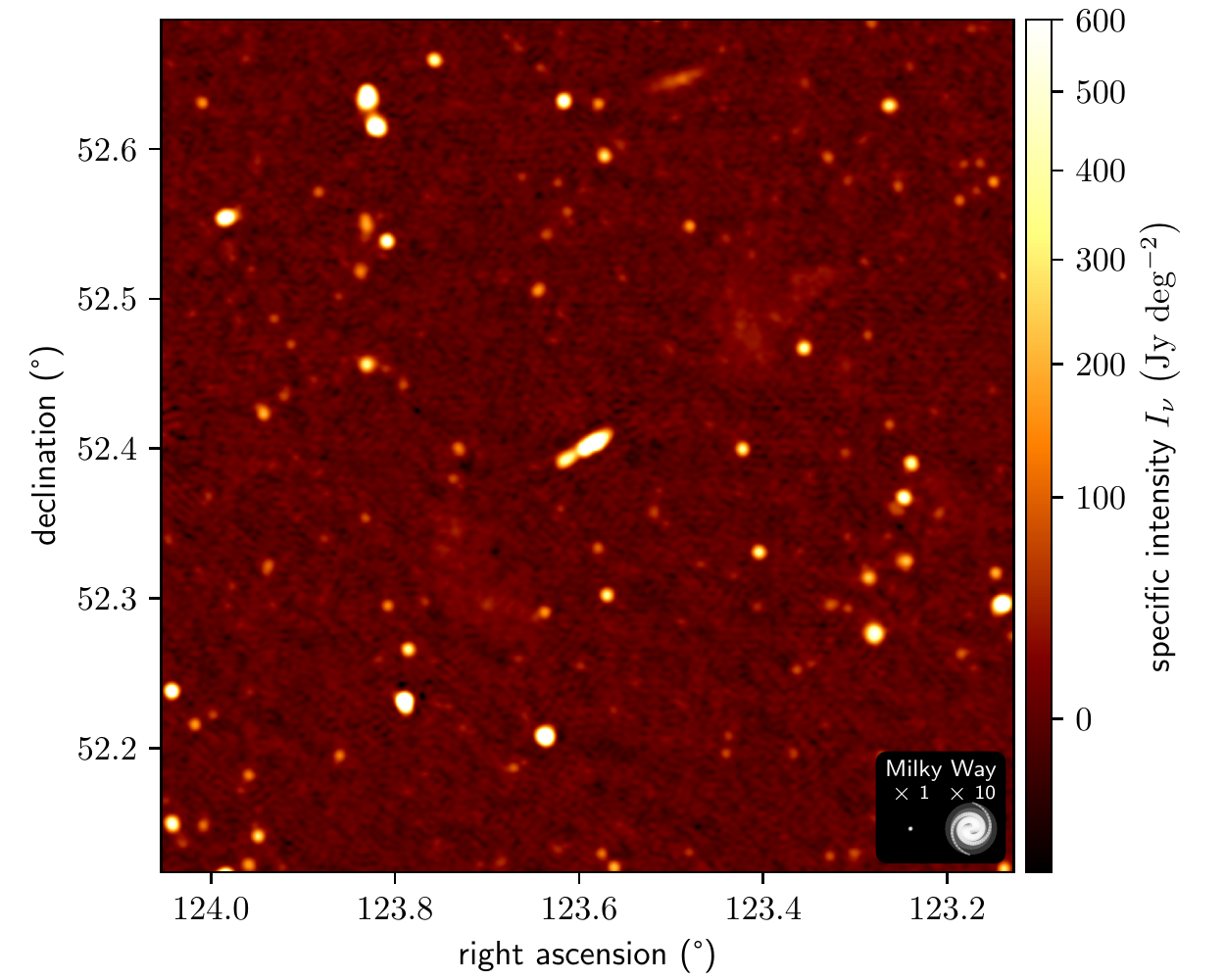}
    \end{subfigure}
    \caption{
    \textbf{Alcyoneus' lobes are easily overlooked in the LoTSS DR2 at its standard resolutions.}
    We show images at central frequency $\nu_\mathrm{c} = 144\ \mathrm{MHz}$ and resolutions $\theta_\mathrm{FWHM} = 6''$ (top) and $\theta_\mathrm{FWHM} = 20''$ (bottom), centred around host galaxy J081421.68+522410.0.
    }
    \label{fig:LoTSS6And20Arcsec}
\end{figure}\noindent
Thus, in \textbf{Section}~\ref{sec:dataMethods}, we describe a reprocessing of the LOFAR Two-metre Sky Survey (LoTSS) Data Release 2 (DR2) aimed at revealing hitherto unknown RG lobes --- among other goals.
An overview of the reprocessed images, which cover thousands of square degrees, and statistics of the lengths and environments of the GRGs they have revealed, are subjects of future publications.
For now, these images allow us to discover \textit{Alcyoneus}\footnote{Alcyoneus was the son of Ouranos, the Greek primordial god of the sky. According to Ps.-Apollodorus, he was also one of the greatest of the \textit{Gigantes} (Giants), and a challenger to Heracles during the Gigantomachy --- the battle between the Giants and the Olympian gods for supremacy over the Cosmos. The poet Pindar described him as `huge as a mountain', fighting by hurling rocks at his foes.}, a 5 Mpc GRG, whose properties we determine and discuss in \textbf{Section}~\ref{sec:resultsDiscussion}.
\textbf{Figure}~\ref{fig:Alcyoneus} provides a multi-wavelength, multi-resolution view of this giant.
\textbf{Section}~\ref{sec:conclusion} contains our concluding remarks.\\
We assume a concordance inflationary $\Lambda$CDM model with parameters $\mathfrak{M}$ from \textcolor{blue}{\citet{PlanckCollaboration12020}}; i.e. $\mathfrak{M} = \left(h = 0.6766, \Omega_\mathrm{BM,0} = 0.0490, \Omega_\mathrm{M,0} = 0.3111, \Omega_{\Lambda,0} = 0.6889\right)$, where $H_0 \coloneqq h \cdot 100\ \mathrm{km\ s^{-1}\ Mpc^{-1}}$.
We define the spectral index $\alpha$ such that it relates to flux density $F_\nu$ at frequency $\nu$ as $F_\nu \propto \nu^\alpha$.
Regarding terminology, we strictly distinguish between a \emph{radio galaxy}, a radio-bright structure of relativistic particles and magnetic fields (consisting of a core, jets, hotspots and lobes), and the \emph{host galaxy} that generates it.

\section{Data and methods}
\label{sec:dataMethods}
The LoTSS, conducted by the LOFAR High-band Antennae (HBA), is a 120--168 MHz interferometric survey \textcolor{blue}{\citep{Shimwell12017, Shimwell12019, Shimwell12021}} with the ultimate aim to image the full Northern Sky at resolutions of $6''$, $20''$, $60''$ and $90''$.
Its central frequency $\nu_\mathrm{c} = 144\ \mathrm{MHz}$.
The latest data release --- the LoTSS DR2 \textcolor{blue}{\citep{Shimwell12021}} --- covers $27\%$ of the Northern Sky, split over two regions of $4178\ \mathrm{deg}^2$ and $1457\ \mathrm{deg}^2$; the largest of these contains the Sloan Digital Sky Survey (SDSS) DR7 \textcolor{blue}{\citep{Abazajian12009}} area.
By default, the LoTSS DR2 provides imagery at the $6''$ and $20''$ resolutions.
We show these standard products in \textbf{Figure}~\ref{fig:LoTSS6And20Arcsec} for the same sky region as in \textbf{Figure}~\ref{fig:Alcyoneus}.
In terms of total source counts, the LoTSS DR2 is the largest radio survey carried out thus far: its catalogue contains $4.4\cdot10^6$ sources, most of which are considered \emph{compact}.
By contrast, the $60''$ and $90''$ imagery, which we discuss in more detail in \textcolor{blue}{\citet{Oei12022}}, is intended to reveal \emph{extended} structures in the low-frequency radio sky, such as giant radio galaxies, supernova remnants in the Milky Way, radio halos and shocks in galaxy clusters, and --- potentially --- accretion shocks or volume-filling emission from filaments of the Cosmic Web.
To avoid the source confusion limit at these resolutions, following \textcolor{blue}{\citet{vanWeeren12021}}, we used DDFacet \textcolor{blue}{\citep{Tasse12018}} to predict visibilities corresponding to the $20''$ LoTSS DR2 sky model and subtracted these from the data, before imaging at $60''$ and $90''$ with WSClean IDG \textcolor{blue}{\citep{Offringa12014, vanDerTol12018}}.
We used -0.5 Briggs weighting and multiscale CLEAN \textcolor{blue}{\citep{Offringa12017}}, with \texttt{-multiscale-scales 0,4,8,16,32,64}.
Importantly, we did not impose an inner $\left(u,v\right)$-cut.
We imaged each pointing separately, then combined the partially overlapping images into a mosaic by calculating, for each direction, a beam-weighted average.\\
Finally, we visually searched the LoTSS DR2 for GRGs, primarily at $6''$ and $60''$ using the Hierarchical Progressive Survey (HiPS) system in \textit{Aladin Desktop 11.0} \textcolor{blue}{\citep{Bonnarel12000}}.

\section{Results and discussion}
\label{sec:resultsDiscussion}
\subsection{Radio morphology and interpretation}
\label{sec:radioMorphology}
During our LoTSS DR2 search, we identified a three-component radio structure of total angular length $\phi = 20.8'$, visible at all ($6''$, $20''$, $60''$ and $90''$) resolutions.
\textbf{Figure}~\ref{fig:LoTSS6And20Arcsec} provides a sense of our data quality; it shows that the outer components are barely discernible in the LoTSS DR2 at its standard $6''$ and $20''$ resolutions.
Meanwhile, \textbf{Figure}~\ref{fig:Alcyoneus} shows the outer components at $60''$, and the top panel of \textbf{Figure}~\ref{fig:doubleConeModel} shows them at $90''$; at these resolutions, they lie firmly above the noise.
Compared with the outer structures, the central structure is bright and elongated, with a $155''$ major axis and a $20''$ minor axis.
The outer structures lie along the major axis at similar distances from the central structure, are diffuse and amorphous, and feature specific intensity maxima along this axis.\\
In the arcminute-scale vicinity of the outer structures, the DESI Legacy Imaging Surveys \textcolor{blue}{\citep{Dey12019}} DR9 does not reveal galaxy overdensities or low-redshift spiral galaxies, the ROSAT All-sky Survey (RASS) \textcolor{blue}{\citep{Voges11999}} does not show X-ray brightness above the noise, and there is no Planck Sunyaev--Zeldovich catalogue 2 (PSZ2) \textcolor{blue}{\citep{PlanckCollaboration12016}} source nearby.
The outer structures therefore cannot be supernova remnants in low-redshift spiral galaxies or radio relics and radio halos in galaxy clusters.
Instead, the outer structures presumably represent radio galaxy emission.
\begin{figure}
    \centering
    \begin{subfigure}{\columnwidth}
    \includegraphics[width=\columnwidth]{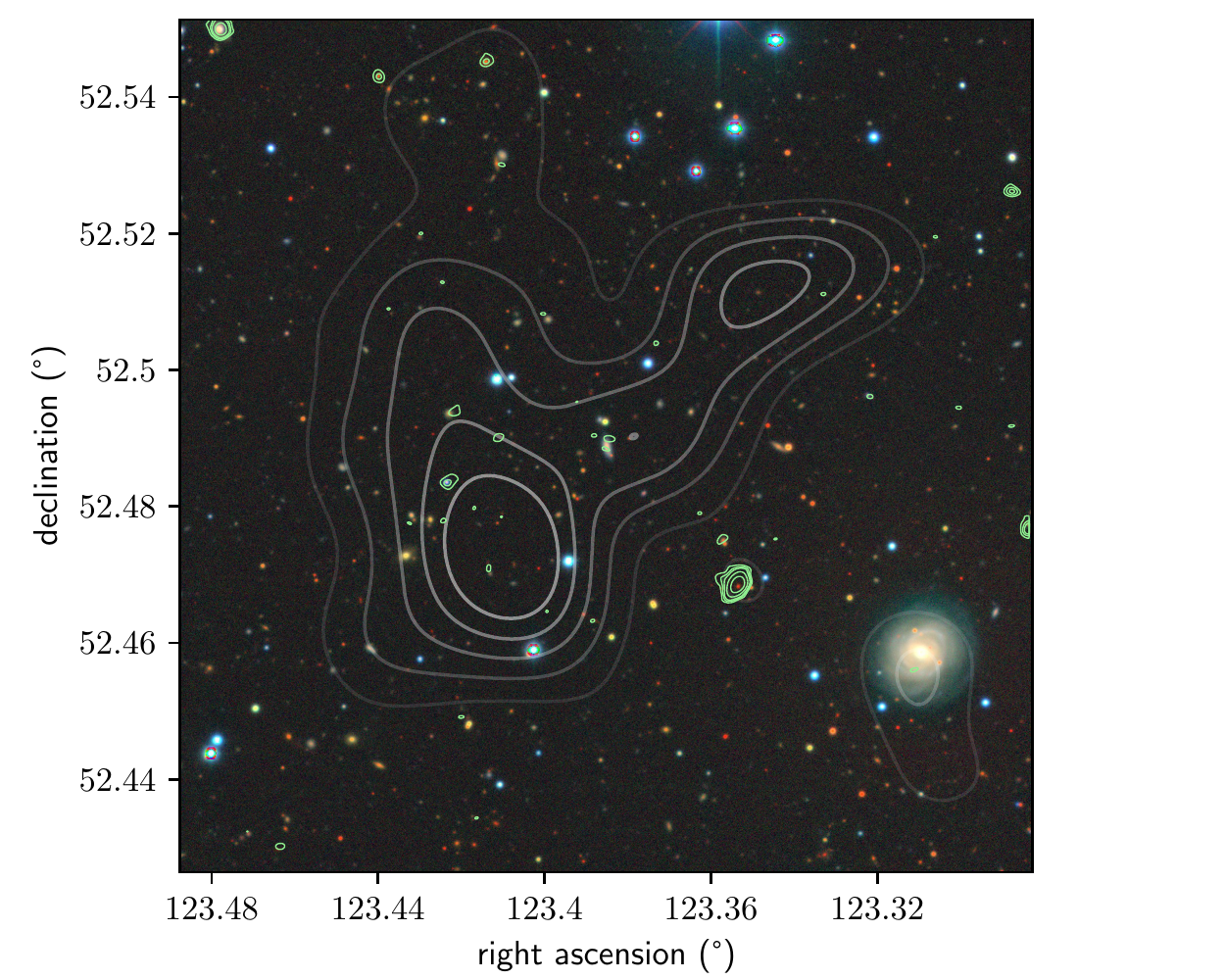}
    \end{subfigure}
    \begin{subfigure}{\columnwidth}
    \includegraphics[width=\columnwidth]{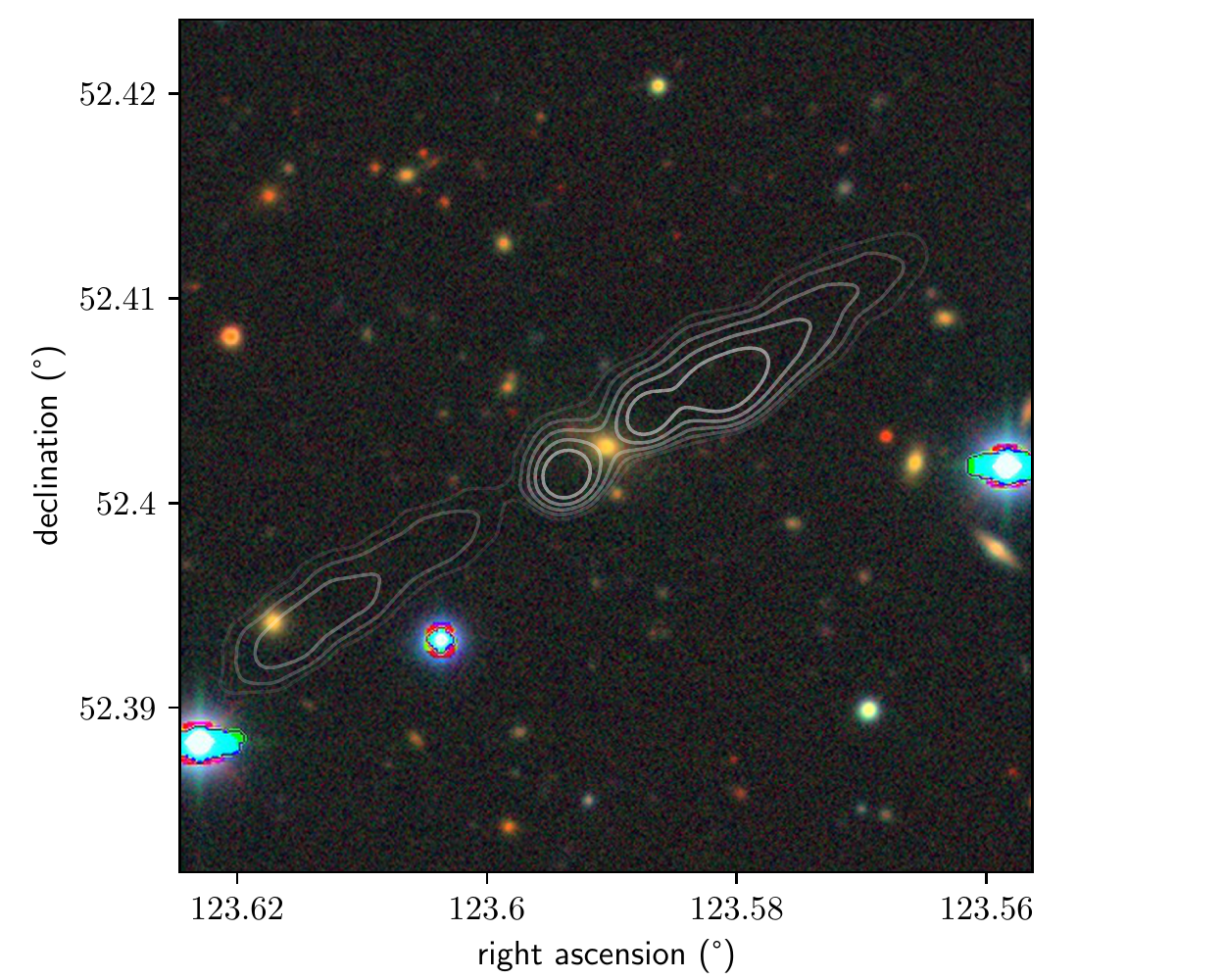}
    \end{subfigure}
    \begin{subfigure}{\columnwidth}
    \includegraphics[width=\columnwidth]{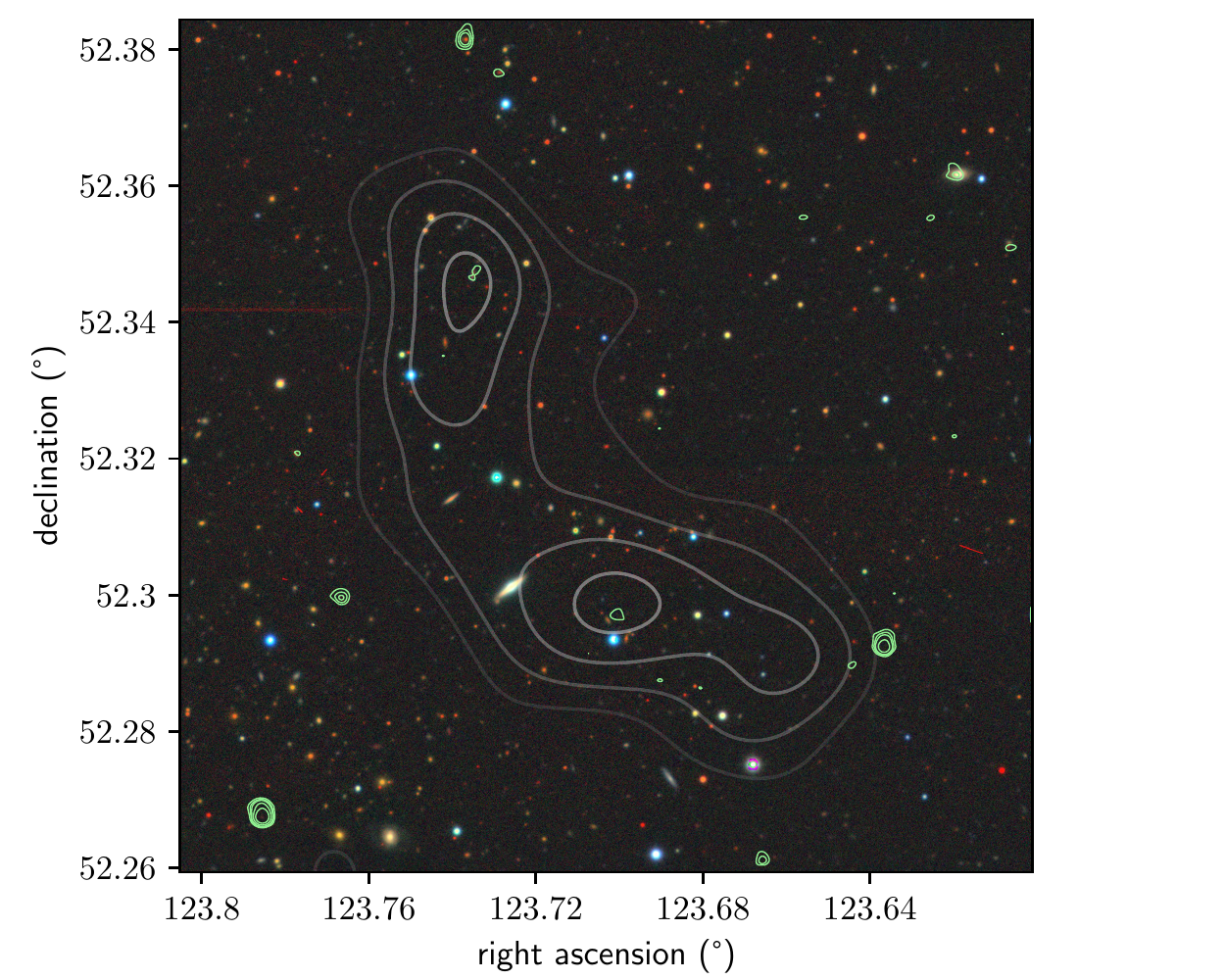}
    \end{subfigure}
    \caption{
    \textbf{Joint radio-optical views show that \textbf{Figure}~\ref{fig:Alcyoneus}'s outer structures are best interpreted as a pair of radio galaxy lobes fed by central jets.}
    On top of DESI Legacy Imaging Surveys DR9 $(g, r, z)$-imagery, we show the LoTSS DR2 at various resolutions through contours at multiples of $\sigma$, where $\sigma$ is the image noise at the relevant resolution.
    The top and bottom panel show translucent white $60''$ contours at $3, 5, 7, 9, 11\sigma$ and solid light green $6''$ contours at $4, 7, 10, 20, 40\sigma$.
    The central panel shows translucent white $6''$ contours at $5, 10, 20, 40, 80\sigma$.
    }
    \label{fig:AlcyoneusHostLegacy}
\end{figure}\noindent
The radio-optical overlays in \textbf{Figure}~\ref{fig:AlcyoneusHostLegacy}'s top and bottom panel show that it is improbable that each outer structure is a radio galaxy of its own, given the lack of significant $6''$ radio emission (solid light green contours) around host galaxy candidates suggested by the morphology of the $60''$ radio emission (translucent white contours).
For these reasons, we interpret the central (jet-like) structure and the outer (lobe-like) structures as components of the \emph{same} radio galaxy.\\
Subsequent analysis --- presented below --- demonstrates that this radio galaxy is the largest hitherto discovered, with a projected proper length of 5.0 Mpc.
We dub this GRG \textit{Alcyoneus}.
\begin{figure}
\centering
\begin{subfigure}{\columnwidth}
\includegraphics[width=\linewidth]{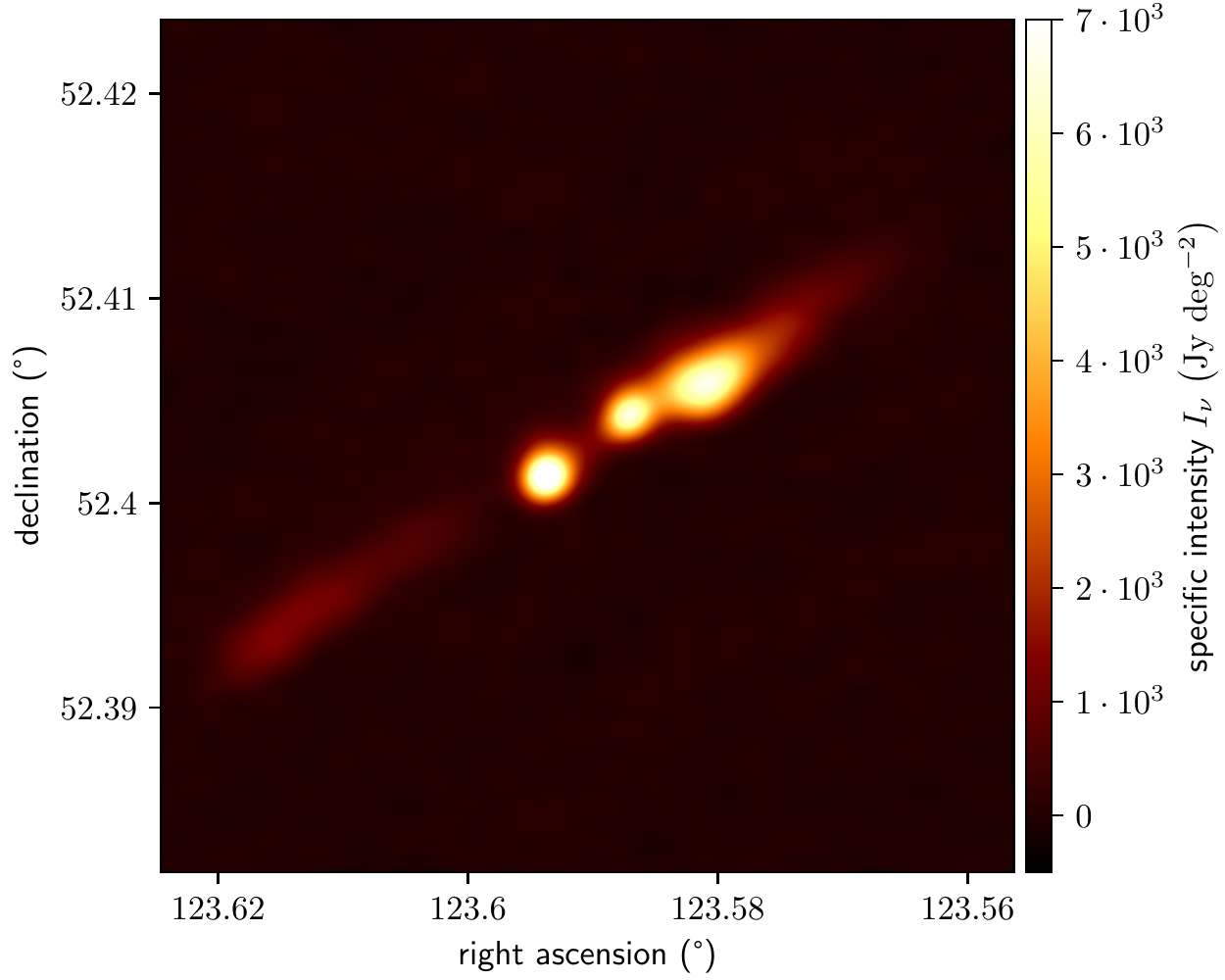}
\end{subfigure}
\begin{subfigure}{\columnwidth}
\includegraphics[width=\linewidth]{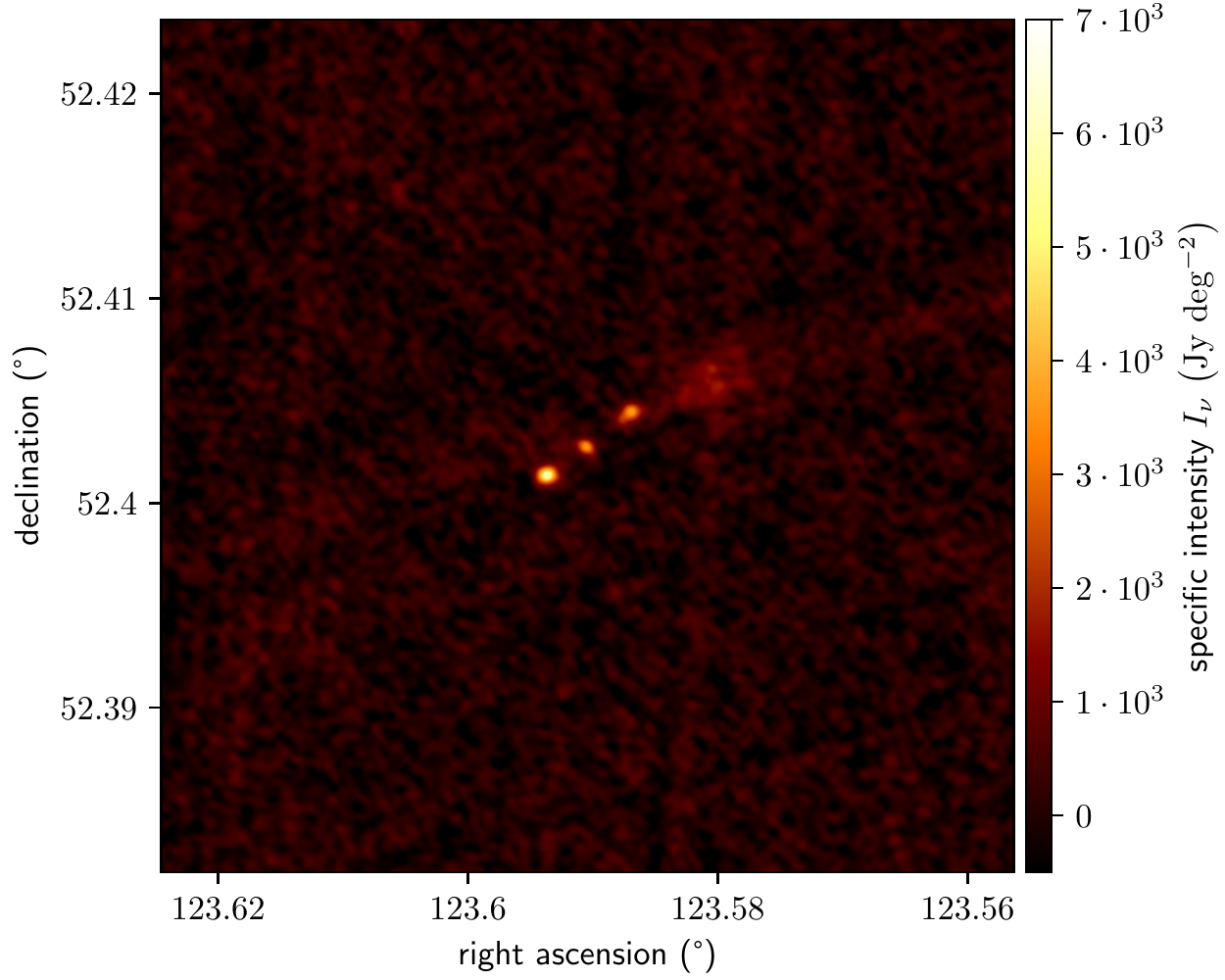}
\end{subfigure}
\begin{subfigure}{\columnwidth}
\includegraphics[width=\linewidth]{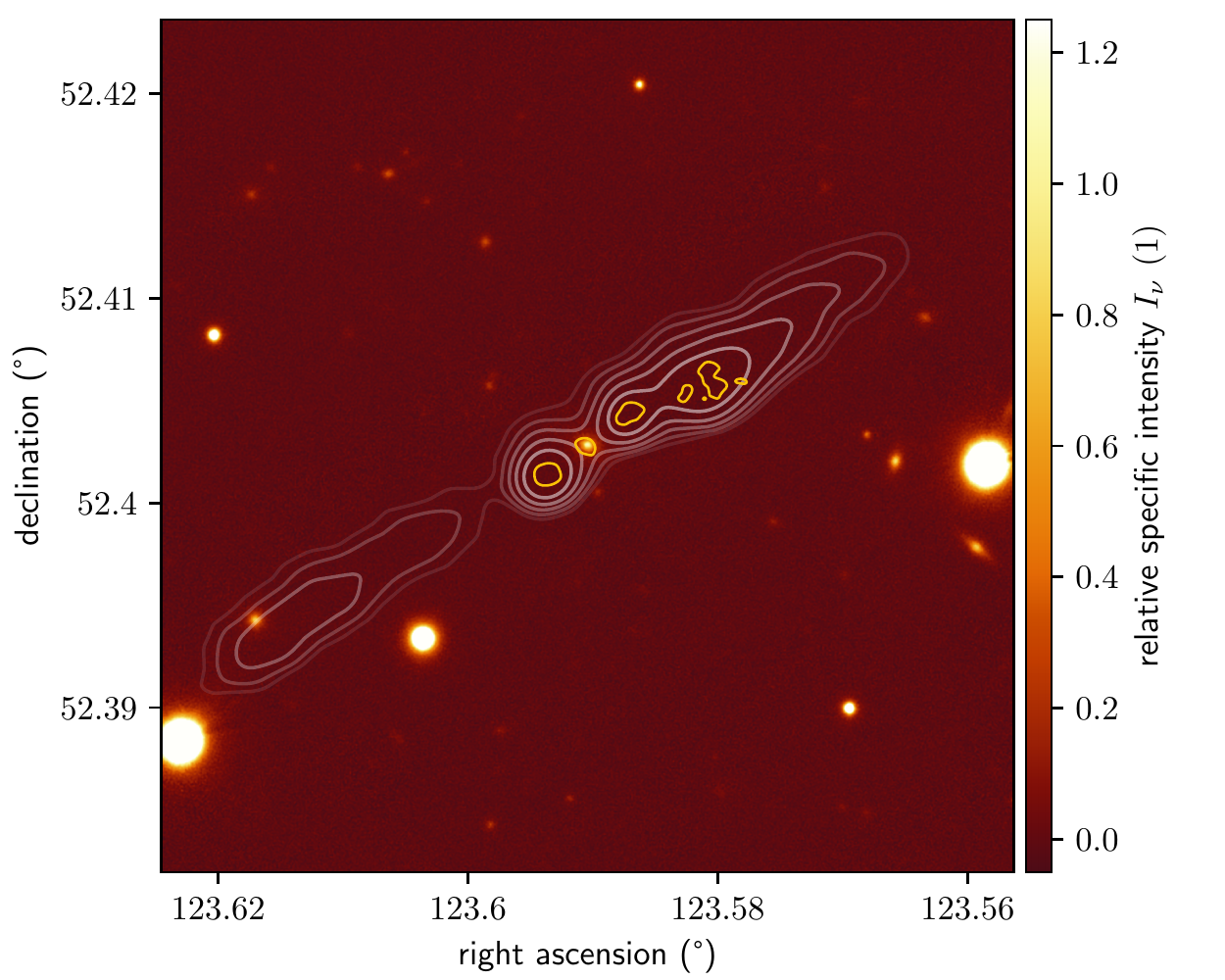}
\end{subfigure}
\caption{
\textbf{The SDSS DR12 source J081421.68+522410.0 is Alcyoneus' host galaxy.}
The panels cover a $2.5' \times 2.5'$ region around J081421.68+522410.0, an elliptical galaxy with spectroscopic redshift $z_\mathrm{spec} = 0.24674\ \pm\ 6 \cdot 10^{-5}$.
From top to bottom, we show the LoTSS DR2 $6''$, the VLASS $2.2''$, and the Pan-STARRS DR1 $i$-band --- relative to the peak specific intensity of J081421.68+522410.0 --- with LoTSS contours (white) as in \textbf{Figure}~\ref{fig:AlcyoneusHostLegacy} and a VLASS contour (gold) at $5\sigma$.
}
\label{fig:AlcyoneusContours}
\end{figure}\noindent
\subsection{Host galaxy identification}
Based on the middle panel of \textbf{Figure}~\ref{fig:AlcyoneusHostLegacy} and an SDSS DR12 \textcolor{blue}{\citep{Alam12015}} spectrum, we identify a source at a J2000 right ascension of $123.590372\degree$, a declination of $52.402795\degree$ and a spectroscopic redshift of $z_\mathrm{spec} = 0.24674\ \pm\ 6 \cdot 10^{-5}$ as Alcyoneus' host.
Like most GRG hosts, this source, with SDSS DR12 name J081421.68+522410.0, is an elliptical galaxy\footnote{Based on the SDSS morphology, \textcolor{blue}{\citet{Kuminski12016}} calculate a probability of $89\%$ that the galaxy is an elliptical.} without a quasar.
From optical contours, we find that the galaxy's minor axis makes a ${\sim}20\degree$ angle with Alcyoneus' jet axis.\\
In \textbf{Figure}~\ref{fig:AlcyoneusContours}, we further explore the connection between J081421.68+522410.0 and Alcyoneus' radio core and jets.
From top to bottom, we show the LoTSS DR2 at $6''$, the Very Large Array Sky Survey (VLASS) \textcolor{blue}{\citep{Lacy12020}} at $2.2''$, and the Panoramic Survey Telescope and Rapid Response System (Pan-STARRS) DR1 \textcolor{blue}{\citep{Chambers12016}} $i$-band.
Two facts confirm that the host identification is highly certain.
First, for both the LoTSS DR2 at $6''$ and the VLASS at $2.2''$, the angular separation between J081421.68+522410.0 and the arc connecting Alcyoneus' two innermost jet features is subarcsecond.
Moreover, the alleged host galaxy is the brightest Pan-STARRS DR1 $i$-band source within a radius of $45''$ of the central VLASS image component.

\subsection{Radiative- or jet-mode active galactic nucleus}
Current understanding \textcolor{blue}{\citep[e.g.][]{Heckman12014}} suggests that the population of active galactic nuclei (AGN) exhibits a dichotomy: AGN seem to be either radiative-mode AGN, which generate high-excitation radio galaxies (HERGs), or jet-mode AGN, which generate low-excitation radio galaxies (LERGs).
Is Alcyoneus a HERG or a LERG?
The SDSS spectrum of the host features very weak emission lines; indeed, the star formation rate (SFR) is just $1.6 \cdot 10^{-2}\ M_\odot\ \mathrm{yr}^{-1}$ \textcolor{blue}{\citep{Chang12015}}.
Following the classification rule of \textcolor{blue}{\citet{Best12012, Best12014, Pracy12016, Williams12018}} based on the strength and equivalent width of the OIII $5007\ \AA$ line, we conclude that Alcyoneus is a LERG.
Moreover, the WISE photometry \textcolor{blue}{\citep{Cutri12012}} at $11.6\ \mu\mathrm{m}$ and $22.1\ \mu\mathrm{m}$ is below the instrumental detection limit.
Following the classification rule of \textcolor{blue}{\citet{Gurkan12014}} based on the $22.1\ \mu\mathrm{m}$ luminosity density, we affirm that Alcyoneus is a LERG.
Through automated classification, \textcolor{blue}{\citet{Best12012}} came to the same conclusion.\\
Being a jet-mode AGN, the supermassive black hole (SMBH) in the centre of Alcyoneus' host galaxy presumably accretes at an efficiency below $1\%$ of the Eddington limit, and is fueled mainly by slowly cooling hot gas.


\subsection{Projected proper length}
We calculate Alcyoneus' projected proper length $l_\mathrm{p}$ through its angular length $\phi$ and spectroscopic redshift $z_\mathrm{spec}$.
We formally determine $\phi = 20.8' \pm 0.15'$ from the compact-source--subtracted $90''$ image (top panel of \textbf{Figure}~\ref{fig:doubleConeModel}) by selecting the largest great-circle distance between all possible pairs of pixels with a specific intensity higher than three sigma-clipped standard deviations above the sigma-clipped median.
We find $l_\mathrm{p} = 4.99 \pm 0.04\ \mathrm{Mpc}$; this makes Alcyoneus the projectively largest radio galaxy known.
For methodology details, and for a probabilistic comparison between the projected proper lengths of Alcyoneus and J1420-0545, see \textbf{Appendix}~\ref{ap:comparisonMachalski}.

\subsection{Radio luminosity densities and kinetic jet powers}
\label{sec:fluxDensities}
\begin{figure}
    \centering
    \includegraphics[width=\columnwidth]{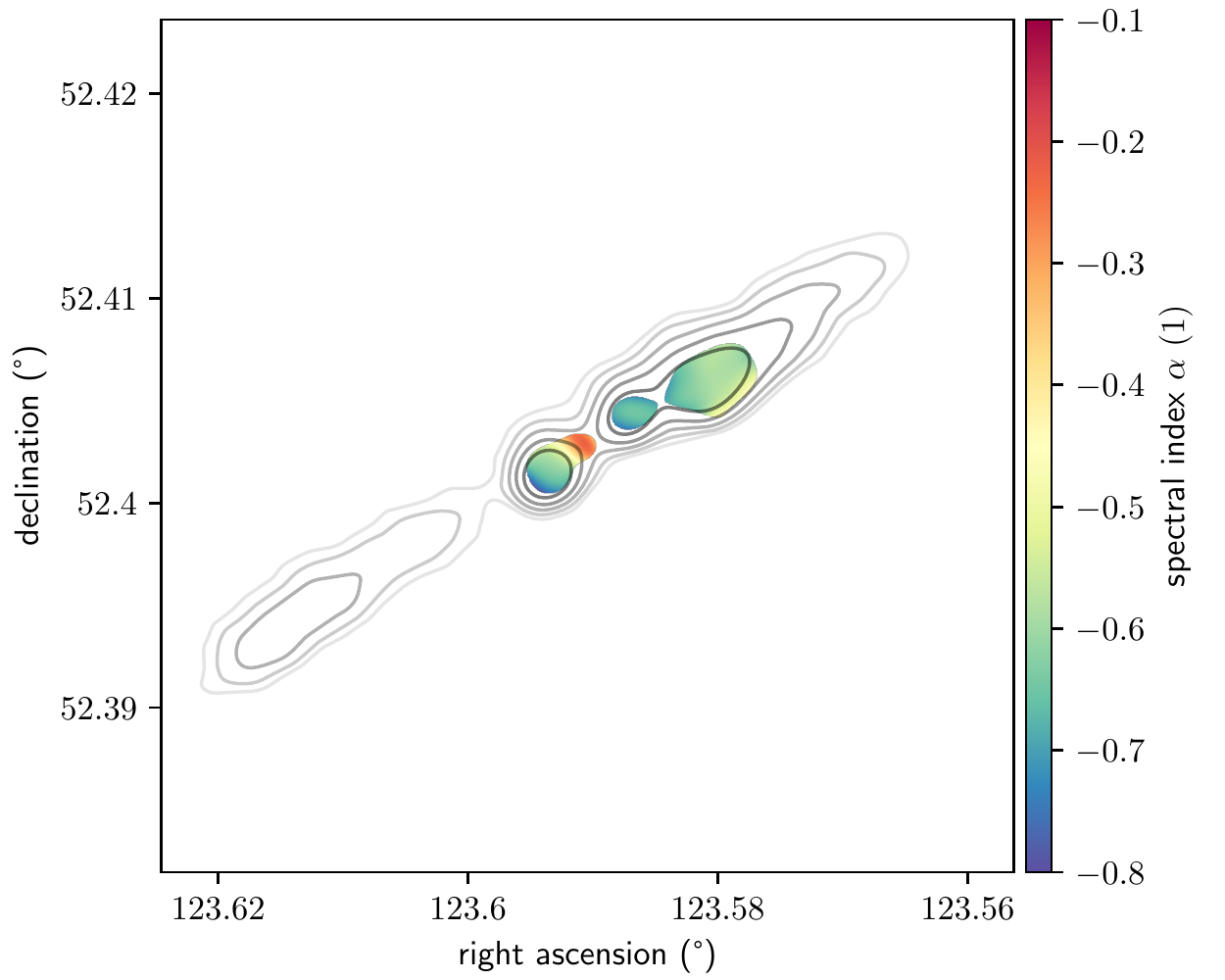}
    \caption{
    \textbf{The LoTSS--VLASS spectral index map reveals Alcyoneus' flat-spectrum core and steeper-spectrum jets.}
    We show all directions where both the LoTSS and VLASS image have at least $5\sigma$ significance.
    In black, we overlay the same LoTSS contours as in \textbf{Figures}~\ref{fig:AlcyoneusHostLegacy} and \ref{fig:AlcyoneusContours}.
    The core spectral index is $\alpha = -0.25 \pm 0.1$ and the combined inner jet spectral index is $\alpha = -0.65 \pm 0.1$.
    }
    \label{fig:spectralIndex}
\end{figure}

\begin{figure}
    \centering
    \includegraphics[width=\columnwidth]{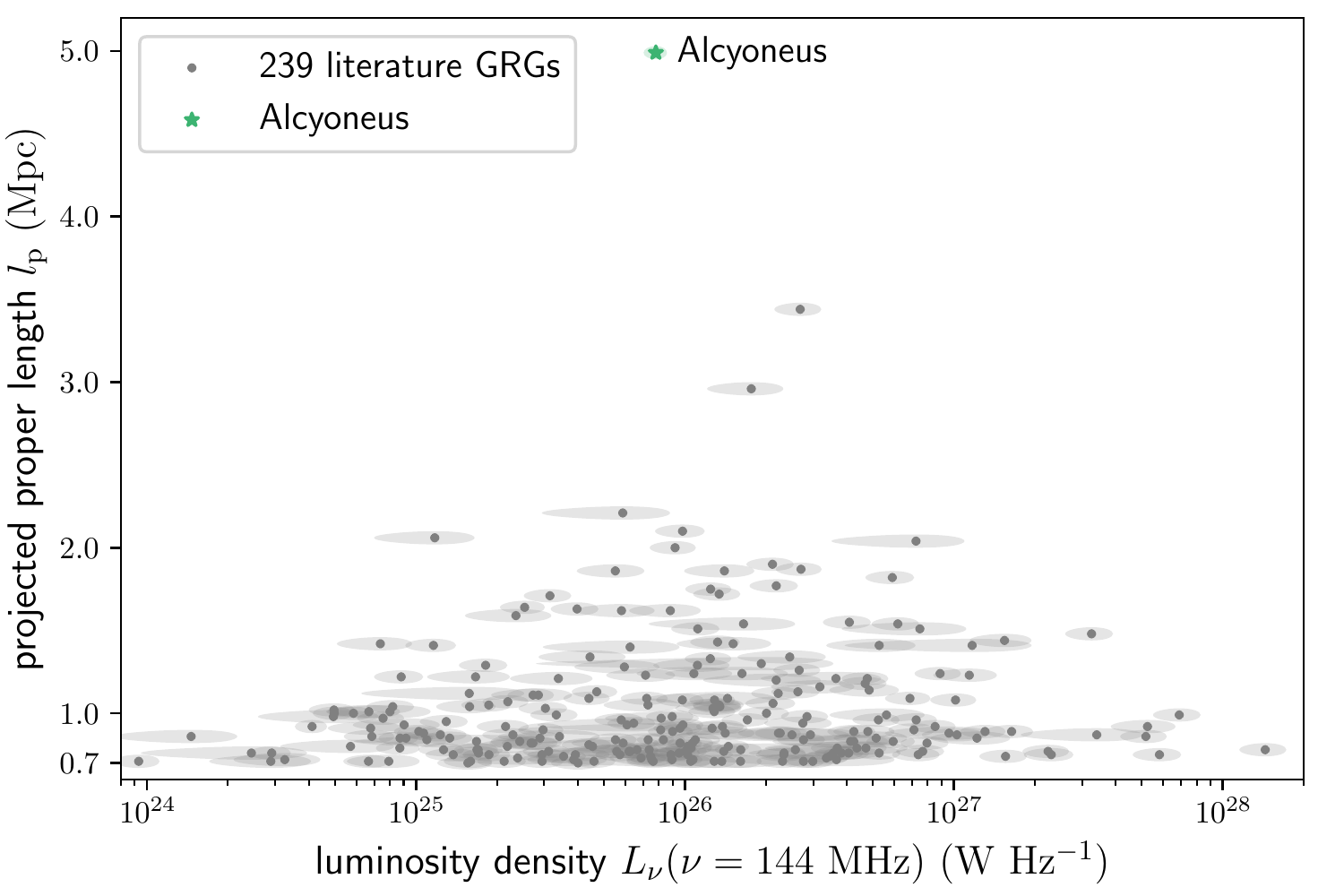}
    \caption{
    \textbf{Alcyoneus has a low-frequency luminosity density typical for GRGs.}
    We explore the relation between GRG projected proper length $l_\mathrm{p}$ and total luminosity density $L_\nu$ at rest-frame frequency $\nu = 144\ \mathrm{MHz}$.
    Total luminosity densities include contributions from all available radio galaxy components (i.e. the core, jets, hotspots and lobes).
    Literature GRGs are from \textcolor{blue}{\citet{Dabhade12020March}}, and are marked with grey disks, while Alcyoneus is marked with a green star.
    Translucent ellipses indicate -1 to +1 standard deviation uncertainties.
    Alcyoneus has a typical luminosity density (percentile $45 \pm 3\%$).
    }
    \label{fig:luminosityDensities}
\end{figure}

From the LoTSS DR2 $6''$ image (top panel of \textbf{Figure}~\ref{fig:AlcyoneusContours}), we measure that two northern jet local maxima occur at angular distances of $9.2 \pm 0.2''$ and $23.7 \pm 0.2''$ from the host, or at projected proper distances of $36.8 \pm 0.8\ \mathrm{kpc}$ and $94.8 \pm 0.8\ \mathrm{kpc}$.
Two southern jet local maxima occur at angular distances of $8.8 \pm 0.2''$ and $62.5 \pm 0.2''$ from the host, or at projected proper distances of $35.2 \pm 0.8\ \mathrm{kpc}$ and $249.9 \pm 0.8\ \mathrm{kpc}$.\\
At the central observing frequency of $\nu_\mathrm{c} = 144\ \mathrm{MHz}$, the northern jet has a flux density $F_\nu = 193 \pm 20\ \mathrm{mJy}$, the southern jet has $F_\nu = 110 \pm 12\ \mathrm{mJy}$, whilst the northern lobe has $F_\nu = 63 \pm 7\ \mathrm{mJy}$ and the southern lobe has $F_\nu = 44 \pm 5\ \mathrm{mJy}$.
To minimise contamination from fore- and background galaxies, we determined the lobe flux densities from the compact-source--subtracted $90''$ image.
The flux density uncertainties are dominated by the $10\%$ flux scale uncertainty inherent to the LoTSS DR2 \textcolor{blue}{\citep{Shimwell12021}}.
The host galaxy flux density is relatively weak, and the corresponding emission has, at $\nu_\mathrm{c} = 144\ \mathrm{MHz}$ and $6''$ resolution, no clear angular separation from the inner jets' emission; we have therefore not determined it.\\
Due to cosmological redshifting, the conversion between flux density and luminosity density depends on the spectral indices $\alpha$ of Alcyoneus' luminous components.
We estimate the spectral indices of the core and jets from the LoTSS DR2 $6''$ and VLASS $2.2''$ images.
After convolving the VLASS image with a Gaussian to the common resolution of $6''$, we calculate the mean spectral index between LoTSS' $\nu_\mathrm{c} = 144\ \mathrm{MHz}$ and VLASS' $\nu_\mathrm{c} = 2.99\ \mathrm{GHz}$.
Using only directions for which both images have a significance of at least $5\sigma$, we deduce a core spectral index $\alpha = -0.25 \pm 0.1$ and a combined inner jet spectral index $\alpha = -0.65 \pm 0.1$.
The spectral index uncertainties are dominated by the LoTSS DR2 and VLASS flux scale uncertainties.
We show the full spectral index map in \textbf{Figure}~\ref{fig:spectralIndex}.
\noindent
We have not determined the spectral index of the lobes, as they are only detected in the LoTSS imagery.\\
The luminosity densities of the northern and southern jet at \emph{rest-frame} frequency $\nu = 144\ \mathrm{MHz}$ are $L_\nu = \left(3.6 \pm 0.4\right) \cdot 10^{25}\ \mathrm{W\ Hz^{-1}}$ and $L_\nu = \left(2.0 \pm 0.2\right) \cdot 10^{25}\ \mathrm{W\ Hz^{-1}}$, respectively.
Following \textcolor{blue}{\citet{Dabhade12020October}}, we estimate the kinetic power of the jets from their luminosity densities and the results of the simulation-based analytical model of \textcolor{blue}{\citet{Hardcastle12018}}.
We find $Q_\mathrm{jet,1} = 1.2 \pm 0.1 \cdot 10^{36}\ \mathrm{W}$ and $Q_\mathrm{jet,2} = 6.6 \pm 0.7 \cdot 10^{35}\ \mathrm{W}$, so that the total kinetic jet power is $Q_\mathrm{jets} \coloneqq Q_\mathrm{jet,1} + Q_\mathrm{jet,2} = 1.9 \pm 0.2 \cdot 10^{36}\ \mathrm{W}$.
Interestingly, this total kinetic jet power is \emph{lower} than the average $Q_\mathrm{jets} = 3.7 \cdot 10^{36}\ \mathrm{W}$, and close to the median $Q_\mathrm{jets} = 2.2 \cdot 10^{36}\ \mathrm{W}$, for low-excitation giant radio galaxies (LEGRGs) in the redshift range $0.18 < z < 0.43$ \textcolor{blue}{\citep{Dabhade12020October}}.\\
Because the lobe spectral indices are unknown, we present luminosity densities for several possible values of $\alpha$ in \textbf{Table}~\ref{tab:luminosityDensities}.\footnote{The inferred luminosity densities have a cosmology-dependence; our results are ${\sim}6\%$ higher than for modern high-$H_0$ cosmologies.}
(Because of electron ageing, $\alpha$ will decrease further away from the core.)
\begin{center}
\captionof{table}{Luminosity densities $L_\nu$ (in $10^{24}\ \mathrm{W\ Hz^{-1}}$) of Alcyoneus' lobes for three potential spectral indices $\alpha$ at \emph{rest-frame} frequency $\nu = 144\ \mathrm{MHz}$, assuming a \textcolor{blue}{\citet{PlanckCollaboration12020}} cosmology.
}
 \begin{tabular}{c | c c c} 
 & $\alpha = -0.8$ & $\alpha = -1.2$ & $\alpha = -1.6$\\
 [3pt] \hline
Northern lobe & $12 \pm 1$ & $13 \pm 1$ & $14 \pm 1$ \\
 [3pt]
Southern lobe & $8.3 \pm 0.8$ & $9.0 \pm 0.9$ & $9.9 \pm 1$ \\
\end{tabular}
\label{tab:luminosityDensities}
\end{center}
Assuming $\alpha = -1.2$, Alcyoneus total luminosity density at $\nu = 144\ \mathrm{MHz}$ is $L_\nu = 7.8 \pm 0.8 \cdot 10^{25}\ \mathrm{W\ Hz^{-1}}$.
In \textbf{Figure}~\ref{fig:luminosityDensities}, we compare this estimate to other GRGs' total luminosity density at the same frequency, as found by \textcolor{blue}{\citet{Dabhade12020March}} through the LoTSS DR1 \textcolor{blue}{\citep{Shimwell12019}}.
Interestingly, Alcyoneus is not particularly luminous: it has a low-frequency luminosity density typical for the currently known GRG population (percentile $45 \pm 3\%$).

\subsection{True proper length: relativistic beaming}
Following \textcolor{blue}{\citet{Hardcastle11998a}}, we simultaneously constrain Alcyoneus' jet speed $u$ and inclination angle $\theta$ from the jets' flux density asymmetry: the northern-to-southern jet flux density ratio $J = 1.78 \pm 0.3$.\footnote{Because $J$ is obtained through division of two independent normal random variables (RVs) with non-zero mean, $J$ is an RV with an uncorrelated noncentral normal ratio distribution.}
We assume that the jets propagate with identical speeds $u$ in exactly opposing directions (making angles with the line-of-sight $\theta$ and $\theta + 180\degree$), and have statistically identical relativistic electron populations, so that they have a common synchrotron spectral index $\alpha$.
Using $\alpha = -0.65 \pm 0.1$ as before, and
\begin{align}
    \beta \coloneqq \frac{u}{c};\ \ \beta \cos{\theta} = \frac{J^\frac{1}{2-\alpha} - 1}{J^\frac{1}{2-\alpha} + 1},
\end{align}
we find $\beta \cos{\theta} = 0.106 \pm 0.03$.
Because $\cos{\theta} \leq 1$, $\beta$ is bounded from below by $\beta_\mathrm{min} = 0.106 \pm 0.03$.\\
From detailed modelling of ten Fanaroff--Riley (FR) I radio galaxies (which have jet luminosities comparable to Alcyoneus'), \textcolor{blue}{\citet{Laing12014}} deduced that initial jet speeds are roughly $\beta = 0.8$, which decrease until roughly $0.6\ r_0$, with $r_0$ being the recollimation distance.
Most of \textcolor{blue}{\citet{Laing12014}}'s ten recollimation distances are between $5$ and $15\ \mathrm{kpc}$, with the largest being that of NGC 315: $r_0 = 35\ \mathrm{kpc}$.
Because the local specific intensity maxima in Alcyoneus' jets closest to the host occur at \emph{projected} proper distances of $36.8 \pm 0.8\ \mathrm{kpc}$ and $35.2 \pm 0.8\ \mathrm{kpc}$, the true proper distances must be even larger.
We conclude that the observed jet emission presumably comes from a region further from the host than $r_0$, so that the initial stage of jet deceleration --- in which the jet speed is typically reduced by several tens of percents of $c$ --- must already be completed.
Thus, $\beta_\mathrm{max} = 0.8$ is a safe upper bound.\\
Taking $\beta_\mathrm{max} = 0.8$, $\theta$ is bounded from above by $\theta_\mathrm{max} = 82.4 \pm 2\degree$ ($\theta \in [0, 90\degree]$), or bounded from below by $180\degree - \theta_\mathrm{max} = 97.6 \pm 2\degree$ ($\theta \in [90\degree, 180\degree]$).\footnote{Taking $\beta_\mathrm{max} = 1$ instead, $\theta$ is bounded from above by $\theta_\mathrm{max} = 83.9 \pm 2\degree$ ($\theta \in [0, 90\degree]$), or bounded from below by $180\degree - \theta_\mathrm{max} = 96.1 \pm 2\degree$ ($\theta \in [90\degree, 180\degree]$).}
If we model Alcyoneus' geometry as a line segment, and assume no jet reorientation, Alcyoneus' true proper length $l$ and projected proper length $l_\mathrm{p}$ relate as
\begin{align}
    l = \frac{l_\mathrm{p}}{\sin{\theta}};\ \ l \geq l_\mathrm{min} = \frac{l_\mathrm{p}}{\sin{\theta_\mathrm{max}}}.
\end{align}
We bound $l$ from below: $l_\mathrm{min} = 5.04 \pm 0.05\ \mathrm{Mpc}$.
A \emph{triangular} prior on $\beta$ between $\beta_\mathrm{min}$ and $\beta_\mathrm{max}$ with the mode at $\beta_\mathrm{max}$ induces a skewed prior on $l$; the 90\% credible interval is $l \in [5.0\ \mathrm{Mpc}, 5.5\ \mathrm{Mpc}]$, with the mean and median being $5.2\ \mathrm{Mpc}$ and $5.1\ \mathrm{Mpc}$, respectively.
A \emph{flat} prior on $\beta$ between $\beta_\mathrm{min}$ and $\beta_\mathrm{max}$ also induces a skewed prior on $l$; the 90\% credible interval is $l \in [5.0\ \mathrm{Mpc}, 7.1\ \mathrm{Mpc}]$, with the mean and median being $5.6\ \mathrm{Mpc}$ and $5.1\ \mathrm{Mpc}$, respectively.
The median of $l$ seems particularly well determined, as it is insensitive to variations of the prior on $\beta$.\\
In \textbf{Appendix}~\ref{ap:challengers}, we explore the inclination angle conditions under which Alcyoneus has the largest true proper length of all known (> 4 Mpc) GRGs.

\subsection{Stellar and supermassive black hole mass}
\begin{figure}
    \centering
    \begin{subfigure}{\columnwidth}
    \includegraphics[width=\linewidth]{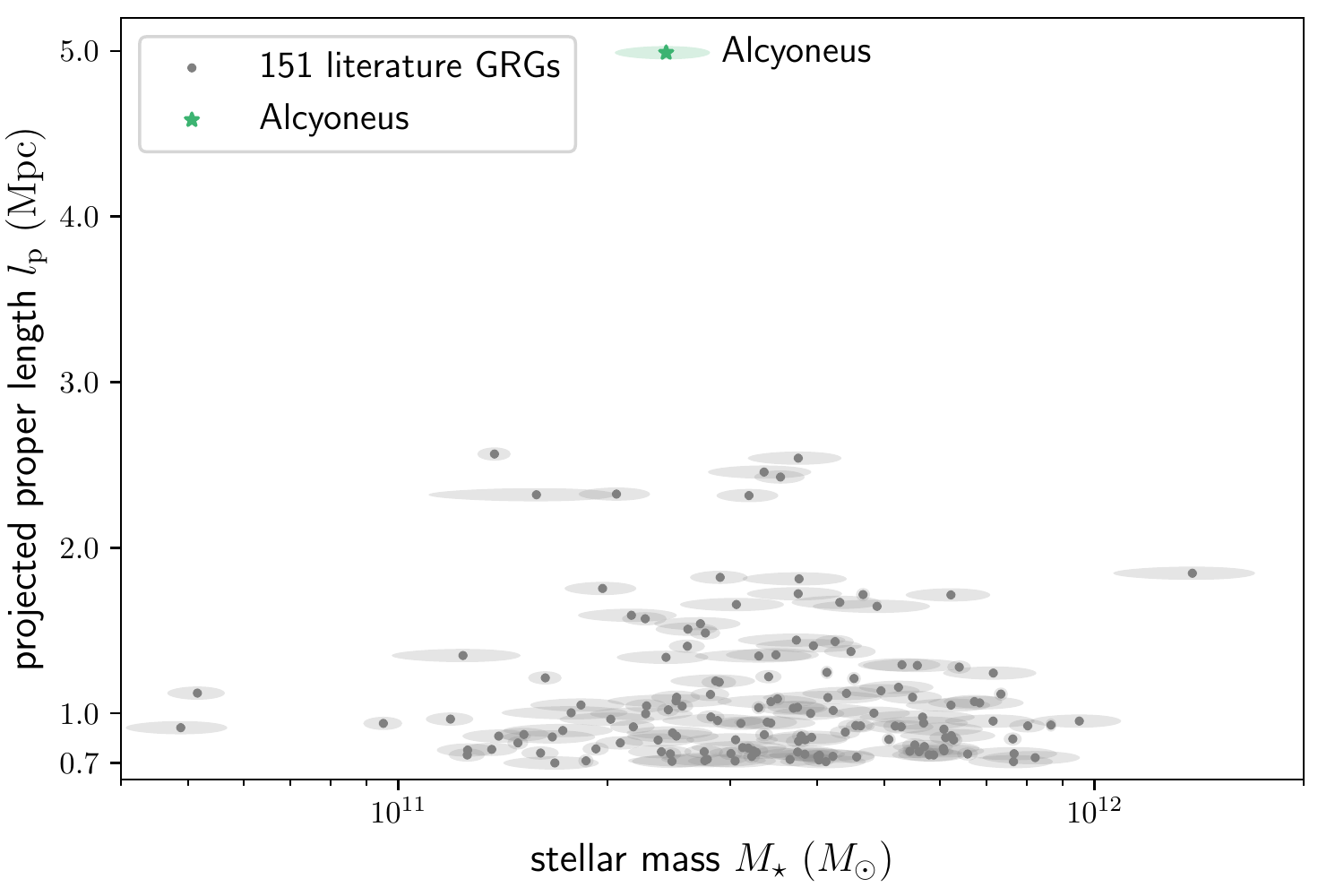}
    \end{subfigure}
    \begin{subfigure}{\columnwidth}
    \includegraphics[width=\linewidth]{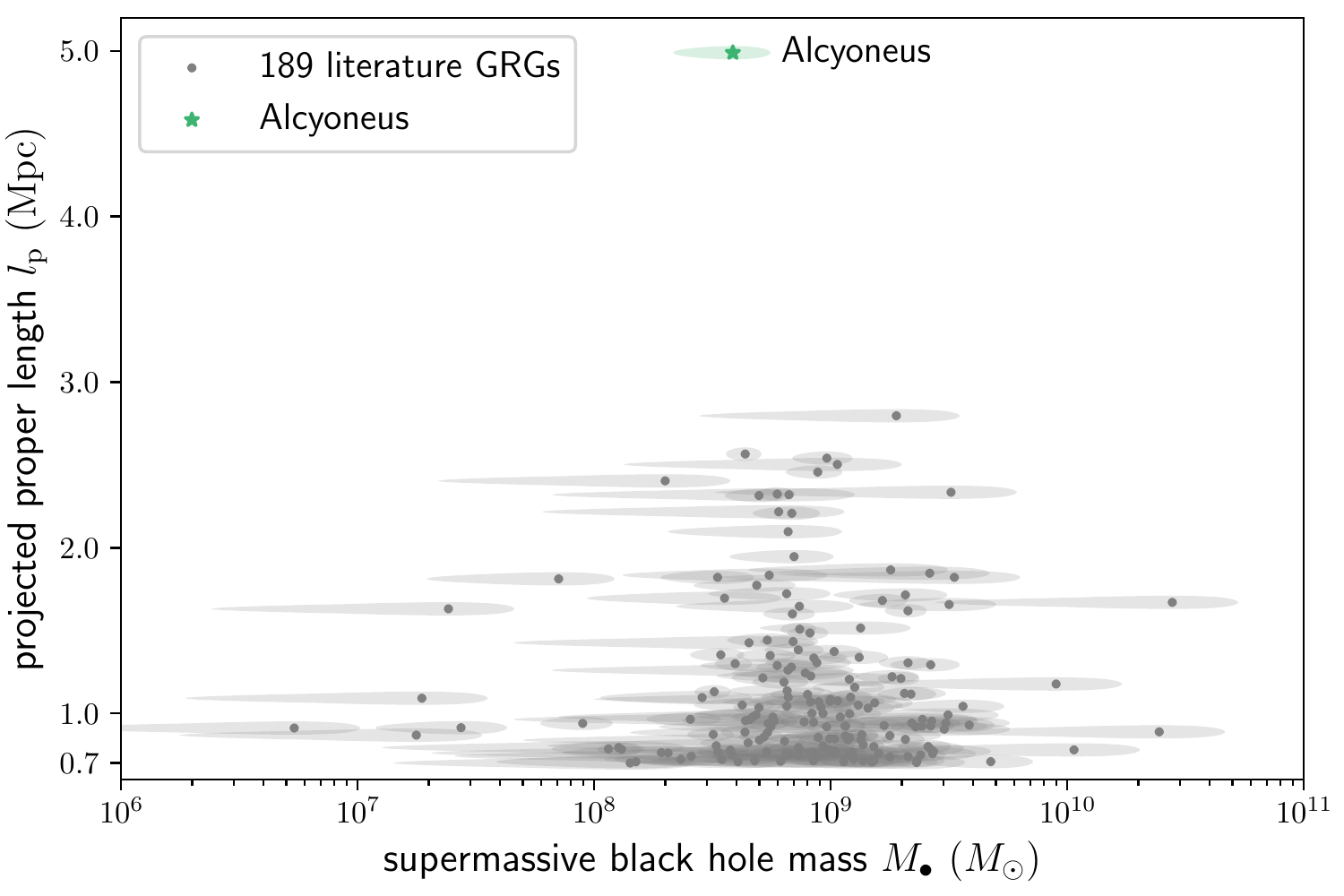}
    \end{subfigure}
    \caption{
    \textbf{Alcyoneus' host has a lower stellar and supermassive black hole mass than most GRG hosts.}
    We explore the relations between GRG projected proper length $l_\mathrm{p}$ and host galaxy stellar mass $M_\star$ (top panel) or host galaxy supermassive black hole mass $M_\bullet$ (bottom panel).
    Our methods allow determining these properties for a small proportion of all literature GRGs only.
    Literature GRGs are marked with grey disks, while Alcyoneus is marked with a green star.
    Translucent ellipses indicate -1 to +1 standard deviation uncertainties.
    Alcyoneus' host has a fairly typical --- though below-average --- stellar mass (percentile $25 \pm 9\%$) and supermassive black hole mass (percentile $23 \pm 11\%$).}
    \label{fig:scatterLengthProjectedMasses}
\end{figure}\noindent
Does a galaxy or its central black hole need to be massive in order to generate a GRG?\\
Alcyoneus' host has a stellar mass $M_\star = 2.4 \pm 0.4 \cdot 10^{11}\ M_\odot$ \textcolor{blue}{\citep{Chang12015}}.
We test whether or not this is a typical stellar mass among the total known GRG population.
We assemble a literature catalogue of 1013 GRGs by merging the compendium of \textcolor{blue}{\citet{Dabhade12020October}}, which is complete up to April 2020, with the GRGs discovered in \textcolor{blue}{\citet{Galvin12020}},
\textcolor{blue}{\citet{IshwaraChandra12020}}, \textcolor{blue}{\citet{Tang12020}}, \textcolor{blue}{\citet{Bassani12021}}, \textcolor{blue}{\citet{Bruggen12021}}, \textcolor{blue}{\citet{Delhaize12021}}, \textcolor{blue}{\citet{Masini12021}}, \textcolor{blue}{\citet{Kuzmicz12021}},
\textcolor{blue}{\citet{Andernach12021}} and \textcolor{blue}{\citet{Mahato12021}}.
We collect stellar masses with uncertainties from \textcolor{blue}{\citet{Chang12015}}, which are based on SDSS and WISE photometry, and from \textcolor{blue}{\citet{Salim12018}}, which are based on GALEX, SDSS and WISE photometry.
We give precedence to the stellar masses by \textcolor{blue}{\citet{Salim12018}} when both are available.
We obtain stellar masses for 151 previously known GRGs.
The typical stellar mass range is $10^{11}$ -- $10^{12}\ M_\odot$, the median $M_\star = 3.5 \cdot 10^{11}\ M_\odot$ and the mean $M_\star = 3.8 \cdot 10^{11}\ M_\odot$.
Strikingly, the top panel of \textbf{Figure}~\ref{fig:scatterLengthProjectedMasses} illustrates that Alcyoneus' host has a fairly \emph{low} (percentile $25 \pm 9\%$) stellar mass compared with the currently known population of GRG hosts.\\
For the GRGs in our literature catalogue, we also estimate SMBH masses via the M-sigma relation.
We collect SDSS DR12 stellar velocity dispersions with uncertainties \textcolor{blue}{\citep{Alam12015}}, and apply the M-sigma relation of \textbf{Equation}~7 in \textcolor{blue}{\citet{Kormendy12013}}.
Alcyoneus' host has a SMBH mass $M_\bullet = 3.9 \pm 1.7 \cdot 10^8\ M_\odot$.
We obtain SMBH masses for 189 previously known GRGs.
The typical SMBH mass range is $10^8$ -- $10^{10}\ M_\odot$, the median $M_\bullet = 7.9 \cdot 10^8\ M_\odot$ and the mean $M_\bullet = 1.5 \cdot 10^9\ M_\odot$.
Strikingly, the bottom panel of \textbf{Figure}~\ref{fig:scatterLengthProjectedMasses} illustrates that Alcyoneus' host has a fairly \emph{low} (percentile $23 \pm 11\%$) SMBH mass compared with the currently known population of GRG hosts.\\
We note that Alcyoneus is the only GRG with $l_\mathrm{p} > 3\ \mathrm{Mpc}$ whose host's stellar mass is known through \textcolor{blue}{\citet{Chang12015}} or \textcolor{blue}{\citet{Salim12018}}, and whose host's SMBH mass can be estimated through its SDSS DR12 velocity dispersion.
These data allow us to state confidently that exceptionally high stellar or SMBH masses are not necessary to generate 5-Mpc--scale GRGs.

\subsection{Surrounding large-scale structure}
\begin{figure}
\begin{subfigure}{\columnwidth}
\centering
    \includegraphics[width=\linewidth]{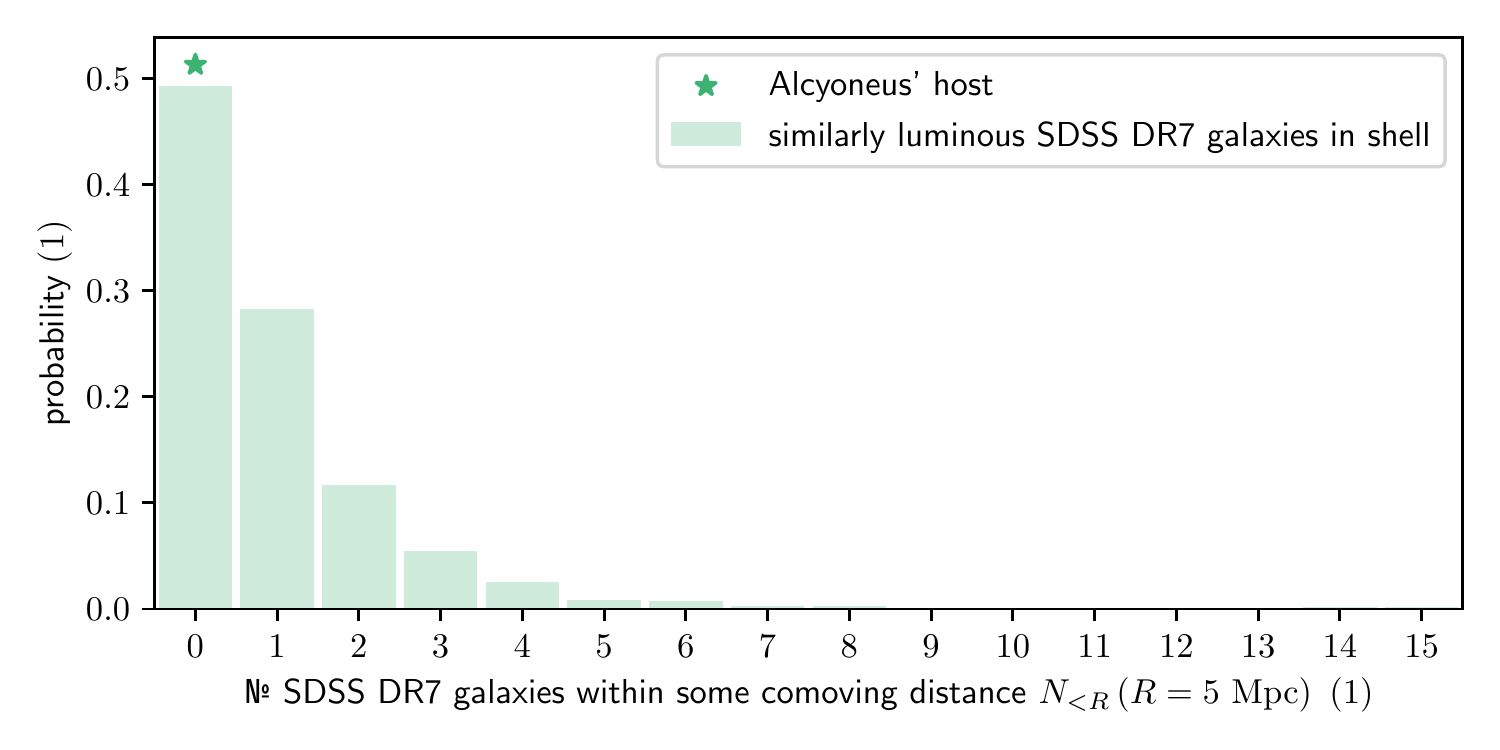}
\end{subfigure}
\begin{subfigure}{\columnwidth}
\centering
    \includegraphics[width=\linewidth]{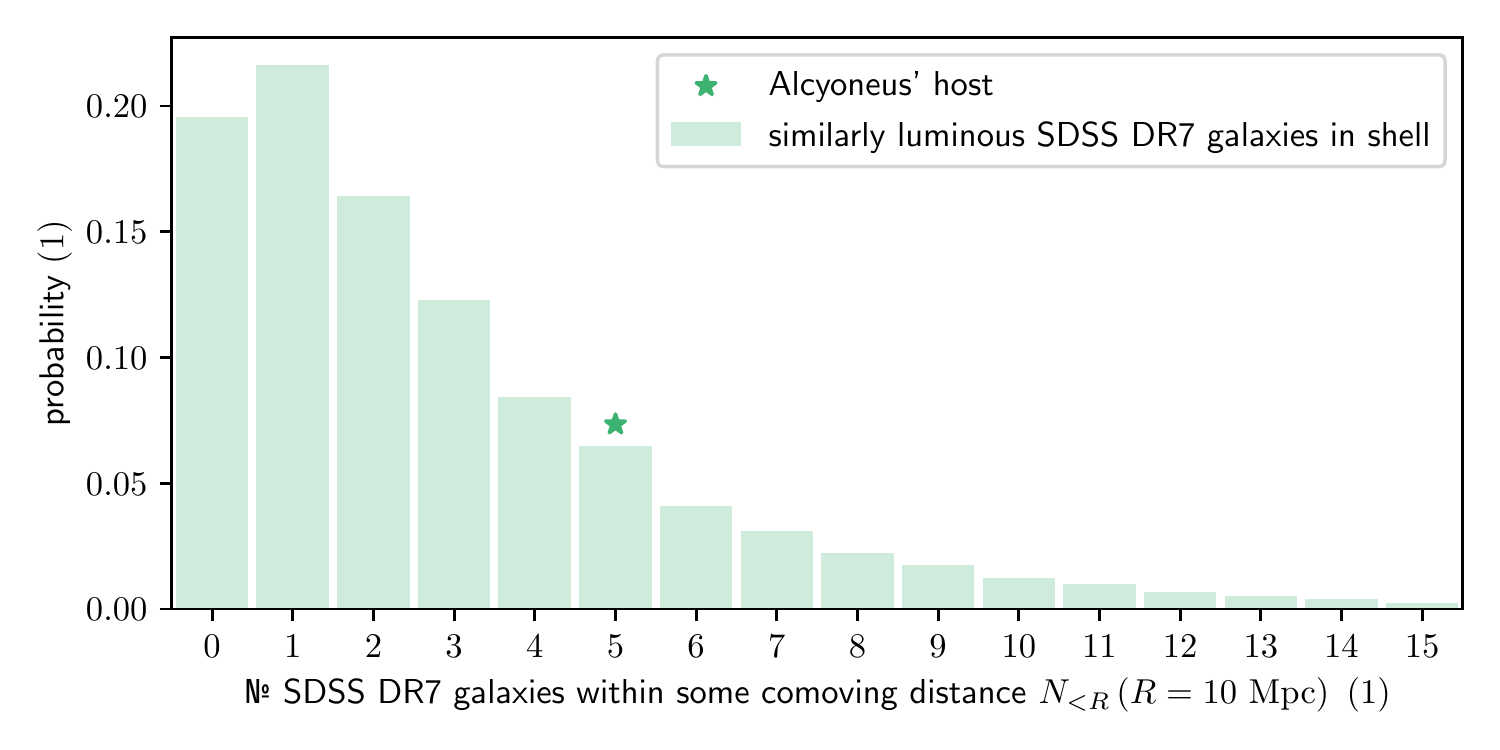}
\end{subfigure}
    \caption{
    \textbf{Like most galaxies of similar $r$-band luminosity density and redshift, Alcyoneus' host has no galactic neighbours in SDSS DR7 within 5 Mpc.
    However, within 10 Mpc, Alcyoneus' host has \emph{more} neighbours than most similar galaxies.}
    For all 9,358 SDSS DR7 galaxies with an $r$-band luminosity density between 75\% and 125\% that of Alcyoneus' host and a comoving radial distance that differs at most $r_0 = 25\ \mathrm{Mpc}$ from Alcyoneus', we count the number of SDSS DR7 galaxies $N_{<R}\left(R\right)$ within a sphere of comoving radius $R = 5\ \mathrm{Mpc}$ (top panel) and $R = 10\ \mathrm{Mpc}$ (bottom panel).
    The top panel indicates that Alcyoneus does not inhabit a galaxy cluster; the bottom panel indicates that Alcyoneus does not inhabit a void.}
    \label{fig:numberOfGalaxies}
\end{figure}

Several approaches to large-scale structure (LSS) classification, such as the T-web scheme \textcolor{blue}{\citep{Hahn12007}}, partition the modern Universe into galaxy clusters, filaments, sheets and voids.
In this section, we determine Alcyoneus' most likely environment type.\\
We conduct a tentative quantitative analysis using the SDSS DR7 spectroscopic galaxy sample \textcolor{blue}{\citep{Abazajian12009}}.
Does Alcyoneus' host have fewer, about equal or more galactic neighbours in SDSS DR7 than a randomly drawn galaxy of similar $r$-band luminosity density and redshift?
Let $r\left(z\right)$ be the comoving radial distance corresponding to cosmological redshift $z$.
We consider a spherical shell with the observer at the centre, inner radius $\max{\{r(z = z_\mathrm{spec}) - r_0, 0\}}$ and outer radius $r(z = z_\mathrm{spec}) + r_0$.
We approximate Alcyoneus' cosmological redshift with $z_\mathrm{spec}$ and choose $r_0 = 25\ \mathrm{Mpc}$.
As all galaxies in the spherical shell have a similar distance to the observer (i.e. distances are at most $2r_0$ different), the SDSS DR7 galaxy number density \emph{completeness} must also be similar throughout the spherical shell.\footnote{For $r_0 = 25\ \mathrm{Mpc}$, this is a good approximation, because the shell is cosmologically thin: $2r_0 = 50\ \mathrm{Mpc}$ roughly amounts to the length of a \emph{single} Cosmic Web filament.}
For each enclosed galaxy with an $r$-band luminosity density between $1-\delta$ and $1+\delta$ times that of Alcyoneus' host, we count the number of SDSS DR7 galaxies $N_{<R}\left(R\right)$ within a sphere of comoving radius $R$ around it --- regardless of luminosity density, and excluding itself.
Alcyoneus' host has an SDSS $r$-band apparent magnitude $m_r = 18.20$; the corresponding luminosity density is $L_\nu\left(\lambda_\mathrm{c} = 623.1\ \mathrm{nm}\right) = 3.75 \cdot 10^{22}\ \mathrm{W\ Hz^{-1}}$.
We choose $\delta = 0.25$; this yields 9,358 such enclosed galaxies.\\
In \textbf{Figure}~\ref{fig:numberOfGalaxies}, we show the distribution of $N_{<R}\left(R\right)$ for $R = 5\ \mathrm{Mpc}$ and $R=10\ \mathrm{Mpc}$.
\noindent
We verify that the distributions are insensitive to reasonable changes in $r_0$ and $\delta$.
Note that there is no SDSS DR7 galaxy within a comoving distance of 5 Mpc from Alcyoneus' host.
The nearest such galaxy, J081323.49+524856.1, occurs at a comoving distance of 7.9 Mpc: the nearest ${\sim}2,000\ \mathrm{Mpc}^3$ of comoving space are free of galactic neighbours with $L_\nu\left(\lambda_\mathrm{c}\right) > 5.57 \cdot 10^{22}\ \mathrm{W\ Hz^{-1}}$.\footnote{This is the luminosity density that corresponds to the SDSS $r$-band apparent magnitude completeness limit $m_r = 17.77$ \textcolor{blue}{\citep{Strauss12002}}.}
In the same way as in \textbf{Section}~\ref{sec:radioMorphology}, we verify that the DESI Legacy Imaging Surveys DR9, RASS and PSZ2 do not contain evidence for a galaxy cluster in the direction of Alcyoneus' host.
The nearest galaxy cluster, according to the SDSS-III cluster catalogue of \textcolor{blue}{\citet{Wen12012}}, instead lies $24'$ away at right ascension $123.19926\degree$, declination $52.72468\degree$ and photometric redshift $z_\mathrm{ph} = 0.2488$.
It has an $R_{200} = 1.1\ \mathrm{Mpc}$ and, according to the DESI cluster catalogue of \textcolor{blue}{\citet{Zou12021}}, a total mass $M = 2.2 \cdot 10^{14}\ M_\odot$.
The comoving distance between the cluster and Alcyoneus' host is $11\ \mathrm{Mpc}$.
All in all, we conclude that Alcyoneus does not reside in a galaxy cluster.
Meanwhile, there are five SDSS DR7 galaxies within a comoving distance of 10 Mpc from Alcyoneus' host: this makes it implausible that Alcyoneus lies in a void.
Finally, one could interpret $N_{<R}\left(R\right)$ as a proxy for the LSS total matter density around a galaxy.
For $R = 10\ \mathrm{Mpc}$, just 17\% of galaxies in the shell with a similar luminosity density as Alcyoneus' host have a higher LSS total matter density.
Being on the high end of the density distribution, but lying outside a cluster, Alcyoneus most probably inhabits a filament of the Cosmic Web.

\subsection{Proper lobe volumes}
\label{sec:BayesianModel}
\begin{figure}
    \centering
    \begin{subfigure}{\columnwidth}
    \includegraphics[width=\columnwidth]{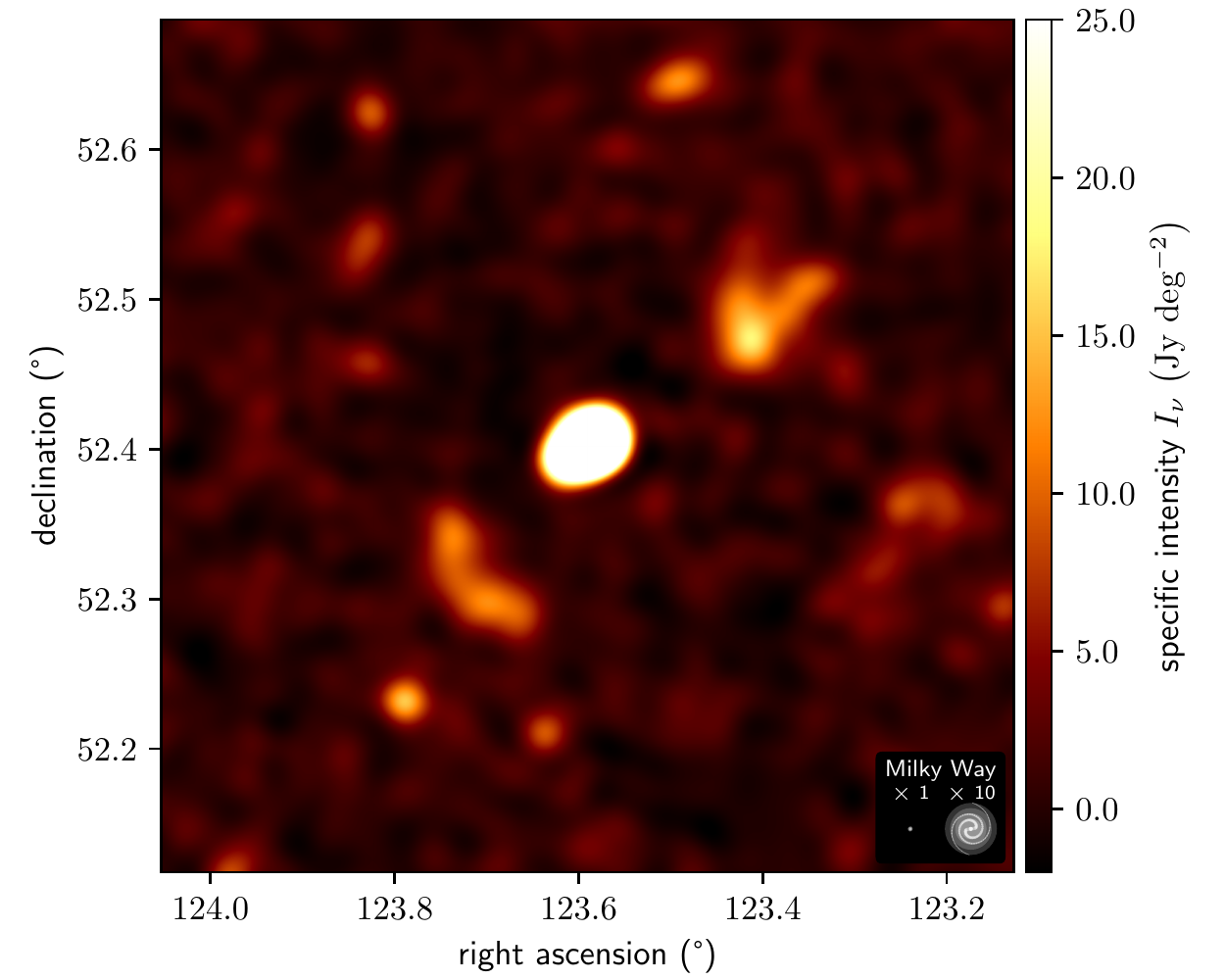}
    \end{subfigure}
    \begin{subfigure}{\columnwidth}
    \includegraphics[width=\columnwidth]{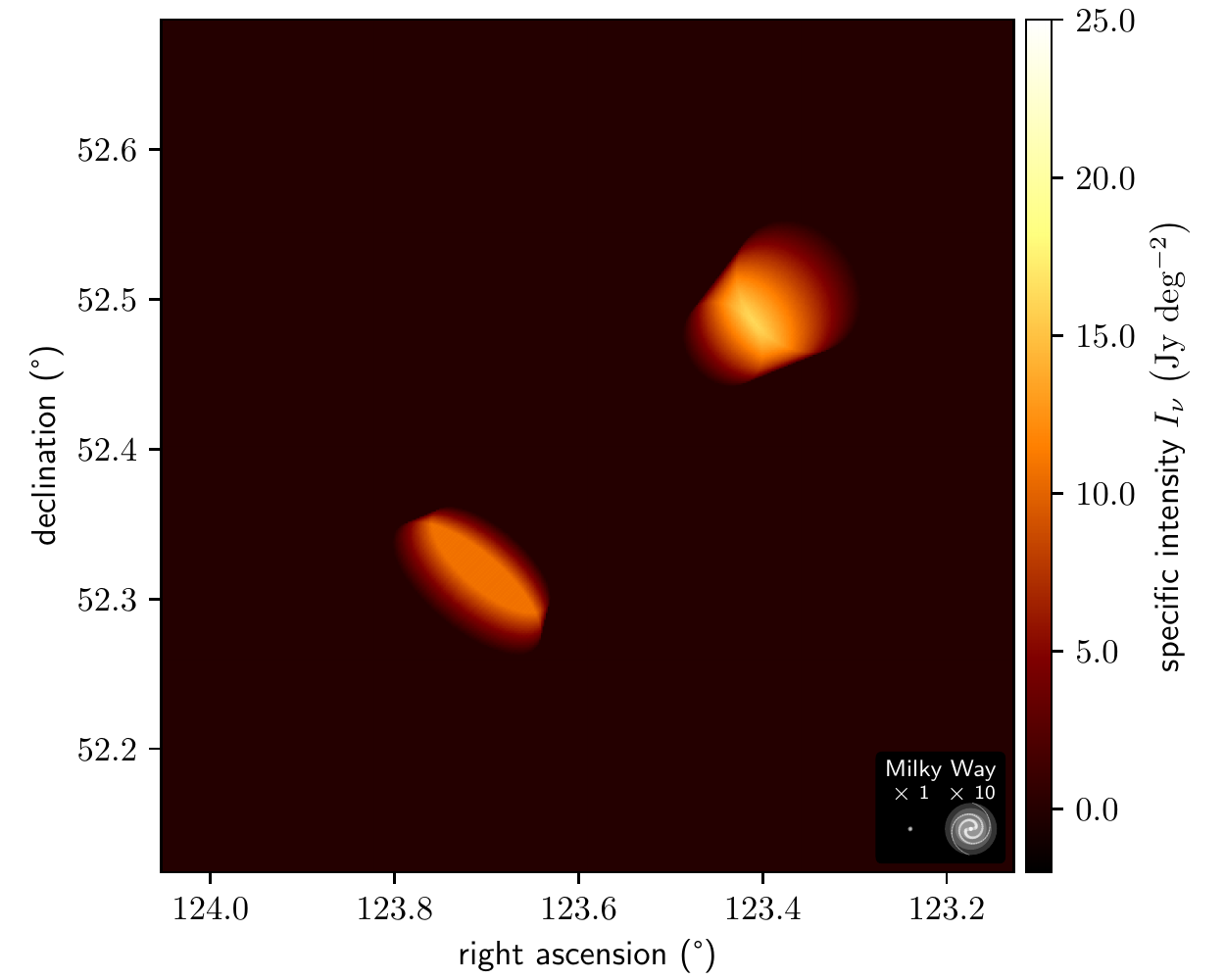}
    \end{subfigure}
    \begin{subfigure}{\columnwidth}
    \includegraphics[width=\columnwidth]{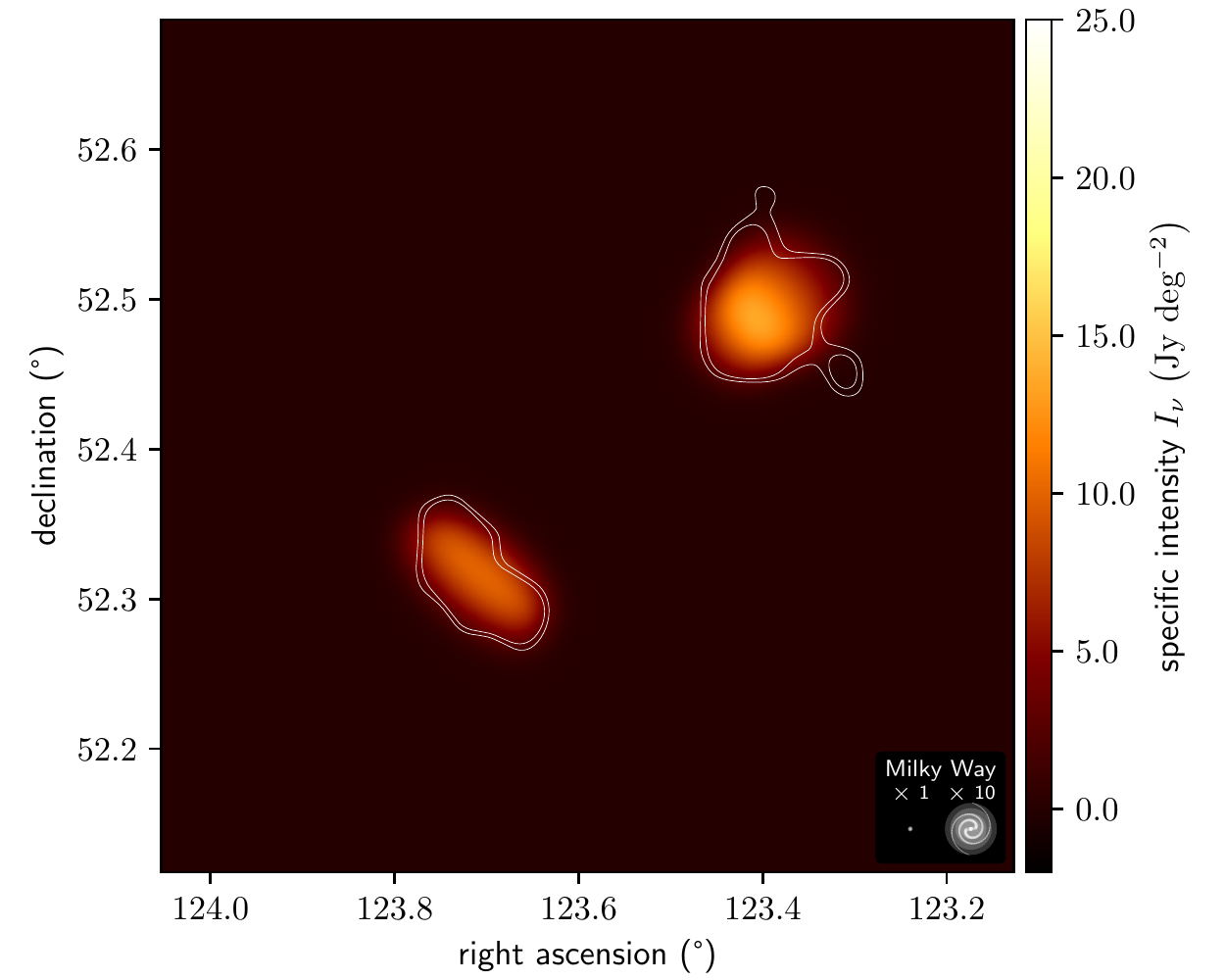}
    \end{subfigure}
    \caption{
    \textbf{Alcyoneus' lobe volumes can be estimated by comparing the observed radio image to modelled radio images.}
    \textbf{Top:}
    LoTSS DR2 compact-source-subtracted $90''$ image of Alcyoneus.
    For scale, we show the stellar Milky Way disk (diameter: 50 kpc) and a 10 times inflated version; the spiral galaxy shape follows \textcolor{blue}{\citet{Ringermacher12009}}.
    \textbf{Middle:} Highest-likelihood model image.
    \textbf{Bottom:}
    The same model image convolved to $90''$ resolution, with $2\sigma$ and $3\sigma$ contours of the observed image overlaid.
    }
    \label{fig:doubleConeModel}
\end{figure}

We determine the proper volumes of Alcyoneus' lobes with a new Bayesian model.
The model describes the lobes through a pair of doubly truncated, optically thin cones, each of which has a spatially constant and isotropic monochromatic emission coefficient (MEC) \textcolor{blue}{\citep{Rybicki11986}}.
We allow the 3D orientations and opening angles of the cones to differ, as the lobes may traverse their way through differently pressured parts of the warm--hot intergalactic medium (WHIM): e.g. the medium near the filament axis, and the medium near the surrounding voids.
By adopting a \emph{spatially constant} MEC, we neglect electron density and magnetic field inhomogeneities as well as spectral-ageing gradients; by adopting an \emph{isotropic} MEC, we assume non-relativistic velocities within the lobe so that beaming effects are negligible.
Numerically, we first generate the GRG's 3D MEC field over a cubical voxel grid, and then calculate the corresponding model image through projection, including expansion-related cosmological effects.
Before comparison with the observed image, we convolve the model image with a Gaussian kernel to the appropriate resolution.
We exploit the approximately Gaussian LoTSS DR2 image noise to formulate the likelihood, and assume a flat prior distribution over the parameters.
Using a Metropolis--Hastings (MH) Markov chain Monte Carlo (MCMC), we sample from the posterior distribution.\footnote{For a detailed description of the model parameters, the MH MCMC and formulae for derived quantities, see \textbf{Appendix}~\ref{ap:doubleConeModel}.}\\
In the top panel of \textbf{Figure}~\ref{fig:doubleConeModel}, we show the LoTSS DR2 compact-source-subtracted $90''$ image of Alcyoneus.
The central region has been excluded from source subtraction, and hence Alcyoneus' core and jets remain.
(However, when we run our MH MCMC on this image, we do mask this central region.)
In the middle panel, we show the highest-likelihood (and thus maximum a posteriori (MAP)) model image \emph{before} convolution.
In the bottom panel, we show the same model image convolved to $90''$ resolution, with $2\sigma$ and $3\sigma$ contours of the observed image overlaid.
We provide the full parameter set that corresponds with this model in \textbf{Table}~\ref{tab:MAPEstimates}.\\
The posterior mean, calculated through the MH MCMC samples after burn-in, suggests the following geometry.
The northern lobe has an opening angle $\gamma_1 = 10 \pm 1\degree$, and the cone truncates at an inner distance $d_\mathrm{i,1} = 2.6 \pm 0.2\ \mathrm{Mpc}$ and at an outer distance $d_\mathrm{o,1} = 4.0 \pm 0.2\ \mathrm{Mpc}$ from the host galaxy.
The southern lobe has a larger opening angle $\gamma_2 = 26 \pm 2\degree$, but its cone truncates at smaller distances of $d_\mathrm{i,2} = 1.5 \pm 0.1\ \mathrm{Mpc}$ and $d_\mathrm{o,2} = 2.0 \pm 0.1\ \mathrm{Mpc}$ from the host galaxy.
These parameters fix the proper volumes of Alcyoneus' northern and southern lobes.
We find $V_1 = 1.5 \pm 0.2\ \mathrm{Mpc}^3$ and $V_2 = 1.0 \pm 0.2\ \mathrm{Mpc}^3$, respectively (see \textbf{Equation}~\ref{eq:modelVolume}).\footnote{As a sanity check, we compare our results to those from a less rigorous, though simpler ellipsoid-based method of estimating volumes.
By fitting ellipses to \textbf{Figure}~\ref{fig:doubleConeModel}'s top panel image, one obtains a semi-minor and semi-major axis; the half-diameter along the ellipsoid's third dimension is assumed to be their mean.
This method suggests a northern lobe volume $V_1 = 1.4 \pm 0.3\ \mathrm{Mpc}^3$ and a southern lobe volume $V_2 = 1.1 \pm 0.3\ \mathrm{Mpc}^3$.
These results agree well with our Bayesian model results.
(If the half-diameter along the third dimension is instead treated as an RV with a uniform distribution between the semi-minor axis and the semi-major axis, the estimates remain the same.)}\\
How are the lobes oriented?
\textbf{Figure}~\ref{fig:Alcyoneus} provides a visual hint that the lobes are subtly non-coaxial.
The posterior indicates that the position angles of the northern and southern lobes are $\varphi_1 = 307 \pm 1\degree$ and $\varphi_2 = 139 \pm 2\degree$, respectively.
The position angle difference is thus $\Delta\varphi = 168 \pm 2\degree$: although close to $\Delta\varphi = 180\degree$, we can reject coaxiality with high significance.
Interestingly, the posterior also constrains the angles that the lobe axes make with the plane of the sky: $\vert\theta_1-90\degree\vert = 51 \pm 2\degree$ and $\vert\theta_2-90\degree\vert = 18 \pm 7\degree$.
Again, the uncertainties imply that the lobes are probably \emph{not} coaxial.
We stress that these inclination angle results are tentative only.
Future model extensions should explore how sensitive they are to the assumed lobe geometry (by testing other shapes than just truncated cones, such as ellipsoids).\\
One way to validate the model is to compare the observed lobe flux densities of \textbf{Section}~\ref{sec:fluxDensities} to the predicted lobe flux densities.
According to the posterior, the MECs of the northern and southern lobes are $j_{\nu,1} = 17 \pm 2\ \mathrm{Jy\ deg^{-2}\ Mpc^{-1}}$ and $j_{\nu,2} = 18 \pm 3\ \mathrm{Jy\ deg^{-2}\ Mpc^{-1}}$.
Combining MECs and volumes, we predict northern and southern lobe flux densities $F_{\nu,1}(\nu_\mathrm{c}) = 63 \pm 4\ \mathrm{mJy}$ and $F_{\nu,2}(\nu_\mathrm{c}) = 45 \pm 5\ \mathrm{mJy}$ (see \textbf{Equation}~\ref{eq:modelFluxDensity}).
We find excellent agreement: the relative differences with the observed results are $0\%$ and $2\%$, respectively.
\noindent

\subsection{Lobe pressures and the local WHIM}
\begin{figure}
    \centering
    \begin{subfigure}{\columnwidth}
    \includegraphics[width=\linewidth]{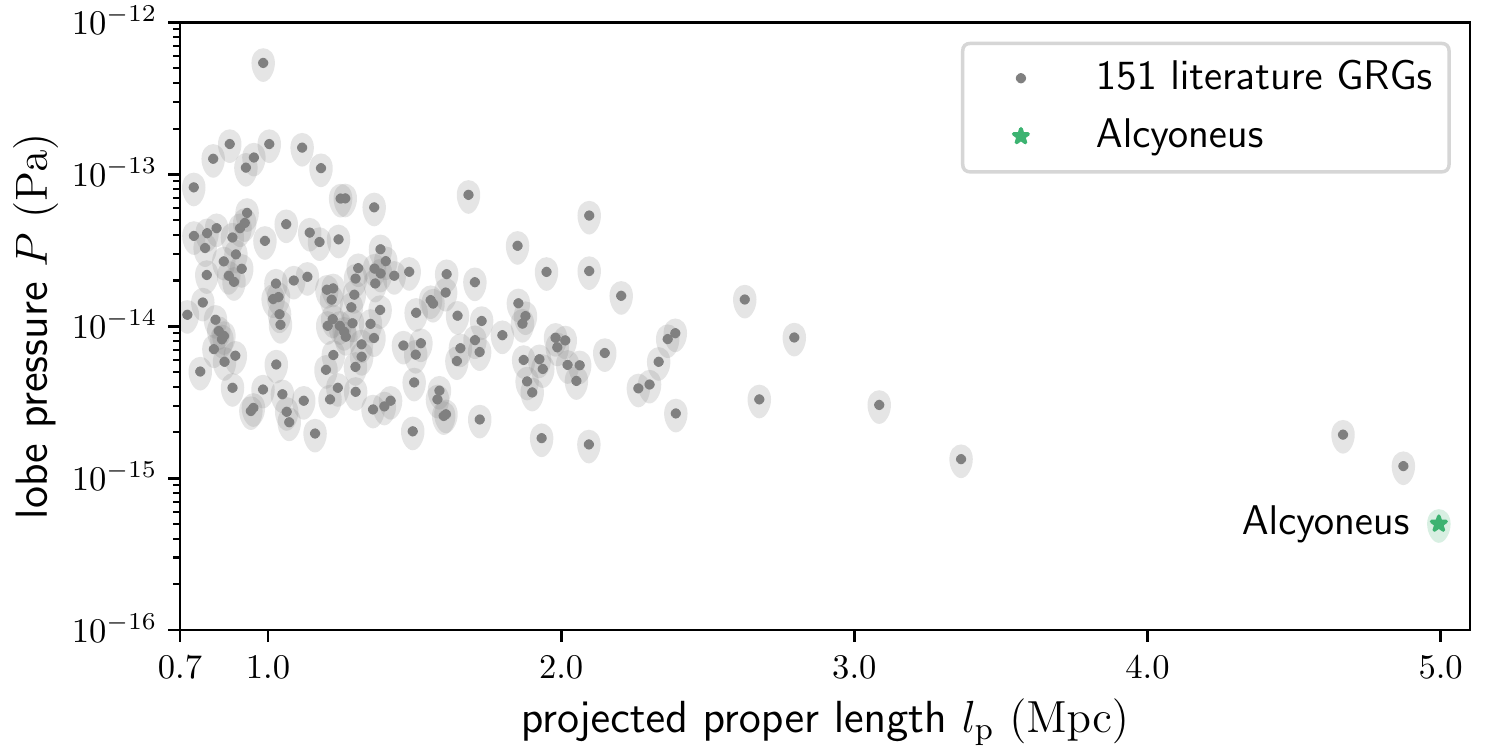}
    \end{subfigure}
    \begin{subfigure}{\columnwidth}
    \includegraphics[width=\linewidth]{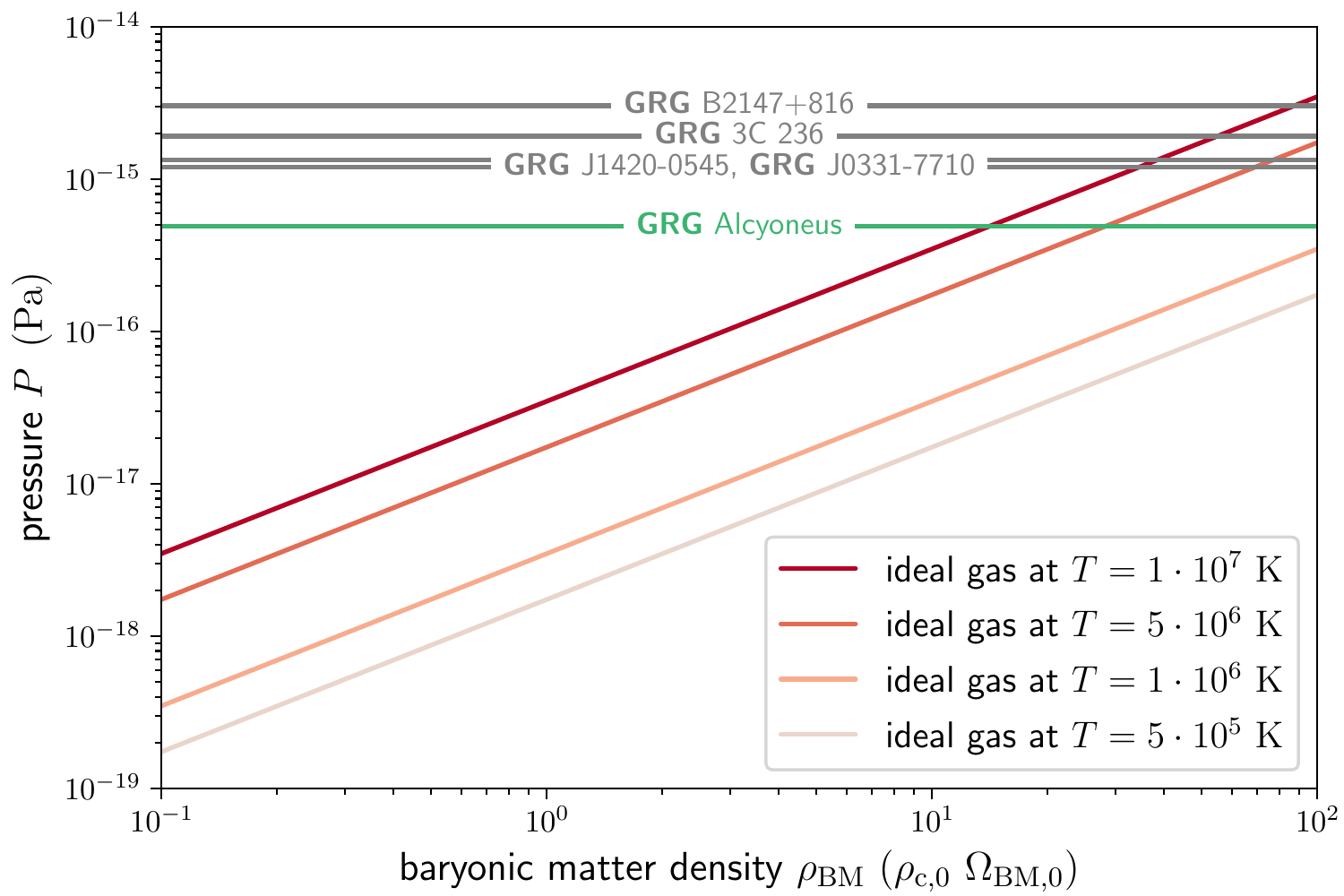}
    \end{subfigure}
    \caption{
    \textbf{Of all GRGs with known lobe pressures, Alcyoneus is the most plausible candidate for pressure equilibrium with the WHIM.}
    In the top panel, we explore the relation between length and lobe pressure for Alcyoneus and 151 literature GRGs.
    In the bottom panel, we compare the lobe pressure of Alcyoneus (green line) with the lobe pressures of the largest four similarly analysed GRGs (grey lines) and with WHIM pressures (red lines).
    }
    \label{fig:pressure}
\end{figure}

From Alcyoneus' lobe flux densities and volumes, we can infer lobe pressures and magnetic field strengths.
We calculate these through \texttt{pysynch}\footnote{The \texttt{pysynch} code is publicly available online: \textcolor{blue}{\url{https://github.com/mhardcastle/pysynch}}.} \textcolor{blue}{\citep{Hardcastle11998b}}, which uses the formulae first proposed by \textcolor{blue}{\citet{Myers11985}} and reexamined by \textcolor{blue}{\citet{Beck12005}}.
Following the notation of \textcolor{blue}{\citet{Hardcastle11998b}}, we assume that the electron energy distribution is a power law in Lorentz factor $\gamma$ with $\gamma_\mathrm{min} = 10$, $\gamma_\mathrm{max} = 10^4$ and exponent $p = -2$; we also assume that the kinetic energy density of protons is vanishingly small compared with that of electrons ($\kappa = 0$), and that the plasma filling factor is unity ($\phi = 1$).
Assuming the minimum-energy condition \textcolor{blue}{\citep{Burbidge11956}}, we find minimum-energy pressures $P_\mathrm{min,1} = 4.8 \pm 0.3 \cdot 10^{-16}\ \mathrm{Pa}$ and $P_\mathrm{min,2} = 4.9 \pm 0.6 \cdot 10^{-16}\ \mathrm{Pa}$ for the northern and southern lobes, respectively.
The corresponding minimum-energy magnetic field strengths are $B_\mathrm{min,1} = 46 \pm 1\ \mathrm{pT}$ and $B_\mathrm{min,2} = 46 \pm 3\ \mathrm{pT}$.
Assuming the equipartition condition \textcolor{blue}{\citep{Pacholczyk11970}}, we find equipartition pressures $P_\mathrm{eq,1} = 4.9 \pm 0.3 \cdot 10^{-16}\ \mathrm{Pa}$ and $P_\mathrm{eq,2} = 4.9 \pm 0.6 \cdot 10^{-16}\ \mathrm{Pa}$ for the northern and southern lobes, respectively.
The corresponding equipartition magnetic field strengths are $B_\mathrm{eq,1} = 43 \pm 2\ \mathrm{pT}$ and $B_\mathrm{eq,2} = 43 \pm 2\ \mathrm{pT}$.
The minimum-energy and equipartition results do not differ significantly.\\
From pressures and volumes, we estimate the internal energy of the lobes $E = 3 P V$.
We find $E_\mathrm{min,1} = 6.2 \pm 0.5 \cdot 10^{52}\ \mathrm{J}$, $E_\mathrm{min,2} = 4.3 \pm 0.6 \cdot 10^{52}\ \mathrm{J}$, $E_\mathrm{eq,1} = 6.3 \pm 0.5 \cdot 10^{52}\ \mathrm{J}$ and $E_\mathrm{eq,2} = 4.4 \pm 0.6 \cdot 10^{52}\ \mathrm{J}$.
Next, we can bound the ages of the lobes from below by neglecting synchrotron losses, and assuming that the jets have been injecting energy in the lobes continuously at the currently observed kinetic jet powers.
Using $\Delta t = E Q_\mathrm{jet}^{-1}$, we find $\Delta t_\mathrm{min,1} = 1.7 \pm 0.2\ \mathrm{Gyr}$, $\Delta t_\mathrm{min,2} = 2.1 \pm 0.4\ \mathrm{Gyr}$, and identical results when assuming the equipartition condition.
Finally, we can obtain a rough estimate of the average expansion speed of the radio galaxy during its lifetime $u = l_\mathrm{p} (\Delta t)^{-1}$.
We find $u = 2.6 \pm 0.3 \cdot 10^3\ \mathrm{km\ s^{-1}}$, or about $1\%$ of the speed of light.\\
Several other authors \textcolor{blue}{\citep{Andernach11992, Lacy11993, Subrahmanyan11996, Parma11996, Mack11998, Schoenmakers11998, Schoenmakers12000, IshwaraChandra11999, Lara12000, Machalski12000, Machalski12001, Saripalli12002, Jamrozy12005, Subrahmanyan12006, Subrahmanyan12008, Saikia12006, Machalski12006July1, Machalski12007, Machalski12008, Safouris12009, Malarecki12013, Tamhane12015, Sebastian12018, Heesen12018, Cantwell12020}} have estimated the minimum-energy or equipartition pressure of the lobes of GRGs embedded in non-cluster environments (i.e. in voids, sheets or filaments of the Cosmic Web).
We compare Alcyoneus to the other 151 GRGs with known lobe pressures in the top panel of \textbf{Figure}~\ref{fig:pressure}.\footnote{We have included all publications that provide pressures, energy densities or magnetic field strengths.
Note that some authors assume $\gamma_\mathrm{min} = 1$, we assume $\gamma_\mathrm{min} = 10$ and \textcolor{blue}{\citet{Malarecki12013}} assume $\gamma_\mathrm{min} = 10^3$.
If possible, angular lengths were updated using the LoTSS DR2 at $6''$ and redshift estimates were updated using the SDSS DR12.
All projected proper lengths have been recalculated using our \textcolor{blue}{\citet{PlanckCollaboration12020}} cosmology.
When authors provided pressures for both lobes, we have taken the average.}
\noindent
Alcyoneus reaffirms the negative correlation between length and lobe pressure \textcolor{blue}{\citep{Jamrozy12002, Machalski12006July2}}, and has the lowest lobe pressures found thus far.
Alcyoneus' lobe pressure is in fact so low, that it is comparable to the pressure in dense and hot parts of the WHIM: for a baryonic matter (BM) density $\rho_\mathrm{WHIM} = 10\ \rho_\mathrm{c,0} \Omega_\mathrm{BM,0}$ and $T_\mathrm{WHIM} = 10^7\ \mathrm{K}$, $P_\mathrm{WHIM} = 4 \cdot 10^{-16}\ \mathrm{Pa}$.
Here, $\rho_\mathrm{c,0}$ is today's critical density, so that $\rho_\mathrm{c,0} \Omega_\mathrm{BM,0}$ is today's mean baryon density.
See the bottom panel of \textbf{Figure}~\ref{fig:pressure} for a more extensive comparison between $P_\mathrm{min}$ (green line) and $P_\mathrm{WHIM}$ (red lines).
For comparison, we also show the lobe pressures of the four other thus-analysed GRGs with $l_\mathrm{p} > 3\ \mathrm{Mpc}$ (grey lines).
These are J1420-0545 of $l_\mathrm{p} = 4.9\ \mathrm{Mpc}$ \textcolor{blue}{\citep{Machalski12008}}, 3C 236 of $l_\mathrm{p} = 4.7\ \mathrm{Mpc}$ \textcolor{blue}{\citep{Schoenmakers12000}}, J0331-7710 of $l_\mathrm{p} = 3.4\ \mathrm{Mpc}$ \textcolor{blue}{\citep{Malarecki12013}} and B2147+816 of $l_\mathrm{p} = 3.1\ \mathrm{Mpc}$ \textcolor{blue}{\citep{Schoenmakers12000}}.\\
Although proposed as probes of WHIM thermodynamics for decades, the bottom panel of \textbf{Figure}~\ref{fig:pressure} demonstrates that even the largest non-cluster literature GRGs are unlikely to be in pressure equilibrium with their environment.
Relying on results from the Overwhelmingly Large Simulations (OWLS) \textcolor{blue}{\citep{Schaye12010}}, \textcolor{blue}{\citet{Malarecki12013}} point out that baryon densities $\rho_\mathrm{BM} > 50\ \rho_\mathrm{c,0}\ \Omega_\mathrm{BM,0}$, which are necessary for pressure equilibrium in these GRGs (see the intersection of grey and red lines in the bottom panel of \textbf{Figure}~\ref{fig:pressure}), occur in only $1\%$ of the WHIM's volume.
By contrast, Alcyoneus can be in pressure equilibrium with the WHIM at baryon densities $\rho_\mathrm{BM} \sim 20\ \rho_\mathrm{c,0}\ \Omega_\mathrm{BM,0}$, and thus represents the most promising intergalactic barometer of its kind yet.\footnote{At Alcyoneus' redshift, this density amounts to a baryon overdensity of ${\sim}10$.}\\
Why do most, if not all, observed non-cluster GRGs have overpressured lobes?
The top panel of \textbf{Figure}~\ref{fig:pressure} suggests that GRGs must grow to several Mpc to approach WHIM pressures in their lobes, and such GRGs are rare.
However, the primary reason is the limited surface brightness sensitivity of all past and current surveys.
Alcyoneus' lobes are visible in the LoTSS, but not in the NRAO VLA Sky Survey (NVSS) \textcolor{blue}{\citep{Condon11998}} or in the Westerbork Northern Sky Survey (WENSS) \textcolor{blue}{\citep{Rengelink11997}}.
Their pressure approaches that of the bulk of the WHIM within an order of magnitude.
Lobes with even lower pressure must be less luminous or more voluminous, and thus will have even lower surface brightness.
It is therefore probable that most GRG lobes that are in true pressure equilibrium with the WHIM still lie hidden in the radio sky.

\section{Conclusion}
\label{sec:conclusion}
   \begin{enumerate}
   \item We reprocess the LoTSS DR2, the latest version of the LOFAR's Northern Sky survey at 144 MHz, by subtracting angularly compact sources and imaging at $60''$ and $90''$ resolution.
   The resulting images \textcolor{blue}{\citep{Oei12022}} allow us to explore a new sensitivity regime for radio galaxy lobes, and thus represent promising data to search for unknown GRGs of large angular length.
   We present a sample in forthcoming work.
      \item We discover the first 5 Mpc GRG, which we dub \textit{Alcyoneus}.
      The projected proper length is $l_\mathrm{p} = 4.99 \pm 0.04\ \mathrm{Mpc}$, while the true proper length is at least $l_\mathrm{min} = 5.04 \pm 0.05\ \mathrm{Mpc}$.
      We confidently associate the $20.8' \pm 0.15'$ radio structure to an elliptical galaxy with a jet-mode AGN detected in the DESI Legacy Imaging Surveys DR9: the SDSS DR12 source J081421.68+522410.0 at J2000 right ascension $123.590372\degree$, declination $52.402795\degree$ and spectroscopic redshift $0.24674 \pm 6 \cdot 10^{-5}$.
      \item
      Alcyoneus has a total luminosity density at $\nu = 144\ \mathrm{MHz}$ of $L_\nu = 8 \pm 1 \cdot 10^{25}\ \mathrm{W\ Hz^{-1}}$, which is typical for GRGs (percentile $45 \pm 3\%$).
      Alcyoneus' host has a fairly low stellar mass and SMBH mass compared with other GRG hosts (percentiles $25 \pm 9\%$ and $23 \pm 11\%$).
      This implies that --- \emph{within} the GRG population --- no strong positive correlation between radio galaxy length and (instantaneous) low-frequency radio power, stellar mass or SMBH mass can exist.
      \item The surrounding sky as imaged by the LoTSS, DESI Legacy Imaging Surveys, RASS and PSZ suggests that Alcyoneus does \emph{not} inhabit a galaxy cluster.
      According to an SDSS-III cluster catalogue, the nearest cluster occurs at a comoving distance of $11\ \mathrm{Mpc}$.
      A local galaxy number density count suggests that Alcyoneus instead inhabits a filament of the Cosmic Web.
      A low-density environment therefore remains a possible explanation for Alcyoneus' formidable size.
      \item 
      We develop a new Bayesian model that parametrises in three dimensions a pair of arbitrarily oriented, optically thin, doubly truncated conical radio galaxy lobes with constant monochromatic emission coefficient.
      We then generate the corresponding specific intensity function, taking into account cosmic expansion, and compare it to data assuming Gaussian image noise.
      We use Metropolis--Hastings Markov chain Monte Carlo to optimise the parameters, and thus determine northern and southern lobe volumes of $1.5\pm0.2\ \mathrm{Mpc}^3$ and $1.0\pm0.2\ \mathrm{Mpc}^3$, respectively.
      In total, the lobes have an internal energy of ${\sim}10^{53}\ \mathrm{J}$, expelled from the host galaxy over a $\mathrm{Gyr}$-scale period.
      The lobe pressures are $4.8 \pm 0.3 \cdot 10^{-16}\ \mathrm{Pa}$ and $4.9 \pm 0.6 \cdot 10^{-16}\ \mathrm{Pa}$, respectively; these are the lowest measured in radio galaxies yet.
      Nevertheless, the lobe pressures still exceed a large range of plausible WHIM pressures.
      Most likely, the lobes are still expanding --- and Alcyoneus' struggle for supremacy of the Cosmos continues.
   \end{enumerate}

\begin{acknowledgements}
M.S.S.L. Oei warmly thanks Frits Sweijen for coding the very useful \textcolor{blue}{\url{https://github.com/tikk3r/legacystamps}}.\\
M.S.S.L. Oei, R.J. van Weeren and A. Botteon acknowledge support from the VIDI research programme with project number 639.042.729, which is financed by The Netherlands Organisation for Scientific Research (NWO).
M. Br\"uggen acknowledges support from the Deutsche Forschungsgemeinschaft under Germany's Excellence Strategy --- EXC 2121 `Quantum Universe' --- 390833306.
W.L. Williams acknowledges support from the CAS--NWO programme for radio astronomy with project number 629.001.024, which is financed by The Netherlands Organisation for Scientific Research (NWO).\\
The LOFAR is the Low-frequency Array designed and constructed by ASTRON.
It has observing, data processing, and data storage facilities in several countries, which are owned by various parties (each with their own funding sources), and which are collectively operated by the ILT Foundation under a joint scientific policy.
The ILT resources have benefited from the following recent major funding sources: CNRS--INSU, Observatoire de Paris and Université d'Orléans, France; BMBF, MIWF--NRW, MPG, Germany; Science Foundation Ireland (SFI), Department of Business, Enterprise and Innovation (DBEI), Ireland; NWO, The Netherlands; the Science and Technology Facilities Council, UK; Ministry of Science and Higher Education, Poland; the Istituto Nazionale di Astrofisica (INAF), Italy.\\
The National Radio Astronomy Observatory is a facility of the National Science Foundation operated under cooperative agreement by Associated Universities, Inc. CIRADA is funded by a grant from the Canada Foundation for Innovation 2017 Innovation Fund (Project 35999), as well as by the Provinces of Ontario, British Columbia, Alberta, Manitoba and Quebec.\\
Funding for SDSS-III has been provided by the Alfred P. Sloan Foundation, the Participating Institutions, the National Science Foundation, and the U.S. Department of Energy Office of Science. The SDSS-III web site is \textcolor{blue}{\url{http://www.sdss3.org/}}.
SDSS-III is managed by the Astrophysical Research Consortium for the Participating Institutions of the SDSS-III Collaboration including the University of Arizona, the Brazilian Participation Group, Brookhaven National Laboratory, Carnegie Mellon University, University of Florida, the French Participation Group, the German Participation Group, Harvard University, the Instituto de Astrofisica de Canarias, the Michigan State/Notre Dame/JINA Participation Group, Johns Hopkins University, Lawrence Berkeley National Laboratory, Max Planck Institute for Astrophysics, Max Planck Institute for Extraterrestrial Physics, New Mexico State University, New York University, Ohio State University, Pennsylvania State University, University of Portsmouth, Princeton University, the Spanish Participation Group, University of Tokyo, University of Utah, Vanderbilt University, University of Virginia, University of Washington, and Yale University.\\
The Pan-STARRS1 Surveys (PS1) and the PS1 public science archive have been made possible through contributions by the Institute for Astronomy, the University of Hawaii, the Pan-STARRS Project Office, the Max-Planck Society and its participating institutes, the Max Planck Institute for Astronomy, Heidelberg and the Max Planck Institute for Extraterrestrial Physics, Garching, The Johns Hopkins University, Durham University, the University of Edinburgh, the Queen's University Belfast, the Harvard-Smithsonian Center for Astrophysics, the Las Cumbres Observatory Global Telescope Network Incorporated, the National Central University of Taiwan, the Space Telescope Science Institute, the National Aeronautics and Space Administration under Grant No. NNX08AR22G issued through the Planetary Science Division of the NASA Science Mission Directorate, the National Science Foundation Grant No. AST-1238877, the University of Maryland, Eotvos Lorand University (ELTE), the Los Alamos National Laboratory, and the Gordon and Betty Moore Foundation.\\
This publication makes use of data products from the Wide-field Infrared Survey Explorer, which is a joint project of the University of California, Los Angeles, and the Jet Propulsion Laboratory/California Institute of Technology, funded by the National Aeronautics and Space Administration.\\
The Legacy Surveys consist of three individual and complementary projects: the Dark Energy Camera Legacy Survey (DECaLS; Proposal ID \#2014B-0404; PIs: David Schlegel and Arjun Dey), the Beijing--Arizona Sky Survey (BASS; NOAO Prop. ID \#2015A-0801; PIs: Zhou Xu and Xiaohui Fan), and the Mayall z-band Legacy Survey (MzLS; Prop. ID \#2016A-0453; PI: Arjun Dey). DECaLS, BASS and MzLS together include data obtained, respectively, at the Blanco telescope, Cerro Tololo Inter-American Observatory, NSF's NOIRLab; the Bok telescope, Steward Observatory, University of Arizona; and the Mayall telescope, Kitt Peak National Observatory, NOIRLab. The Legacy Surveys project is honored to be permitted to conduct astronomical research on Iolkam Du'ag (Kitt Peak), a mountain with particular significance to the Tohono O'odham Nation.
NOIRLab is operated by the Association of Universities for Research in Astronomy (AURA) under a cooperative agreement with the National Science Foundation.
This project used data obtained with the Dark Energy Camera (DECam), which was constructed by the Dark Energy Survey (DES) collaboration. Funding for the DES Projects has been provided by the U.S. Department of Energy, the U.S. National Science Foundation, the Ministry of Science and Education of Spain, the Science and Technology Facilities Council of the United Kingdom, the Higher Education Funding Council for England, the National Center for Supercomputing Applications at the University of Illinois at Urbana-Champaign, the Kavli Institute of Cosmological Physics at the University of Chicago, Center for Cosmology and Astro-Particle Physics at the Ohio State University, the Mitchell Institute for Fundamental Physics and Astronomy at Texas A\&M University, Financiadora de Estudos e Projetos, Fundacao Carlos Chagas Filho de Amparo, Financiadora de Estudos e Projetos, Fundacao Carlos Chagas Filho de Amparo a Pesquisa do Estado do Rio de Janeiro, Conselho Nacional de Desenvolvimento Cientifico e Tecnologico and the Ministerio da Ciencia, Tecnologia e Inovacao, the Deutsche Forschungsgemeinschaft and the Collaborating Institutions in the Dark Energy Survey. The Collaborating Institutions are Argonne National Laboratory, the University of California at Santa Cruz, the University of Cambridge, Centro de Investigaciones Energeticas, Medioambientales y Tecnologicas-Madrid, the University of Chicago, University College London, the DES-Brazil Consortium, the University of Edinburgh, the Eidgen\"ossische Technische Hochschule (ETH) Z\"urich, Fermi National Accelerator Laboratory, the University of Illinois at Urbana-Champaign, the Institut de Ciencies de l'Espai (IEEC/CSIC), the Institut de Fisica d'Altes Energies, Lawrence Berkeley National Laboratory, the Ludwig Maximilians Universit\"at M\"unchen and the associated Excellence Cluster Universe, the University of Michigan, NSF's NOIRLab, the University of Nottingham, the Ohio State University, the University of Pennsylvania, the University of Portsmouth, SLAC National Accelerator Laboratory, Stanford University, the University of Sussex, and Texas A\&M University.
BASS is a key project of the Telescope Access Program (TAP), which has been funded by the National Astronomical Observatories of China, the Chinese Academy of Sciences (the Strategic Priority Research Program “The Emergence of Cosmological Structures” Grant \# XDB09000000), and the Special Fund for Astronomy from the Ministry of Finance. The BASS is also supported by the External Cooperation Program of Chinese Academy of Sciences (Grant \# 114A11KYSB20160057), and Chinese National Natural Science Foundation (Grant \# 11433005).
The Legacy Survey team makes use of data products from the Near-Earth Object Wide-field Infrared Survey Explorer (NEOWISE), which is a project of the Jet Propulsion Laboratory/California Institute of Technology. NEOWISE is funded by the National Aeronautics and Space Administration.
The Legacy Surveys imaging of the DESI footprint is supported by the Director, Office of Science, Office of High Energy Physics of the U.S. Department of Energy under Contract No. DE-AC02-05CH1123, by the National Energy Research Scientific Computing Center, a DOE Office of Science User Facility under the same contract; and by the U.S. National Science Foundation, Division of Astronomical Sciences under Contract No. AST-0950945 to NOAO.
\end{acknowledgements}

{\footnotesize\bibliography{cite}}

\begin{thebibliography}{96}
\expandafter\ifx\csname natexlab\endcsname\relax\def\natexlab#1{#1}\fi

\bibitem[{{Abazajian} {et~al.}(2009){Abazajian}, {Adelman-McCarthy},
  {Ag{\"u}eros}, {Allam}, {Allende Prieto}, {An}, {Anderson}, {Anderson},
  {Annis}, {Bahcall}, {Bailer-Jones}, {Barentine}, {Bassett}, {Becker},
  {Beers}, {Bell}, {Belokurov}, {Berlind}, {Berman}, {Bernardi}, {Bickerton},
  {Bizyaev}, {Blakeslee}, {Blanton}, {Bochanski}, {Boroski}, {Brewington},
  {Brinchmann}, {Brinkmann}, {Brunner}, {Budav{\'a}ri}, {Carey}, {Carliles},
  {Carr}, {Castander}, {Cinabro}, {Connolly}, {Csabai}, {Cunha}, {Czarapata},
  {Davenport}, {de Haas}, {Dilday}, {Doi}, {Eisenstein}, {Evans}, {Evans},
  {Fan}, {Friedman}, {Frieman}, {Fukugita}, {G{\"a}nsicke}, {Gates},
  {Gillespie}, {Gilmore}, {Gonzalez}, {Gonzalez}, {Grebel}, {Gunn},
  {Gy{\"o}ry}, {Hall}, {Harding}, {Harris}, {Harvanek}, {Hawley}, {Hayes},
  {Heckman}, {Hendry}, {Hennessy}, {Hindsley}, {Hoblitt}, {Hogan}, {Hogg},
  {Holtzman}, {Hyde}, {Ichikawa}, {Ichikawa}, {Im}, {Ivezi{\'c}}, {Jester},
  {Jiang}, {Johnson}, {Jorgensen}, {Juri{\'c}}, {Kent}, {Kessler}, {Kleinman},
  {Knapp}, {Konishi}, {Kron}, {Krzesinski}, {Kuropatkin}, {Lampeitl},
  {Lebedeva}, {Lee}, {Lee}, {French Leger}, {L{\'e}pine}, {Li}, {Lima}, {Lin},
  {Long}, {Loomis}, {Loveday}, {Lupton}, {Magnier}, {Malanushenko},
  {Malanushenko}, {Mand elbaum}, {Margon}, {Marriner}, {Mart{\'\i}nez-Delgado},
  {Matsubara}, {McGehee}, {McKay}, {Meiksin}, {Morrison}, {Mullally}, {Munn},
  {Murphy}, {Nash}, {Nebot}, {Neilsen}, {Newberg}, {Newman}, {Nichol},
  {Nicinski}, {Nieto-Santisteban}, {Nitta}, {Okamura}, {Oravetz}, {Ostriker},
  {Owen}, {Padmanabhan}, {Pan}, {Park}, {Pauls}, {Peoples}, {Percival}, {Pier},
  {Pope}, {Pourbaix}, {Price}, {Purger}, {Quinn}, {Raddick}, {Re Fiorentin},
  {Richards}, {Richmond}, {Riess}, {Rix}, {Rockosi}, {Sako}, {Schlegel},
  {Schneider}, {Scholz}, {Schreiber}, {Schwope}, {Seljak}, {Sesar}, {Sheldon},
  {Shimasaku}, {Sibley}, {Simmons}, {Sivarani}, {Allyn Smith}, {Smith},
  {Smol{\v{c}}i{\'c}}, {Snedden}, {Stebbins}, {Steinmetz}, {Stoughton},
  {Strauss}, {SubbaRao}, {Suto}, {Szalay}, {Szapudi}, {Szkody}, {Tanaka},
  {Tegmark}, {Teodoro}, {Thakar}, {Tremonti}, {Tucker}, {Uomoto}, {Vanden
  Berk}, {Vandenberg}, {Vidrih}, {Vogeley}, {Voges}, {Vogt}, {Wadadekar},
  {Watters}, {Weinberg}, {West}, {White}, {Wilhite}, {Wonders}, {Yanny},
  {Yocum}, {York}, {Zehavi}, {Zibetti}, \& {Zucker}}]{Abazajian12009}
{Abazajian}, K.~N., {Adelman-McCarthy}, J.~K., {Ag{\"u}eros}, M.~A., {et~al.}
  2009, \apjs, 182, 543

\bibitem[{Alam {et~al.}(2015)Alam, Albareti, Prieto, Anders, Anderson,
  Anderton, Andrews, Armengaud, Aubourg, Bailey, Basu, Bautista, Beaton, Beers,
  Bender, Berlind, Beutler, Bhardwaj, Bird, Bizyaev, Blake, Blanton, Blomqvist,
  Bochanski, Bolton, Bovy, Bradley, Brandt, Brauer, Brinkmann, Brown,
  Brownstein, Burden, Burtin, Busca, Cai, Capozzi, Rosell, Carr, Carrera,
  Chambers, Chaplin, Chen, Chiappini, Chojnowski, Chuang, Clerc, Comparat,
  Covey, Croft, Cuesta, Cunha, da~Costa, Rio, Davenport, Dawson, Lee, Delubac,
  Deshpande, Dhital, Dutra-Ferreira, Dwelly, Ealet, Ebelke, Edmondson,
  Eisenstein, Ellsworth, Elsworth, Epstein, Eracleous, Escoffier, Esposito,
  Evans, Fan, Fern{\'{a}}ndez-Alvar, Feuillet, Ak, Finley, Finoguenov,
  Flaherty, Fleming, Font-Ribera, Foster, Frinchaboy, Galbraith-Frew,
  Garc{\'{\i}}a, Garc{\'{\i}}a-Hern{\'{a}}ndez, P{\'{e}}rez, Gaulme, Ge,
  G{\'{e}}nova-Santos, Georgakakis, Ghezzi, Gillespie, Girardi, Goddard,
  Gontcho, Hern{\'{a}}ndez, Grebel, Green, Grieb, Grieves, Gunn, Guo, Harding,
  Hasselquist, Hawley, Hayden, Hearty, Hekker, Ho, Hogg, Holley-Bockelmann,
  Holtzman, Honscheid, Huber, Huehnerhoff, Ivans, Jiang, Johnson, Kinemuchi,
  Kirkby, Kitaura, Klaene, Knapp, Kneib, Koenig, Lam, Lan, Lang, Laurent, Goff,
  Leauthaud, Lee, Lee, Licquia, Liu, Long, L{\'{o}}pez-Corredoira,
  Lorenzo-Oliveira, Lucatello, Lundgren, Lupton, III, Mahadevan, Maia,
  Majewski, Malanushenko, Malanushenko, Manchado, Manera, Mao, Maraston,
  Marchwinski, Margala, Martell, Martig, Masters, Mathur, McBride, McGehee,
  McGreer, McMahon, M{\'{e}}nard, Menzel, Merloni, M{\'{e}}sz{\'{a}}ros,
  Miller, Miralda-Escud{\'{e}}, Miyatake, Montero-Dorta, More, Morganson,
  Morice-Atkinson, Morrison, Mosser, Muna, Myers, Nandra, Newman, Neyrinck,
  Nguyen, Nichol, Nidever, Noterdaeme, Nuza, O'Connell, O'Connell, O'Connell,
  Ogando, Olmstead, Oravetz, Oravetz, Osumi, Owen, Padgett, Padmanabhan,
  Paegert, Palanque-Delabrouille, Pan, Parejko, P{\^{a}}ris, Park,
  Pattarakijwanich, Pellejero-Ibanez, Pepper, Percival, P{\'{e}}rez-Fournon,
  Pe{\textasciiacute}rez-Ra{\textasciigrave}fols, Petitjean, Pieri,
  Pinsonneault, de~Mello, Prada, Prakash, Price-Whelan, Protopapas, Raddick,
  Rahman, Reid, Rich, Rix, Robin, Rockosi, Rodrigues, Rodr{\'{\i}}guez-Torres,
  Roe, Ross, Ross, Rossi, Ruan, Rubi{\~{n}}o-Mart{\'{\i}}n, Rykoff,
  Salazar-Albornoz, Salvato, Samushia, S{\'{a}}nchez, Santiago, Sayres,
  Schiavon, Schlegel, Schmidt, Schneider, Schultheis, Schwope, Sc{\'{o}}ccola,
  Scott, Sellgren, Seo, Serenelli, Shane, Shen, Shetrone, Shu, Aguirre,
  Sivarani, Skrutskie, Slosar, Smith, Sobreira, Souto, Stassun, Steinmetz,
  Stello, Strauss, Streblyanska, Suzuki, Swanson, Tan, Tayar, Terrien, Thakar,
  Thomas, Thomas, Thompson, Tinker, Tojeiro, Troup, Vargas-Maga{\~{n}}a,
  Vazquez, Verde, Viel, Vogt, Wake, Wang, Weaver, Weinberg, Weiner, White,
  Wilson, Wisniewski, Wood-Vasey, Ye{\textasciigrave}che, York, Zakamska,
  Zamora, Zasowski, Zehavi, Zhao, Zheng, (周旭), (周志民), (邹虎), \&
  Zhu}]{Alam12015}
Alam, S., Albareti, F.~D., Prieto, C.~A., {et~al.} 2015, ApJSS, 219, 12

\bibitem[{{Andernach} {et~al.}(1992){Andernach}, {Feretti}, {Giovannini},
  {Klein}, {Rossetti}, \& {Schnaubelt}}]{Andernach11992}
{Andernach}, H., {Feretti}, L., {Giovannini}, G., {et~al.} 1992, \aaps, 93, 331

\bibitem[{Andernach {et~al.}(2021)Andernach, Jim{\'e}nez-Andrade, \&
  Willis}]{Andernach12021}
Andernach, H., Jim{\'e}nez-Andrade, E.~F., \& Willis, A.~G. 2021, Galaxies, 9

\bibitem[{{Bassani} {et~al.}(2021){Bassani}, {Ursini}, {Malizia}, {Bruni},
  {Panessa}, {Masetti}, {Saviane}, {Monaco}, {Venturi}, {Dallacasa}, {Bazzano},
  \& {Ubertini}}]{Bassani12021}
{Bassani}, L., {Ursini}, F., {Malizia}, A., {et~al.} 2021, \mnras, 500, 3111

\bibitem[{{Beck} \& {Krause}(2005)}]{Beck12005}
{Beck}, R. \& {Krause}, M. 2005, Astronomische Nachrichten, 326, 414

\bibitem[{{Best} \& {Heckman}(2012)}]{Best12012}
{Best}, P.~N. \& {Heckman}, T.~M. 2012, \mnras, 421, 1569

\bibitem[{{Best} {et~al.}(2014){Best}, {Ker}, {Simpson}, {Rigby}, \&
  {Sabater}}]{Best12014}
{Best}, P.~N., {Ker}, L.~M., {Simpson}, C., {Rigby}, E.~E., \& {Sabater}, J.
  2014, \mnras, 445, 955

\bibitem[{{Blandford} \& {Rees}(1974)}]{Blandford11974}
{Blandford}, R.~D. \& {Rees}, M.~J. 1974, \mnras, 169, 395

\bibitem[{{Bonnarel} {et~al.}(2000){Bonnarel}, {Fernique}, {Bienaym{\'e}},
  {Egret}, {Genova}, {Louys}, {Ochsenbein}, {Wenger}, \&
  {Bartlett}}]{Bonnarel12000}
{Bonnarel}, F., {Fernique}, P., {Bienaym{\'e}}, O., {et~al.} 2000, \aaps, 143,
  33

\bibitem[{{Boxelaar} {et~al.}(2021){Boxelaar}, {van Weeren}, \&
  {Botteon}}]{Boxelaar12021}
{Boxelaar}, J.~M., {van Weeren}, R.~J., \& {Botteon}, A. 2021, Astronomy and
  Computing, 35, 100464

\bibitem[{{Br{\"u}ggen} {et~al.}(2021){Br{\"u}ggen}, {Reiprich}, {Bulbul},
  {Koribalski}, {Andernach}, {Rudnick}, {Hoang}, {Wilber}, {Duchesne},
  {Veronica}, {Pacaud}, {Hopkins}, {Norris}, {Johnston-Hollitt}, {Brown},
  {Bonafede}, {Brunetti}, {Collier}, {Sanders}, {Vardoulaki}, {Venturi},
  {Kapinska}, \& {Marvil}}]{Bruggen12021}
{Br{\"u}ggen}, M., {Reiprich}, T.~H., {Bulbul}, E., {et~al.} 2021, \aap, 647,
  A3

\bibitem[{{Burbidge}(1956)}]{Burbidge11956}
{Burbidge}, G.~R. 1956, \apj, 124, 416

\bibitem[{{Cantwell} {et~al.}(2020){Cantwell}, {Bray}, {Croston}, {Scaife},
  {Mulcahy}, {Best}, {Br{\"u}ggen}, {Brunetti}, {Callingham}, {Clarke},
  {Hardcastle}, {Harwood}, {Heald}, {Heesen}, {Iacobelli}, {Jamrozy},
  {Morganti}, {Orr{\'u}}, {O'Sullivan}, {Riseley}, {R{\"o}ttgering},
  {Shulevski}, {Sridhar}, {Tasse}, \& {Van Eck}}]{Cantwell12020}
{Cantwell}, T.~M., {Bray}, J.~D., {Croston}, J.~H., {et~al.} 2020, \mnras, 495,
  143

\bibitem[{{Chambers} {et~al.}(2016){Chambers}, {Magnier}, {Metcalfe},
  {Flewelling}, {Huber}, {Waters}, {Denneau}, {Draper}, {Farrow}, {Finkbeiner},
  {Holmberg}, {Koppenhoefer}, {Price}, {Rest}, {Saglia}, {Schlafly}, {Smartt},
  {Sweeney}, {Wainscoat}, {Burgett}, {Chastel}, {Grav}, {Heasley}, {Hodapp},
  {Jedicke}, {Kaiser}, {Kudritzki}, {Luppino}, {Lupton}, {Monet}, {Morgan},
  {Onaka}, {Shiao}, {Stubbs}, {Tonry}, {White}, {Ba{\~n}ados}, {Bell},
  {Bender}, {Bernard}, {Boegner}, {Boffi}, {Botticella}, {Calamida},
  {Casertano}, {Chen}, {Chen}, {Cole}, {Deacon}, {Frenk}, {Fitzsimmons},
  {Gezari}, {Gibbs}, {Goessl}, {Goggia}, {Gourgue}, {Goldman}, {Grant},
  {Grebel}, {Hambly}, {Hasinger}, {Heavens}, {Heckman}, {Henderson}, {Henning},
  {Holman}, {Hopp}, {Ip}, {Isani}, {Jackson}, {Keyes}, {Koekemoer}, {Kotak},
  {Le}, {Liska}, {Long}, {Lucey}, {Liu}, {Martin}, {Masci}, {McLean}, {Mindel},
  {Misra}, {Morganson}, {Murphy}, {Obaika}, {Narayan}, {Nieto-Santisteban},
  {Norberg}, {Peacock}, {Pier}, {Postman}, {Primak}, {Rae}, {Rai}, {Riess},
  {Riffeser}, {Rix}, {R{\"o}ser}, {Russel}, {Rutz}, {Schilbach}, {Schultz},
  {Scolnic}, {Strolger}, {Szalay}, {Seitz}, {Small}, {Smith}, {Soderblom},
  {Taylor}, {Thomson}, {Taylor}, {Thakar}, {Thiel}, {Thilker}, {Unger},
  {Urata}, {Valenti}, {Wagner}, {Walder}, {Walter}, {Watters}, {Werner},
  {Wood-Vasey}, \& {Wyse}}]{Chambers12016}
{Chambers}, K.~C., {Magnier}, E.~A., {Metcalfe}, N., {et~al.} 2016, arXiv
  e-prints, arXiv:1612.05560

\bibitem[{{Chang} {et~al.}(2015){Chang}, {van der Wel}, {da Cunha}, \&
  {Rix}}]{Chang12015}
{Chang}, Y.-Y., {van der Wel}, A., {da Cunha}, E., \& {Rix}, H.-W. 2015, \apjs,
  219, 8

\bibitem[{{Condon} {et~al.}(2012){Condon}, {Cotton}, {Fomalont}, {Kellermann},
  {Miller}, {Perley}, {Scott}, {Vernstrom}, \& {Wall}}]{Condon12012}
{Condon}, J.~J., {Cotton}, W.~D., {Fomalont}, E.~B., {et~al.} 2012, \apj, 758,
  23

\bibitem[{{Condon} {et~al.}(1998){Condon}, {Cotton}, {Greisen}, {Yin},
  {Perley}, {Taylor}, \& {Broderick}}]{Condon11998}
{Condon}, J.~J., {Cotton}, W.~D., {Greisen}, E.~W., {et~al.} 1998, \aj, 115,
  1693

\bibitem[{{Cutri} \& {et al.}(2012)}]{Cutri12012}
{Cutri}, R.~M. \& {et al.} 2012, VizieR Online Data Catalog, II/311

\bibitem[{{Dabhade} {et~al.}(2020{\natexlab{a}}){Dabhade}, {Mahato}, {Bagchi},
  {Saikia}, {Combes}, {Sankhyayan}, {R{\"o}ttgering}, {Ho}, {Gaikwad},
  {Raychaudhury}, {Vaidya}, \& {Guiderdoni}}]{Dabhade12020October}
{Dabhade}, P., {Mahato}, M., {Bagchi}, J., {et~al.} 2020{\natexlab{a}}, \aap,
  642, A153

\bibitem[{{Dabhade} {et~al.}(2020{\natexlab{b}}){Dabhade}, {R{\"o}ttgering},
  {Bagchi}, {Shimwell}, {Hardcastle}, {Sankhyayan}, {Morganti}, {Jamrozy},
  {Shulevski}, \& {Duncan}}]{Dabhade12020March}
{Dabhade}, P., {R{\"o}ttgering}, H.~J.~A., {Bagchi}, J., {et~al.}
  2020{\natexlab{b}}, \aap, 635, A5

\bibitem[{{Delhaize} {et~al.}(2021){Delhaize}, {Heywood}, {Prescott}, {Jarvis},
  {Delvecchio}, {Whittam}, {White}, {Hardcastle}, {Hale}, {Afonso}, {Ao},
  {Brienza}, {Br{\"u}ggen}, {Collier}, {Daddi}, {Glowacki}, {Maddox},
  {Morabito}, {Prandoni}, {Randriamanakoto}, {Sekhar}, {An}, {Adams}, {Blyth},
  {Bowler}, {Leeuw}, {Marchetti}, {Randriamampandry}, {Thorat}, {Seymour},
  {Smirnov}, {Taylor}, {Tasse}, \& {Vaccari}}]{Delhaize12021}
{Delhaize}, J., {Heywood}, I., {Prescott}, M., {et~al.} 2021, \mnras, 501, 3833

\bibitem[{Dey {et~al.}(2019)Dey, Schlegel, Lang, Blum, Burleigh, Fan, Findlay,
  Finkbeiner, Herrera, Juneau, Landriau, Levi, McGreer, Meisner, Myers,
  Moustakas, Nugent, Patej, Schlafly, Walker, Valdes, Weaver, Y{\`{e}}che, Zou,
  Zhou, Abareshi, Abbott, Abolfathi, Aguilera, Alam, Allen, Alvarez, Annis,
  Ansarinejad, Aubert, Beechert, Bell, BenZvi, Beutler, Bielby, Bolton,
  Brice{\~{n}}o, Buckley-Geer, Butler, Calamida, Carlberg, Carter, Casas,
  Castander, Choi, Comparat, Cukanovaite, Delubac, DeVries, Dey, Dhungana,
  Dickinson, Ding, Donaldson, Duan, Duckworth, Eftekharzadeh, Eisenstein,
  Etourneau, Fagrelius, Farihi, Fitzpatrick, Font-Ribera, Fulmer, G{\"a}nsicke,
  Gaztanaga, George, Gerdes, Gontcho, Gorgoni, Green, Guy, Harmer, Hernandez,
  Honscheid, Huang, James, Jannuzi, Jiang, Joyce, Karcher, Karkar, Kehoe,
  Jean-Paul, Kueter-Young, Lan, Lauer, Guillou, Suu, Lee, Lesser, Levasseur,
  Li, Mann, Marshall, Mart{\'{\i}}nez-V{\'{a}}zquez, Martini, du~Mas~des
  Bourboux, McManus, Meier, M{\'{e}}nard, Metcalfe,
  Mu{\~{n}}oz-Guti{\'{e}}rrez, Najita, Napier, Narayan, Newman, Nie, Nord,
  Norman, Olsen, Paat, Palanque-Delabrouille, Peng, Poppett, Poremba, Prakash,
  Rabinowitz, Raichoor, Rezaie, Robertson, Roe, Ross, Ross, Rudnick, Safonova,
  Saha, S{\'{a}}nchez, Savary, Schweiker, Scott, Seo, Shan, Silva, Slepian,
  Soto, Sprayberry, Staten, Stillman, Stupak, Summers, Tie, Tirado,
  Vargas-Maga{\~{n}}a, Vivas, Wechsler, Williams, Yang, Yang, Yapici, Zaritsky,
  Zenteno, Zhang, Zhang, Zhou, \& Zhou}]{Dey12019}
Dey, A., Schlegel, D.~J., Lang, D., {et~al.} 2019, AJ, 157, 168

\bibitem[{{Galvin} {et~al.}(2020){Galvin}, {Huynh}, {Norris}, {Wang},
  {Hopkins}, {Polsterer}, {Ralph}, {O'Brien}, \& {Heald}}]{Galvin12020}
{Galvin}, T.~J., {Huynh}, M.~T., {Norris}, R.~P., {et~al.} 2020, \mnras, 497,
  2730

\bibitem[{{G{\"u}rkan} {et~al.}(2014){G{\"u}rkan}, {Hardcastle}, \&
  {Jarvis}}]{Gurkan12014}
{G{\"u}rkan}, G., {Hardcastle}, M.~J., \& {Jarvis}, M.~J. 2014, \mnras, 438,
  1149

\bibitem[{{Hahn} {et~al.}(2007){Hahn}, {Porciani}, {Carollo}, \&
  {Dekel}}]{Hahn12007}
{Hahn}, O., {Porciani}, C., {Carollo}, C.~M., \& {Dekel}, A. 2007, \mnras, 375,
  489

\bibitem[{{Hardcastle}(2018)}]{Hardcastle12018}
{Hardcastle}, M.~J. 2018, \mnras, 475, 2768

\bibitem[{{Hardcastle} {et~al.}(1998{\natexlab{a}}){Hardcastle}, {Alexander},
  {Pooley}, \& {Riley}}]{Hardcastle11998a}
{Hardcastle}, M.~J., {Alexander}, P., {Pooley}, G.~G., \& {Riley}, J.~M.
  1998{\natexlab{a}}, \mnras, 296, 445

\bibitem[{{Hardcastle} \& {Croston}(2020)}]{Hardcastle12020}
{Hardcastle}, M.~J. \& {Croston}, J.~H. 2020, \nar, 88, 101539

\bibitem[{{Hardcastle} {et~al.}(1998{\natexlab{b}}){Hardcastle}, {Worrall}, \&
  {Birkinshaw}}]{Hardcastle11998b}
{Hardcastle}, M.~J., {Worrall}, D.~M., \& {Birkinshaw}, M. 1998{\natexlab{b}},
  \mnras, 296, 1098

\bibitem[{{Heckman} \& {Best}(2014)}]{Heckman12014}
{Heckman}, T.~M. \& {Best}, P.~N. 2014, \araa, 52, 589

\bibitem[{{Heesen} {et~al.}(2018){Heesen}, {Croston}, {Morganti}, {Hardcastle},
  {Stewart}, {Best}, {Broderick}, {Br{\"u}ggen}, {Brunetti}, {Chy{\.Z}y},
  {Harwood}, {Haverkorn}, {Hess}, {Intema}, {Jamrozy}, {Kunert-Bajraszewska},
  {McKean}, {Orr{\'u}}, {R{\"o}ttgering}, {Shimwell}, {Shulevski}, {White},
  {Wilcots}, \& {Williams}}]{Heesen12018}
{Heesen}, V., {Croston}, J.~H., {Morganti}, R., {et~al.} 2018, \mnras, 474,
  5049

\bibitem[{{Ishwara-Chandra} \& {Saikia}(1999)}]{IshwaraChandra11999}
{Ishwara-Chandra}, C.~H. \& {Saikia}, D.~J. 1999, \mnras, 309, 100

\bibitem[{{Ishwara-Chandra} {et~al.}(2020){Ishwara-Chandra}, {Taylor}, {Green},
  {Stil}, {Vaccari}, \& {Ocran}}]{IshwaraChandra12020}
{Ishwara-Chandra}, C.~H., {Taylor}, A.~R., {Green}, D.~A., {et~al.} 2020,
  \mnras, 497, 5383

\bibitem[{{Jamrozy} {et~al.}(2008){Jamrozy}, {Konar}, {Machalski}, \&
  {Saikia}}]{Jamrozy12008}
{Jamrozy}, M., {Konar}, C., {Machalski}, J., \& {Saikia}, D.~J. 2008, \mnras,
  385, 1286

\bibitem[{{Jamrozy} \& {Machalski}(2002)}]{Jamrozy12002}
{Jamrozy}, M. \& {Machalski}, J. 2002, in Astronomical Society of the Pacific
  Conference Series, Vol. 284, IAU Colloq. 184: AGN Surveys, ed. R.~F. {Green},
  E.~Y. {Khachikian}, \& D.~B. {Sanders}, 295

\bibitem[{{Jamrozy} {et~al.}(2005){Jamrozy}, {Machalski}, {Mack}, \&
  {Klein}}]{Jamrozy12005}
{Jamrozy}, M., {Machalski}, J., {Mack}, K.~H., \& {Klein}, U. 2005, \aap, 433,
  467

\bibitem[{{Kormendy} \& {Ho}(2013)}]{Kormendy12013}
{Kormendy}, J. \& {Ho}, L.~C. 2013, \araa, 51, 511

\bibitem[{{Kuminski} \& {Shamir}(2016)}]{Kuminski12016}
{Kuminski}, E. \& {Shamir}, L. 2016, \apjs, 223, 20

\bibitem[{{Ku{\'z}micz} \& {Jamrozy}(2021)}]{Kuzmicz12021}
{Ku{\'z}micz}, A. \& {Jamrozy}, M. 2021, \apjs, 253, 25

\bibitem[{{Ku{\'z}micz} {et~al.}(2018){Ku{\'z}micz}, {Jamrozy}, {Bronarska},
  {Janda-Boczar}, \& {Saikia}}]{Kuzmicz12018}
{Ku{\'z}micz}, A., {Jamrozy}, M., {Bronarska}, K., {Janda-Boczar}, K., \&
  {Saikia}, D.~J. 2018, \apjs, 238, 9

\bibitem[{{Lacy} {et~al.}(2020){Lacy}, {Baum}, {Chandler}, {Chatterjee},
  {Clarke}, {Deustua}, {English}, {Farnes}, {Gaensler}, {Gugliucci},
  {Hallinan}, {Kent}, {Kimball}, {Law}, {Lazio}, {Marvil}, {Mao}, {Medlin},
  {Mooley}, {Murphy}, {Myers}, {Osten}, {Richards}, {Rosolowsky}, {Rudnick},
  {Schinzel}, {Sivakoff}, {Sjouwerman}, {Taylor}, {White}, {Wrobel},
  {Andernach}, {Beasley}, {Berger}, {Bhatnager}, {Birkinshaw}, {Bower},
  {Brandt}, {Brown}, {Burke-Spolaor}, {Butler}, {Comerford}, {Demorest}, {Fu},
  {Giacintucci}, {Golap}, {G{\"u}th}, {Hales}, {Hiriart}, {Hodge}, {Horesh},
  {Ivezi{\'c}}, {Jarvis}, {Kamble}, {Kassim}, {Liu}, {Loinard}, {Lyons},
  {Masters}, {Mezcua}, {Moellenbrock}, {Mroczkowski}, {Nyland}, {O'Dea},
  {O'Sullivan}, {Peters}, {Radford}, {Rao}, {Robnett}, {Salcido}, {Shen},
  {Sobotka}, {Witz}, {Vaccari}, {van Weeren}, {Vargas}, {Williams}, \&
  {Yoon}}]{Lacy12020}
{Lacy}, M., {Baum}, S.~A., {Chandler}, C.~J., {et~al.} 2020, \pasp, 132, 035001

\bibitem[{{Lacy} {et~al.}(1993){Lacy}, {Rawlings}, {Saunders}, \&
  {Warner}}]{Lacy11993}
{Lacy}, M., {Rawlings}, S., {Saunders}, R., \& {Warner}, P.~J. 1993, \mnras,
  264, 721

\bibitem[{{Laing} \& {Bridle}(2014)}]{Laing12014}
{Laing}, R.~A. \& {Bridle}, A.~H. 2014, \mnras, 437, 3405

\bibitem[{{Lara} {et~al.}(2000){Lara}, {Mack}, {Lacy}, {Klein}, {Cotton},
  {Feretti}, {Giovannini}, \& {Murgia}}]{Lara12000}
{Lara}, L., {Mack}, K.~H., {Lacy}, M., {et~al.} 2000, \aap, 356, 63

\bibitem[{{Machalski}(2011)}]{Machalski12011}
{Machalski}, J. 2011, \mnras, 413, 2429

\bibitem[{{Machalski} \& {Jamrozy}(2000)}]{Machalski12000}
{Machalski}, J. \& {Jamrozy}, M. 2000, \aap, 363, L17

\bibitem[{{Machalski} \& {Jamrozy}(2006)}]{Machalski12006July2}
{Machalski}, J. \& {Jamrozy}, M. 2006, \aap, 454, 95

\bibitem[{{Machalski} {et~al.}(2001){Machalski}, {Jamrozy}, \&
  {Zola}}]{Machalski12001}
{Machalski}, J., {Jamrozy}, M., \& {Zola}, S. 2001, \aap, 371, 445

\bibitem[{{Machalski} {et~al.}(2006){Machalski}, {Jamrozy}, {Zola}, \&
  {Koziel}}]{Machalski12006July1}
{Machalski}, J., {Jamrozy}, M., {Zola}, S., \& {Koziel}, D. 2006, \aap, 454, 85

\bibitem[{{Machalski} {et~al.}(2007){Machalski}, {Koziel-Wierzbowska}, \&
  {Jamrozy}}]{Machalski12007}
{Machalski}, J., {Koziel-Wierzbowska}, D., \& {Jamrozy}, M. 2007, AcA, 57, 227

\bibitem[{{Machalski} {et~al.}(2008){Machalski}, {Kozie{\l}-Wierzbowska},
  {Jamrozy}, \& {Saikia}}]{Machalski12008}
{Machalski}, J., {Kozie{\l}-Wierzbowska}, D., {Jamrozy}, M., \& {Saikia}, D.~J.
  2008, \apj, 679, 149

\bibitem[{{Mack} {et~al.}(1998){Mack}, {Klein}, {O'Dea}, {Willis}, \&
  {Saripalli}}]{Mack11998}
{Mack}, K.~H., {Klein}, U., {O'Dea}, C.~P., {Willis}, A.~G., \& {Saripalli}, L.
  1998, \aap, 329, 431

\bibitem[{{Mahato} {et~al.}(2021){Mahato}, {Dabhade}, {Saikia}, {Combes},
  {Bagchi}, {Ho}, \& {Raychaudhury}}]{Mahato12021}
{Mahato}, M., {Dabhade}, P., {Saikia}, D.~J., {et~al.} 2021, arXiv e-prints,
  arXiv:2111.11905

\bibitem[{{Malarecki} {et~al.}(2013){Malarecki}, {Staveley-Smith}, {Saripalli},
  {Subrahmanyan}, {Jones}, {Duffy}, \& {Rioja}}]{Malarecki12013}
{Malarecki}, J.~M., {Staveley-Smith}, L., {Saripalli}, L., {et~al.} 2013,
  \mnras, 432, 200

\bibitem[{{Masini} {et~al.}(2021){Masini}, {Celotti}, {Grandi}, {Moravec}, \&
  {Williams}}]{Masini12021}
{Masini}, A., {Celotti}, A., {Grandi}, P., {Moravec}, E., \& {Williams}, W.~L.
  2021, \aap, 650, A51

\bibitem[{{Myers} \& {Spangler}(1985)}]{Myers11985}
{Myers}, S.~T. \& {Spangler}, S.~R. 1985, \apj, 291, 52

\bibitem[{{Oei} {et~al.}(in prep.){Oei}, {van Weeren}, {de Gasperin},
  {Botteon}, {Hardcastle}, {Shimwell}, \& {R\"ottgering}}]{Oei12022}
{Oei}, M. S. S.~L., {van Weeren}, R.~J., {de Gasperin}, F., {et~al.} in prep.

\bibitem[{{Offringa} {et~al.}(2014){Offringa}, {McKinley}, {Hurley-Walker},
  {Briggs}, {Wayth}, {Kaplan}, {Bell}, {Feng}, {Neben}, {Hughes}, {Rhee},
  {Murphy}, {Bhat}, {Bernardi}, {Bowman}, {Cappallo}, {Corey}, {Deshpande},
  {Emrich}, {Ewall-Wice}, {Gaensler}, {Goeke}, {Greenhill}, {Hazelton},
  {Hindson}, {Johnston-Hollitt}, {Jacobs}, {Kasper}, {Kratzenberg}, {Lenc},
  {Lonsdale}, {Lynch}, {McWhirter}, {Mitchell}, {Morales}, {Morgan},
  {Kudryavtseva}, {Oberoi}, {Ord}, {Pindor}, {Procopio}, {Prabu}, {Riding},
  {Roshi}, {Shankar}, {Srivani}, {Subrahmanyan}, {Tingay}, {Waterson},
  {Webster}, {Whitney}, {Williams}, \& {Williams}}]{Offringa12014}
{Offringa}, A.~R., {McKinley}, B., {Hurley-Walker}, N., {et~al.} 2014, \mnras,
  444, 606

\bibitem[{{Offringa} \& {Smirnov}(2017)}]{Offringa12017}
{Offringa}, A.~R. \& {Smirnov}, O. 2017, \mnras, 471, 301

\bibitem[{{Pacholczyk}(1970)}]{Pacholczyk11970}
{Pacholczyk}, A.~G. 1970, {Radio astrophysics. Nonthermal processes in galactic
  and extragalactic sources}

\bibitem[{{Parma} {et~al.}(1996){Parma}, {de Ruiter}, {Mack}, {van Breugel},
  {Dey}, {Fanti}, \& {Klein}}]{Parma11996}
{Parma}, P., {de Ruiter}, H.~R., {Mack}, K.~H., {et~al.} 1996, \aap, 311, 49

\bibitem[{{Planck Collaboration} {et~al.}(2016){Planck Collaboration}, {Ade},
  {Aghanim}, {Arnaud}, {Ashdown}, {Aumont}, {Baccigalupi}, {Banday},
  {Barreiro}, {Barrena}, {Bartlett}, {Bartolo}, {Battaner}, {Battye},
  {Benabed}, {Beno{\^\i}t}, {Benoit-L{\'e}vy}, {Bernard}, {Bersanelli},
  {Bielewicz}, {Bikmaev}, {B{\"o}hringer}, {Bonaldi}, {Bonavera}, {Bond},
  {Borrill}, {Bouchet}, {Bucher}, {Burenin}, {Burigana}, {Butler}, {Calabrese},
  {Cardoso}, {Carvalho}, {Catalano}, {Challinor}, {Chamballu}, {Chary},
  {Chiang}, {Chon}, {Christensen}, {Clements}, {Colombi}, {Colombo}, {Combet},
  {Comis}, {Couchot}, {Coulais}, {Crill}, {Curto}, {Cuttaia}, {Dahle},
  {Danese}, {Davies}, {Davis}, {de Bernardis}, {de Rosa}, {de Zotti},
  {Delabrouille}, {D{\'e}sert}, {Dickinson}, {Diego}, {Dolag}, {Dole},
  {Donzelli}, {Dor{\'e}}, {Douspis}, {Ducout}, {Dupac}, {Efstathiou},
  {Eisenhardt}, {Elsner}, {En{\ss}lin}, {Eriksen}, {Falgarone}, {Fergusson},
  {Feroz}, {Ferragamo}, {Finelli}, {Forni}, {Frailis}, {Fraisse}, {Franceschi},
  {Frejsel}, {Galeotta}, {Galli}, {Ganga}, {G{\'e}nova-Santos}, {Giard},
  {Giraud-H{\'e}raud}, {Gjerl{\o}w}, {Gonz{\'a}lez-Nuevo}, {G{\'o}rski},
  {Grainge}, {Gratton}, {Gregorio}, {Gruppuso}, {Gudmundsson}, {Hansen},
  {Hanson}, {Harrison}, {Hempel}, {Henrot-Versill{\'e}},
  {Hern{\'a}ndez-Monteagudo}, {Herranz}, {Hildebrandt}, {Hivon}, {Hobson},
  {Holmes}, {Hornstrup}, {Hovest}, {Huffenberger}, {Hurier}, {Jaffe}, {Jaffe},
  {Jin}, {Jones}, {Juvela}, {Keih{\"a}nen}, {Keskitalo}, {Khamitov}, {Kisner},
  {Kneissl}, {Knoche}, {Kunz}, {Kurki-Suonio}, {Lagache}, {Lamarre}, {Lasenby},
  {Lattanzi}, {Lawrence}, {Leonardi}, {Lesgourgues}, {Levrier}, {Liguori},
  {Lilje}, {Linden-V{\o}rnle}, {L{\'o}pez-Caniego}, {Lubin},
  {Mac{\'\i}as-P{\'e}rez}, {Maggio}, {Maino}, {Mak}, {Mandolesi}, {Mangilli},
  {Martin}, {Mart{\'\i}nez-Gonz{\'a}lez}, {Masi}, {Matarrese}, {Mazzotta},
  {McGehee}, {Mei}, {Melchiorri}, {Melin}, {Mendes}, {Mennella}, {Migliaccio},
  {Mitra}, {Miville-Desch{\^e}nes}, {Moneti}, {Montier}, {Morgante},
  {Mortlock}, {Moss}, {Munshi}, {Murphy}, {Naselsky}, {Nastasi}, {Nati},
  {Natoli}, {Netterfield}, {N{\o}rgaard-Nielsen}, {Noviello}, {Novikov},
  {Novikov}, {Olamaie}, {Oxborrow}, {Paci}, {Pagano}, {Pajot}, {Paoletti},
  {Pasian}, {Patanchon}, {Pearson}, {Perdereau}, {Perotto}, {Perrott},
  {Perrotta}, {Pettorino}, {Piacentini}, {Piat}, {Pierpaoli}, {Pietrobon},
  {Plaszczynski}, {Pointecouteau}, {Polenta}, {Pratt}, {Pr{\'e}zeau}, {Prunet},
  {Puget}, {Rachen}, {Reach}, {Rebolo}, {Reinecke}, {Remazeilles}, {Renault},
  {Renzi}, {Ristorcelli}, {Rocha}, {Rosset}, {Rossetti}, {Roudier}, {Rozo},
  {Rubi{\~n}o-Mart{\'\i}n}, {Rumsey}, {Rusholme}, {Rykoff}, {Sandri}, {Santos},
  {Saunders}, {Savelainen}, {Savini}, {Schammel}, {Scott}, {Seiffert},
  {Shellard}, {Shimwell}, {Spencer}, {Stanford}, {Stern}, {Stolyarov},
  {Stompor}, {Streblyanska}, {Sudiwala}, {Sunyaev}, {Sutton}, {Suur-Uski},
  {Sygnet}, {Tauber}, {Terenzi}, {Toffolatti}, {Tomasi}, {Tramonte},
  {Tristram}, {Tucci}, {Tuovinen}, {Umana}, {Valenziano}, {Valiviita}, {Van
  Tent}, {Vielva}, {Villa}, {Wade}, {Wandelt}, {Wehus}, {White}, {Wright},
  {Yvon}, {Zacchei}, \& {Zonca}}]{PlanckCollaboration12016}
{Planck Collaboration}, {Ade}, P.~A.~R., {Aghanim}, N., {et~al.} 2016, \aap,
  594, A27

\bibitem[{{Planck Collaboration} {et~al.}(2020){Planck Collaboration},
  {Aghanim}, {Akrami}, {Ashdown}, {Aumont}, {Baccigalupi}, {Ballardini},
  {Banday}, {Barreiro}, {Bartolo}, {Basak}, {Battye}, {Benabed}, {Bernard},
  {Bersanelli}, {Bielewicz}, {Bock}, {Bond}, {Borrill}, {Bouchet}, {Boulanger},
  {Bucher}, {Burigana}, {Butler}, {Calabrese}, {Cardoso}, {Carron},
  {Challinor}, {Chiang}, {Chluba}, {Colombo}, {Combet}, {Contreras}, {Crill},
  {Cuttaia}, {de Bernardis}, {de Zotti}, {Delabrouille}, {Delouis}, {Di
  Valentino}, {Diego}, {Dor{\'e}}, {Douspis}, {Ducout}, {Dupac}, {Dusini},
  {Efstathiou}, {Elsner}, {En{\ss}lin}, {Eriksen}, {Fantaye}, {Farhang},
  {Fergusson}, {Fernandez-Cobos}, {Finelli}, {Forastieri}, {Frailis},
  {Fraisse}, {Franceschi}, {Frolov}, {Galeotta}, {Galli}, {Ganga},
  {G{\'e}nova-Santos}, {Gerbino}, {Ghosh}, {Gonz{\'a}lez-Nuevo}, {G{\'o}rski},
  {Gratton}, {Gruppuso}, {Gudmundsson}, {Hamann}, {Handley}, {Hansen},
  {Herranz}, {Hildebrandt}, {Hivon}, {Huang}, {Jaffe}, {Jones}, {Karakci},
  {Keih{\"a}nen}, {Keskitalo}, {Kiiveri}, {Kim}, {Kisner}, {Knox},
  {Krachmalnicoff}, {Kunz}, {Kurki-Suonio}, {Lagache}, {Lamarre}, {Lasenby},
  {Lattanzi}, {Lawrence}, {Le Jeune}, {Lemos}, {Lesgourgues}, {Levrier},
  {Lewis}, {Liguori}, {Lilje}, {Lilley}, {Lindholm}, {L{\'o}pez-Caniego},
  {Lubin}, {Ma}, {Mac{\'\i}as-P{\'e}rez}, {Maggio}, {Maino}, {Mandolesi},
  {Mangilli}, {Marcos-Caballero}, {Maris}, {Martin}, {Martinelli},
  {Mart{\'\i}nez-Gonz{\'a}lez}, {Matarrese}, {Mauri}, {McEwen}, {Meinhold},
  {Melchiorri}, {Mennella}, {Migliaccio}, {Millea}, {Mitra},
  {Miville-Desch{\^e}nes}, {Molinari}, {Montier}, {Morgante}, {Moss}, {Natoli},
  {N{\o}rgaard-Nielsen}, {Pagano}, {Paoletti}, {Partridge}, {Patanchon},
  {Peiris}, {Perrotta}, {Pettorino}, {Piacentini}, {Polastri}, {Polenta},
  {Puget}, {Rachen}, {Reinecke}, {Remazeilles}, {Renzi}, {Rocha}, {Rosset},
  {Roudier}, {Rubi{\~n}o-Mart{\'\i}n}, {Ruiz-Granados}, {Salvati}, {Sandri},
  {Savelainen}, {Scott}, {Shellard}, {Sirignano}, {Sirri}, {Spencer},
  {Sunyaev}, {Suur-Uski}, {Tauber}, {Tavagnacco}, {Tenti}, {Toffolatti},
  {Tomasi}, {Trombetti}, {Valenziano}, {Valiviita}, {Van Tent}, {Vibert},
  {Vielva}, {Villa}, {Vittorio}, {Wandelt}, {Wehus}, {White}, {White},
  {Zacchei}, \& {Zonca}}]{PlanckCollaboration12020}
{Planck Collaboration}, {Aghanim}, N., {Akrami}, Y., {et~al.} 2020, \aap, 641,
  A6

\bibitem[{{Pracy} {et~al.}(2016){Pracy}, {Ching}, {Sadler}, {Croom}, {Baldry},
  {Bland-Hawthorn}, {Brough}, {Brown}, {Couch}, {Davis}, {Drinkwater},
  {Hopkins}, {Jarvis}, {Jelliffe}, {Jurek}, {Loveday}, {Pimbblet}, {Prescott},
  {Wisnioski}, \& {Woods}}]{Pracy12016}
{Pracy}, M.~B., {Ching}, J. H.~Y., {Sadler}, E.~M., {et~al.} 2016, \mnras, 460,
  2

\bibitem[{{Rengelink} {et~al.}(1997){Rengelink}, {Tang}, {de Bruyn}, {Miley},
  {Bremer}, {Roettgering}, \& {Bremer}}]{Rengelink11997}
{Rengelink}, R.~B., {Tang}, Y., {de Bruyn}, A.~G., {et~al.} 1997, \aaps, 124,
  259

\bibitem[{{Ringermacher} \& {Mead}(2009)}]{Ringermacher12009}
{Ringermacher}, H.~I. \& {Mead}, L.~R. 2009, \mnras, 397, 164

\bibitem[{{Rybicki} \& {Lightman}(1986)}]{Rybicki11986}
{Rybicki}, G.~B. \& {Lightman}, A.~P. 1986, {Radiative Processes in
  Astrophysics}

\bibitem[{{Safouris} {et~al.}(2009){Safouris}, {Subrahmanyan}, {Bicknell}, \&
  {Saripalli}}]{Safouris12009}
{Safouris}, V., {Subrahmanyan}, R., {Bicknell}, G.~V., \& {Saripalli}, L. 2009,
  \mnras, 393, 2

\bibitem[{{Saikia} {et~al.}(2006){Saikia}, {Konar}, \&
  {Kulkarni}}]{Saikia12006}
{Saikia}, D.~J., {Konar}, C., \& {Kulkarni}, V.~K. 2006, \mnras, 366, 1391

\bibitem[{{Salim} {et~al.}(2018){Salim}, {Boquien}, \& {Lee}}]{Salim12018}
{Salim}, S., {Boquien}, M., \& {Lee}, J.~C. 2018, \apj, 859, 11

\bibitem[{{Saripalli} {et~al.}(2002){Saripalli}, {Subrahmanyan}, \& {Udaya
  Shankar}}]{Saripalli12002}
{Saripalli}, L., {Subrahmanyan}, R., \& {Udaya Shankar}, N. 2002, \apj, 565,
  256

\bibitem[{{Schaye} {et~al.}(2010){Schaye}, {Dalla Vecchia}, {Booth}, {Wiersma},
  {Theuns}, {Haas}, {Bertone}, {Duffy}, {McCarthy}, \& {van de
  Voort}}]{Schaye12010}
{Schaye}, J., {Dalla Vecchia}, C., {Booth}, C.~M., {et~al.} 2010, \mnras, 402,
  1536

\bibitem[{{Schoenmakers} {et~al.}(2000){Schoenmakers}, {Mack}, {de Bruyn},
  {R{\"o}ttgering}, {Klein}, \& {van der Laan}}]{Schoenmakers12000}
{Schoenmakers}, A.~P., {Mack}, K.~H., {de Bruyn}, A.~G., {et~al.} 2000, \aaps,
  146, 293

\bibitem[{{Schoenmakers} {et~al.}(1998){Schoenmakers}, {Mack}, {Lara},
  {R{\"o}ttgering}, {de Bruyn}, {van der Laan}, \&
  {Giovannini}}]{Schoenmakers11998}
{Schoenmakers}, A.~P., {Mack}, K.~H., {Lara}, L., {et~al.} 1998, \aap, 336, 455

\bibitem[{{Scott} \& {Tout}(1989)}]{Scott11989}
{Scott}, D. \& {Tout}, C.~A. 1989, \mnras, 241, 109

\bibitem[{{Sebastian} {et~al.}(2018){Sebastian}, {Ishwara-Chandra}, {Joshi}, \&
  {Wadadekar}}]{Sebastian12018}
{Sebastian}, B., {Ishwara-Chandra}, C.~H., {Joshi}, R., \& {Wadadekar}, Y.
  2018, \mnras, 473, 4926

\bibitem[{{Shimwell} {et~al.}(in prep.){Shimwell}, {Hardcastle}, \&
  {Tasse}}]{Shimwell12021}
{Shimwell}, T.~W., {Hardcastle}, M.~J., \& {Tasse}, C. in prep.

\bibitem[{{Shimwell} {et~al.}(2017){Shimwell}, {R{\"o}ttgering}, {Best},
  {Williams}, {Dijkema}, {de Gasperin}, {Hardcastle}, {Heald}, {Hoang},
  {Horneffer}, {Intema}, {Mahony}, {Mandal}, {Mechev}, {Morabito}, {Oonk},
  {Rafferty}, {Retana-Montenegro}, {Sabater}, {Tasse}, {van Weeren},
  {Br{\"u}ggen}, {Brunetti}, {Chy{\.z}y}, {Conway}, {Haverkorn}, {Jackson},
  {Jarvis}, {McKean}, {Miley}, {Morganti}, {White}, {Wise}, {van Bemmel},
  {Beck}, {Brienza}, {Bonafede}, {Calistro Rivera}, {Cassano}, {Clarke},
  {Cseh}, {Deller}, {Drabent}, {van Driel}, {Engels}, {Falcke}, {Ferrari},
  {Fr{\"o}hlich}, {Garrett}, {Harwood}, {Heesen}, {Hoeft}, {Horellou},
  {Israel}, {Kapi{\'n}ska}, {Kunert-Bajraszewska}, {McKay}, {Mohan},
  {Orr{\'u}}, {Pizzo}, {Prandoni}, {Schwarz}, {Shulevski}, {Sipior}, {Smith},
  {Sridhar}, {Steinmetz}, {Stroe}, {Varenius}, {van der Werf}, {Zensus}, \&
  {Zwart}}]{Shimwell12017}
{Shimwell}, T.~W., {R{\"o}ttgering}, H.~J.~A., {Best}, P.~N., {et~al.} 2017,
  \aap, 598, A104

\bibitem[{{Shimwell} {et~al.}(2019){Shimwell}, {Tasse}, {Hardcastle}, {Mechev},
  {Williams}, {Best}, {R{\"o}ttgering}, {Callingham}, {Dijkema}, {de Gasperin},
  {Hoang}, {Hugo}, {Mirmont}, {Oonk}, {Prandoni}, {Rafferty}, {Sabater},
  {Smirnov}, {van Weeren}, {White}, {Atemkeng}, {Bester}, {Bonnassieux},
  {Br{\"u}ggen}, {Brunetti}, {Chy{\.z}y}, {Cochrane}, {Conway}, {Croston},
  {Danezi}, {Duncan}, {Haverkorn}, {Heald}, {Iacobelli}, {Intema}, {Jackson},
  {Jamrozy}, {Jarvis}, {Lakhoo}, {Mevius}, {Miley}, {Morabito}, {Morganti},
  {Nisbet}, {Orr{\'u}}, {Perkins}, {Pizzo}, {Schrijvers}, {Smith}, {Vermeulen},
  {Wise}, {Alegre}, {Bacon}, {van Bemmel}, {Beswick}, {Bonafede}, {Botteon},
  {Bourke}, {Brienza}, {Calistro Rivera}, {Cassano}, {Clarke}, {Conselice},
  {Dettmar}, {Drabent}, {Dumba}, {Emig}, {En{\ss}lin}, {Ferrari}, {Garrett},
  {G{\'e}nova-Santos}, {Goyal}, {G{\"u}rkan}, {Hale}, {Harwood}, {Heesen},
  {Hoeft}, {Horellou}, {Jackson}, {Kokotanekov}, {Kondapally},
  {Kunert-Bajraszewska}, {Mahatma}, {Mahony}, {Mandal}, {McKean}, {Merloni},
  {Mingo}, {Miskolczi}, {Mooney}, {Nikiel-Wroczy{\'n}ski}, {O'Sullivan},
  {Quinn}, {Reich}, {Roskowi{\'n}ski}, {Rowlinson}, {Savini}, {Saxena},
  {Schwarz}, {Shulevski}, {Sridhar}, {Stacey}, {Urquhart}, {van der Wiel},
  {Varenius}, {Webster}, \& {Wilber}}]{Shimwell12019}
{Shimwell}, T.~W., {Tasse}, C., {Hardcastle}, M.~J., {et~al.} 2019, \aap, 622,
  A1

\bibitem[{{Soltan}(1982)}]{Soltan11982}
{Soltan}, A. 1982, \mnras, 200, 115

\bibitem[{{Strauss} {et~al.}(2002){Strauss}, {Weinberg}, {Lupton}, {Narayanan},
  {Annis}, {Bernardi}, {Blanton}, {Burles}, {Connolly}, {Dalcanton}, {Doi},
  {Eisenstein}, {Frieman}, {Fukugita}, {Gunn}, {Ivezi{\'c}}, {Kent}, {Kim},
  {Knapp}, {Kron}, {Munn}, {Newberg}, {Nichol}, {Okamura}, {Quinn}, {Richmond},
  {Schlegel}, {Shimasaku}, {SubbaRao}, {Szalay}, {Vanden Berk}, {Vogeley},
  {Yanny}, {Yasuda}, {York}, \& {Zehavi}}]{Strauss12002}
{Strauss}, M.~A., {Weinberg}, D.~H., {Lupton}, R.~H., {et~al.} 2002, \aj, 124,
  1810

\bibitem[{{Subrahmanyan} {et~al.}(2006){Subrahmanyan}, {Hunstead}, {Cox}, \&
  {McIntyre}}]{Subrahmanyan12006}
{Subrahmanyan}, R., {Hunstead}, R.~W., {Cox}, N.~L.~J., \& {McIntyre}, V. 2006,
  \apj, 636, 172

\bibitem[{{Subrahmanyan} {et~al.}(1996){Subrahmanyan}, {Saripalli}, \&
  {Hunstead}}]{Subrahmanyan11996}
{Subrahmanyan}, R., {Saripalli}, L., \& {Hunstead}, R.~W. 1996, \mnras, 279,
  257

\bibitem[{{Subrahmanyan} {et~al.}(2008){Subrahmanyan}, {Saripalli}, {Safouris},
  \& {Hunstead}}]{Subrahmanyan12008}
{Subrahmanyan}, R., {Saripalli}, L., {Safouris}, V., \& {Hunstead}, R.~W. 2008,
  \apj, 677, 63

\bibitem[{{Tamhane} {et~al.}(2015){Tamhane}, {Wadadekar}, {Basu}, {Singh},
  {Ishwara-Chandra}, {Beelen}, \& {Sirothia}}]{Tamhane12015}
{Tamhane}, P., {Wadadekar}, Y., {Basu}, A., {et~al.} 2015, \mnras, 453, 2438

\bibitem[{{Tang} {et~al.}(2020){Tang}, {Scaife}, {Wong}, {Kapi{\'n}ska},
  {Rudnick}, {Shabala}, {Seymour}, \& {Norris}}]{Tang12020}
{Tang}, H., {Scaife}, A.~M.~M., {Wong}, O.~I., {et~al.} 2020, \mnras, 499, 68

\bibitem[{{Tasse} {et~al.}(2018){Tasse}, {Hugo}, {Mirmont}, {Smirnov},
  {Atemkeng}, {Bester}, {Hardcastle}, {Lakhoo}, {Perkins}, \&
  {Shimwell}}]{Tasse12018}
{Tasse}, C., {Hugo}, B., {Mirmont}, M., {et~al.} 2018, \aap, 611, A87

\bibitem[{{van der Tol} {et~al.}(2018){van der Tol}, {Veenboer}, \&
  {Offringa}}]{vanDerTol12018}
{van der Tol}, S., {Veenboer}, B., \& {Offringa}, A.~R. 2018, \aap, 616, A27

\bibitem[{{van Haarlem} {et~al.}(2013){van Haarlem}, {Wise}, {Gunst}, {Heald},
  {McKean}, {Hessels}, {de Bruyn}, {Nijboer}, {Swinbank}, {Fallows},
  {Brentjens}, {Nelles}, {Beck}, {Falcke}, {Fender}, {H{\"o}randel},
  {Koopmans}, {Mann}, {Miley}, {R{\"o}ttgering}, {Stappers}, {Wijers},
  {Zaroubi}, {van den Akker}, {Alexov}, {Anderson}, {Anderson}, {van Ardenne},
  {Arts}, {Asgekar}, {Avruch}, {Batejat}, {B{\"a}hren}, {Bell}, {Bell}, {van
  Bemmel}, {Bennema}, {Bentum}, {Bernardi}, {Best}, {B{\^\i}rzan}, {Bonafede},
  {Boonstra}, {Braun}, {Bregman}, {Breitling}, {van de Brink}, {Broderick},
  {Broekema}, {Brouw}, {Br{\"u}ggen}, {Butcher}, {van Cappellen}, {Ciardi},
  {Coenen}, {Conway}, {Coolen}, {Corstanje}, {Damstra}, {Davies}, {Deller},
  {Dettmar}, {van Diepen}, {Dijkstra}, {Donker}, {Doorduin}, {Dromer}, {Drost},
  {van Duin}, {Eisl{\"o}ffel}, {van Enst}, {Ferrari}, {Frieswijk}, {Gankema},
  {Garrett}, {de Gasperin}, {Gerbers}, {de Geus}, {Grie{\ss}meier}, {Grit},
  {Gruppen}, {Hamaker}, {Hassall}, {Hoeft}, {Holties}, {Horneffer}, {van der
  Horst}, {van Houwelingen}, {Huijgen}, {Iacobelli}, {Intema}, {Jackson},
  {Jelic}, {de Jong}, {Juette}, {Kant}, {Karastergiou}, {Koers}, {Kollen},
  {Kondratiev}, {Kooistra}, {Koopman}, {Koster}, {Kuniyoshi}, {Kramer},
  {Kuper}, {Lambropoulos}, {Law}, {van Leeuwen}, {Lemaitre}, {Loose}, {Maat},
  {Macario}, {Markoff}, {Masters}, {McFadden}, {McKay-Bukowski}, {Meijering},
  {Meulman}, {Mevius}, {Middelberg}, {Millenaar}, {Miller-Jones}, {Mohan},
  {Mol}, {Morawietz}, {Morganti}, {Mulcahy}, {Mulder}, {Munk}, {Nieuwenhuis},
  {van Nieuwpoort}, {Noordam}, {Norden}, {Noutsos}, {Offringa}, {Olofsson},
  {Omar}, {Orr{\'u}}, {Overeem}, {Paas}, {Pandey-Pommier}, {Pandey}, {Pizzo},
  {Polatidis}, {Rafferty}, {Rawlings}, {Reich}, {de Reijer}, {Reitsma},
  {Renting}, {Riemers}, {Rol}, {Romein}, {Roosjen}, {Ruiter}, {Scaife}, {van
  der Schaaf}, {Scheers}, {Schellart}, {Schoenmakers}, {Schoonderbeek},
  {Serylak}, {Shulevski}, {Sluman}, {Smirnov}, {Sobey}, {Spreeuw}, {Steinmetz},
  {Sterks}, {Stiepel}, {Stuurwold}, {Tagger}, {Tang}, {Tasse}, {Thomas},
  {Thoudam}, {Toribio}, {van der Tol}, {Usov}, {van Veelen}, {van der Veen},
  {ter Veen}, {Verbiest}, {Vermeulen}, {Vermaas}, {Vocks}, {Vogt}, {de Vos},
  {van der Wal}, {van Weeren}, {Weggemans}, {Weltevrede}, {White}, {Wijnholds},
  {Wilhelmsson}, {Wucknitz}, {Yatawatta}, {Zarka}, {Zensus}, \& {van
  Zwieten}}]{vanHaarlem12013}
{van Haarlem}, M.~P., {Wise}, M.~W., {Gunst}, A.~W., {et~al.} 2013, \aap, 556,
  A2

\bibitem[{{van Weeren} {et~al.}(2021){van Weeren}, {Shimwell}, {Botteon},
  {Brunetti}, {Br{\"u}ggen}, {Boxelaar}, {Cassano}, {Di Gennaro},
  {Andrade-Santos}, {Bonnassieux}, {Bonafede}, {Cuciti}, {Dallacasa}, {de
  Gasperin}, {Gastaldello}, {Hardcastle}, {Hoeft}, {Kraft}, {Mandal},
  {Rossetti}, {R{\"o}ttgering}, {Tasse}, \& {Wilber}}]{vanWeeren12021}
{van Weeren}, R.~J., {Shimwell}, T.~W., {Botteon}, A., {et~al.} 2021, \aap,
  651, A115

\bibitem[{{Voges} {et~al.}(1999){Voges}, {Aschenbach}, {Boller},
  {Br{\"a}uninger}, {Briel}, {Burkert}, {Dennerl}, {Englhauser}, {Gruber},
  {Haberl}, {Hartner}, {Hasinger}, {K{\"u}rster}, {Pfeffermann}, {Pietsch},
  {Predehl}, {Rosso}, {Schmitt}, {Tr{\"u}mper}, \& {Zimmermann}}]{Voges11999}
{Voges}, W., {Aschenbach}, B., {Boller}, T., {et~al.} 1999, \aap, 349, 389

\bibitem[{{Wen} {et~al.}(2012){Wen}, {Han}, \& {Liu}}]{Wen12012}
{Wen}, Z.~L., {Han}, J.~L., \& {Liu}, F.~S. 2012, \apjs, 199, 34

\bibitem[{{Williams} {et~al.}(2018){Williams}, {Calistro Rivera}, {Best},
  {Hardcastle}, {R{\"o}ttgering}, {Duncan}, {de Gasperin}, {Jarvis}, {Miley},
  {Mahony}, {Morabito}, {Nisbet}, {Prandoni}, {Smith}, {Tasse}, \&
  {White}}]{Williams12018}
{Williams}, W.~L., {Calistro Rivera}, G., {Best}, P.~N., {et~al.} 2018, \mnras,
  475, 3429

\bibitem[{{Willis} {et~al.}(1974){Willis}, {Strom}, \& {Wilson}}]{Willis11974}
{Willis}, A.~G., {Strom}, R.~G., \& {Wilson}, A.~S. 1974, \nat, 250, 625

\bibitem[{{Zou} {et~al.}(2021){Zou}, {Gao}, {Xu}, {Zhou}, {Ma}, {Zhou},
  {Zhang}, {Nie}, {Wang}, \& {Xue}}]{Zou12021}
{Zou}, H., {Gao}, J., {Xu}, X., {et~al.} 2021, \apjs, 253, 56

\end{thebibliography}

\appendix

\section{J1420-0545 comparison}
\label{ap:comparisonMachalski}
\begin{figure}
    \centering
    \includegraphics[width=\linewidth]{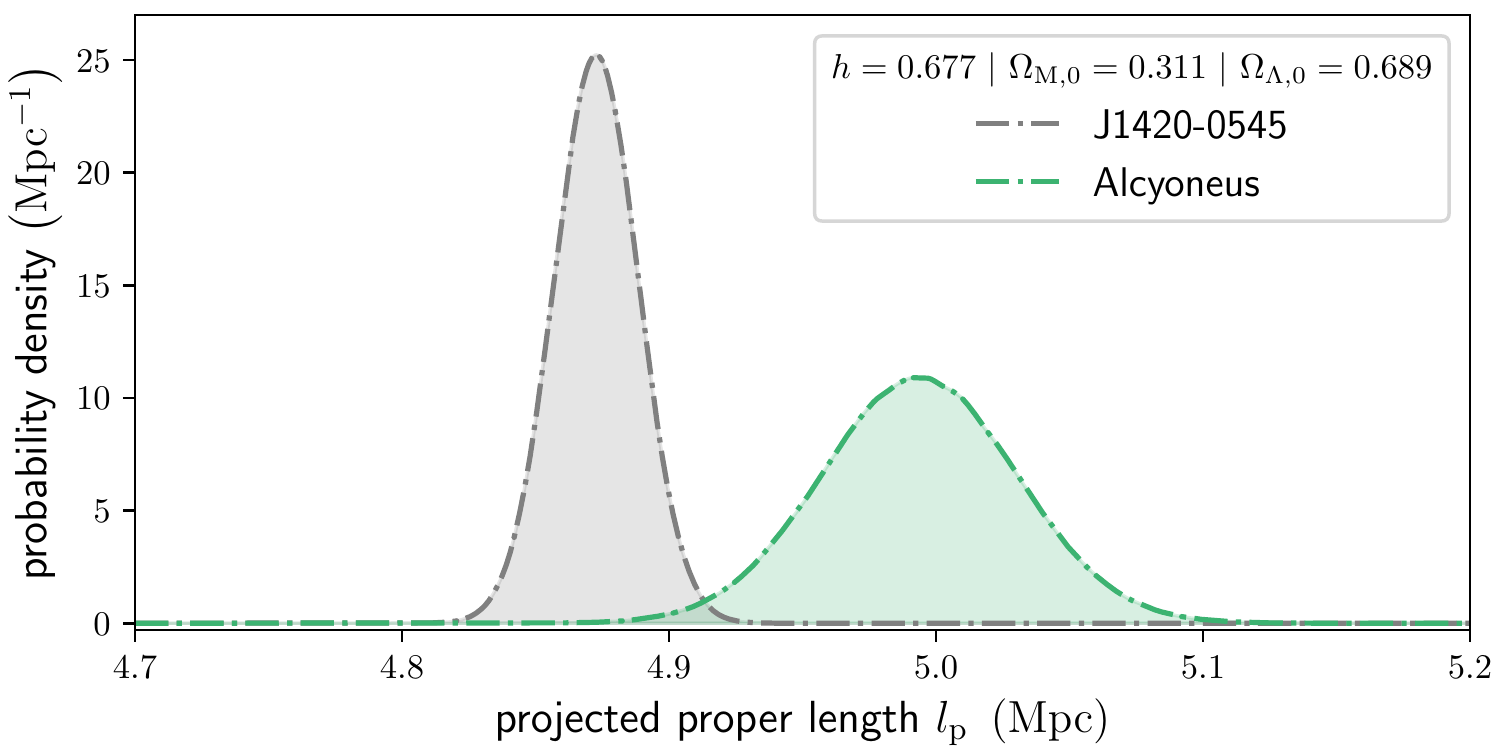}
    \caption{
    \textbf{Alcyoneus' projected proper length just exceeds that of J1420-0545.}
    The probability that Alcyoneus (green) has a larger projected proper length than J1420-0545 (grey) \textcolor{blue}{\citep{Machalski12008}} is $99.9\%$.
    For both GRGs, we take into account uncertainty in angular length and spectroscopic redshift, as well as the possibility of peculiar motion along the line of sight.
    }
    \label{fig:uncertaintyAnalysis}
\end{figure}

We verify that Alcyoneus is the largest known radio galaxy (RG) in projection by comparing it with J1420-0545 \textcolor{blue}{\citep{Machalski12008}}, the literature's record holder.\\
The angular lengths of Alcyoneus and J1420-0545 are $\phi = 20.8' \pm 0.15'$ and $\phi = 17.4' \pm 0.05'$, respectively.
For J1420-0545, we adopt the angular length reported by \textcolor{blue}{\citet{Machalski12008}} because it lies outside the LoTSS DR2 coverage.
The spectroscopic redshifts of Alcyoneus and J1420-0545 are $z_\mathrm{spec} = 0.24674 \pm 6 \cdot 10^{-5}$ and $z_\mathrm{spec} = 0.3067 \pm 5 \cdot 10^{-4}$, respectively.
For both giants, we assume the peculiar velocity along the line of sight $u_\mathrm{p}$ to be a Gaussian random variable (RV) with mean 0 and standard deviation $100\ \mathrm{km\ s^{-1}}$, similar to conditions in low-mass galaxy clusters.\\
\textbf{Equations}~\ref{eq:redshiftCosmological} describe how to calculate the cosmological redshift RV $z$ via the peculiar velocity redshift RV $z_\mathrm{p}$:
\begin{align}
    \beta_\mathrm{p} \coloneqq \frac{u_\mathrm{p}}{c};\ \ z_\mathrm{p} = \sqrt{\frac{1+\beta_\mathrm{p}}{1-\beta_\mathrm{p}}} - 1;\ \ z = \frac{1 + z_\mathrm{spec}}{1 + z_\mathrm{p}} - 1.
\label{eq:redshiftCosmological}
\end{align}
Here, $c$ is the speed of light \textit{in vacuo}.
Finally, we calculate the projected proper length RV $l_\mathrm{p} = r_\phi\left(z, \mathfrak{M}\right) \cdot \phi$.
Here, $r_\phi$ is the angular diameter distance RV, which depends on cosmological model parameters $\mathfrak{M}$.
Propagating the uncertainties in angular length $\phi$, spectroscopic redshift $z_\mathrm{spec}$ and peculiar velocity along the line of sight $u_\mathrm{p}$ through Monte Carlo simulation, the projected proper lengths of Alcyoneus and J1420-0545 are $l_\mathrm{p} = 4.99 \pm 0.04\ \mathrm{Mpc}$ and $l_\mathrm{p} = 4.87 \pm 0.02\ \mathrm{Mpc}$, respectively.\\
We show the two projected proper length distributions in \textbf{Figure}~\ref{fig:uncertaintyAnalysis}.
\noindent
The probability that Alcyoneus has the largest projected proper length is $99.9\%$.
This result is insensitive to plausible changes in cosmological parameters; for example, the high-$H_0$ (i.e. $H_0 > 70\ \mathrm{km\ s^{-1}\ Mpc^{-1}}$) cosmology with $\mathfrak{M} = \left(h = 0.7020, \Omega_\mathrm{BM,0} = 0.0455, \Omega_\mathrm{M,0} = 0.2720, \Omega_\mathrm{\Lambda,0} = 0.7280\right)$ yields a probability of $99.8\%$.
\begin{figure}
    \centering
    \includegraphics[width=\linewidth]{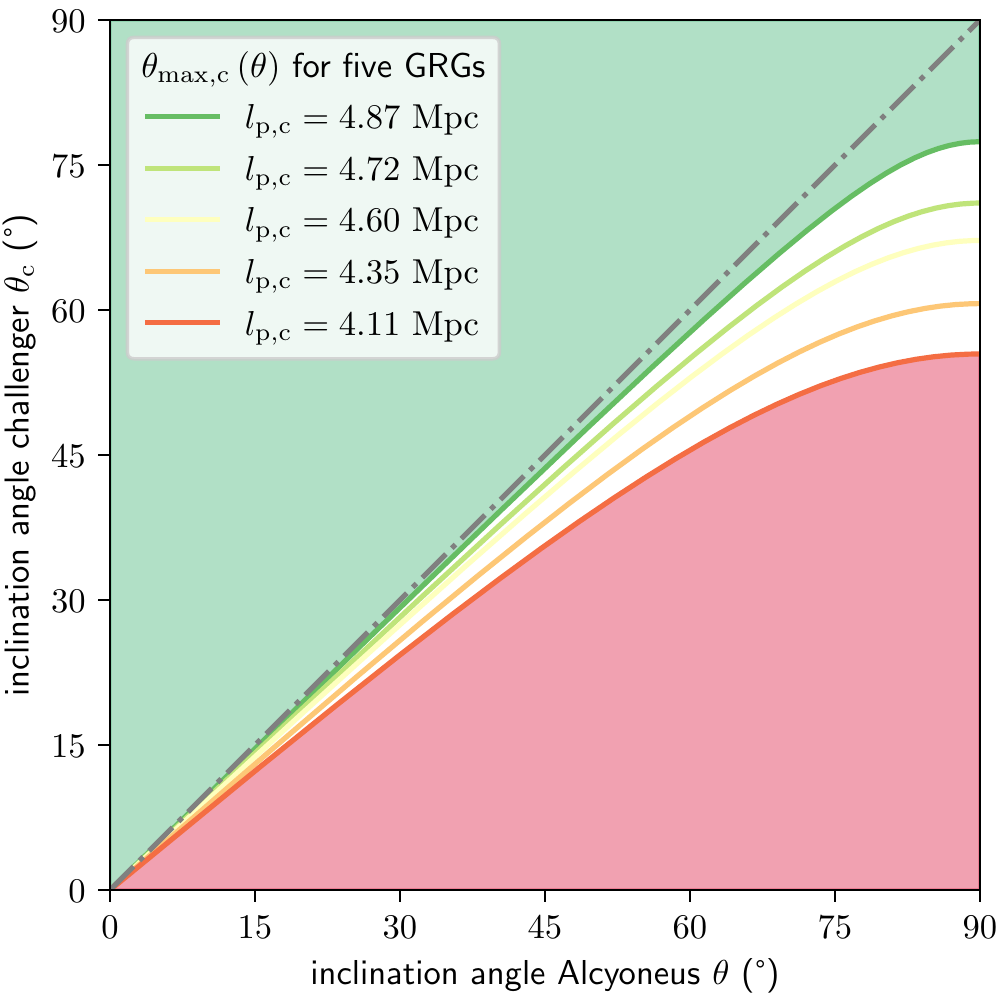}
    \caption{\textbf{When is Alcyoneus not only the largest GRG in the plane of the sky, but also in three dimensions?}
    Alcyoneus' inclination angle $\theta$ is not well determined, and therefore the full range of possibilities is shown on the horizontal axis.
    To surpass Alcyoneus in true proper length, a challenger must have an inclination angle (vertical axis) of at most Alcyoneus' (grey dotted line).
    More specifically, as a function of $\theta$, we show the \emph{maximum} inclination angle for which challengers with a projected proper length $l_\mathrm{p,c} > 4\ \mathrm{Mpc}$ trump Alcyoneus (coloured curves).
    The shaded areas of parameter space represent regimes with a particularly straightforward interpretation.
    One can imagine populating the graph with five points (located along the same vertical line), representing the ground-truth inclination angles of Alcyoneus and its five challengers.
    If \emph{any} of these points fall in the red-shaded area, Alcyoneus is not the largest GRG in 3D.
    If \emph{all} points fall in the green-shaded area, Alcyoneus is the largest GRG in 3D.}
    \label{fig:inclinationAnglesAlcyoneusChallengers}
\end{figure}
\section{Inclination angle comparison}
\label{ap:challengers}
Under what conditions is Alcyoneus not only the largest GRG in the plane of the sky, but also in three dimensions?
To answer this question, we compare Alcyoneus to the five previously known GRGs with projected proper lengths above 4 Mpc, which we dub \textit{challengers}.
A challenger surpasses Alcyoneus in true proper length when
\begin{align}
    l_\mathrm{c} > l,\ \ \mathrm{or}\ \ \frac{l_\mathrm{p,c}}{\sin{\theta_\mathrm{c}}} > \frac{l_\mathrm{p}}{\sin{\theta}},\ \ \mathrm{or}\ \ \sin{\theta_\mathrm{c}} < \frac{l_\mathrm{p,c}}{l_\mathrm{p}}\sin{\theta},
\end{align}
where $l_\mathrm{c}$, $l_\mathrm{p,c}$ and $\theta_\mathrm{c}$ are the challenger's true proper length, projected proper length and inclination angle, respectively.
Because the arcsine is a monotonically increasing function, a challenger surpasses Alcyoneus if its inclination angle obeys
\begin{align}
    \theta_\mathrm{c} < \theta_\mathrm{max,c}\left(\theta\right),\ \mathrm{where}\ \theta_\mathrm{max,c}\left(\theta\right) \coloneqq \arcsin{\left(\frac{l_\mathrm{p,c}}{l_\mathrm{p}}\sin{\theta}\right)}.
\end{align}
In \textbf{Figure}~\ref{fig:inclinationAnglesAlcyoneusChallengers}, we show $\theta_\mathrm{max,c}\left(\theta\right)$ for the five challengers with $l_\mathrm{p,c} \in \{4.11\ \mathrm{Mpc},\ 4.35\ \mathrm{Mpc},\ 4.60\ \mathrm{Mpc},\ 4.72\ \mathrm{Mpc},\ 4.87\ \mathrm{Mpc}\}$ (coloured curves).
Alcyoneus is least likely to be the longest GRG in 3D when its true proper length equals its projected proper length; i.e. when $\theta = 90\degree$.
The challengers then surpass Alcyoneus in true proper length when their inclination angles are less than 55$\degree$, 61$\degree$, 67$\degree$, 71$\degree$ and 77$\degree$, respectively.
For $\theta < 90\degree$, the conditions are more stringent.\\
The third and fourth largest challengers, whose respective SDSS DR12 host names are J100601.73+345410.5 and J093139.03+320400.1, harbour quasars in their host galaxies.
If small inclination angles distinguish quasars from non-quasar AGN, as proposed by the unification model \textcolor{blue}{\citep[e.g.][]{Hardcastle12020}}, these two challengers may well be the longest radio galaxies in three dimensions.

\section{Lobe volumes with truncated double cone model}
\label{ap:doubleConeModel}
\subsection{Synopsis}
We build a Metropolis--Hastings Markov chain Monte Carlo (MH MCMC) model, similar in spirit to the model of \textcolor{blue}{\citet{Boxelaar12021}} for galaxy cluster halos, in order to formalise the determination of RG lobe volumes from a radio image.
To this end, we introduce a parametrisation of a pair of 3D radio galaxy lobes, and explore the corresponding parameter space via the Metropolis algorithm.\footnote{The more general Metropolis\emph{--Hastings} variant need not be considered, as we work with a symmetric proposal distribution.}
For each parameter tuple encountered during exploration, we first calculate the monochromatic emission coefficient (MEC) function of the lobes on a uniform 3D grid representing a proper (rather than comoving) cubical volume.
The RG is assumed to be far enough from the observer that the conversion to a 2D image through ray tracing simplifies to summing up the cube's voxels along one dimension, and applying a cosmological attenuation factor.
This factor depends on the galaxy's cosmological redshift, which is a hyperparameter.
We blur the model image to the resolution of the observed image, which is also a hyperparameter.
Next, we calculate the likelihood that the observed image is a noisy version of the proposed model image.
The imaged sky region is divided into patches with a solid angle equal to the PSF solid angle; the noise per patch is then assumed to be an independent Gaussian RV.
These RVs have zero mean and share the same variance, which is another hyperparameter --- typically obtained from the observed image.
We choose a uniform prior over the full physically realisable part of parameter space.
The resulting posterior, which contains both geometric and radiative parameters, allows one to calculate probability distributions for many interesting quantities, such as the RG's lobe volumes and inclination angle.
The inferences depend weakly on cosmological parameters $\mathfrak{M}$.
Furthermore, their reliability depends significantly on the validity of the model assumptions.

\subsection{Model}
\subsubsection{Geometry}
We model each lobe in 3D with a truncated right circular cone with apex $\mathcal{O} \in \mathbb{R}^3$, central axis unit vector $\hat{a} \in \mathbb{S}^2$ and opening angle $\gamma \in [0, \frac{\pi}{2}]$, as in \textbf{Figure}~\ref{fig:doubleConeModel}.
The lobes share the same $\mathcal{O}$, which is the RG host location.
Each central axis unit vector can be parametrised through a position angle $\varphi \in [0, 2\pi)$ and an inclination angle $\theta \in [0, \pi]$.
Each cone is truncated twice, through planes that intersect the cone perpendicularly to its central axis.
Thus, each truncation is parametrised by the distance from the apex to the point where the plane intersects the central axis.
The two inner ($d_{\mathrm{i},1}, d_{\mathrm{i},2} \in \mathbb{R}_{\geq 0}$) and two outer ($d_{\mathrm{o},1},d_{\mathrm{o},2} \in \mathbb{R}_{\geq 0}$) truncation distances are parameters that we allow to vary independently, with the only constraint that each \emph{inner} truncation distance cannot exceed the corresponding \emph{outer} truncation distance.

\subsubsection{Radiative processes}
The radiative formulation of our model is among the simplest possible.
The radio emission from the lobes is synchrotron radiation.
We consider the lobes to be perfectly optically thin: we neglect synchrotron self-absorption.
The proper MEC is assumed spatially constant throughout a lobe, though possibly different among lobes; this leads to parameters $j_{\nu,1}, j_{\nu,2} \in \mathbb{R}_{\geq 0}$.
The relationship between the specific intensity $I_\nu$ (in direction $\hat{r}$ at central frequency $\nu_\mathrm{c}$) and the MEC $j_\nu$ (in direction $\hat{r}$ at cosmological redshift $z$ and rest-frame frequency $\nu = \nu_\mathrm{c}\left(1+z\right)$) is
\begin{align}
    I_\nu\left(\hat{r},\nu_\mathrm{c}\right) = \int_0^\infty \frac{j_\nu\left(\hat{r},z\left(l\right), \nu_\mathrm{c}\left(1+z\left(l\right)\right)\right)}{\left(1+z\left(l\right)\right)^3}\ \mathrm{d}l \approx \frac{j_\nu\left(\nu\right)\Delta l\left(\hat{r}\right)}{\left(1+z\right)^3},
\end{align}
where $l$ represents proper length.
The approximation is valid for a lobe with a spatially constant MEC that is small enough to assume a constant redshift for it.
$\Delta l(\hat{r})$ is the proper length of the line of sight through the lobe in direction $\hat{r}$.
The inferred MECs $j_{\nu,1}\left(\nu\right), j_{\nu,2}\left(\nu\right)$ thus correspond to \emph{rest-frame} frequency $\nu$.


\subsection{Proposal distribution}
In order to explore the posterior distribution on the parameter space, we follow the Metropolis algorithm.
The Metropolis algorithm assumes a symmetric proposal distribution.
\subsubsection{Radio galaxy axis direction}
To propose a new RG axis direction given the current one whilst satisfying the symmetry assumption, we perform a trick.
We populate the unit sphere with $N \in \mathbb{N}_{\geq 1}$ points (interpreted as directions) drawn from a uniform distribution.
Of these $N$ directions, the proposed axis direction is taken to be the one closest to the current axis direction (in the great-circle distance sense).
Note that this approach evidently satisfies the criterion that proposing the new direction given the old one is equally likely as proposing the old direction given the new one.
Also note that the distribution of the angular distance between current and proposed axis directions is determined solely by $N$.\\
In the following paragraphs, we first review how to perform uniform sampling of the unit two-sphere.
More explicitly than in \textcolor{blue}{\citet{Scott11989}}, we then derive the distribution of the angular distance between a reference point and the nearest of $N$ uniformly drawn other points.
The result is a continuous univariate distribution with a single parameter $N$ and finite support $\left(0, \pi\right)$.
Finally, we present the mode, median and maximum likelihood estimator of $N$.
As far as we know, these properties are new to the literature.

\paragraph{\textbf{Uniform sampling of $\mathbb{S}^2$}}
Let us place a number of points uniformly on the celestial sphere $\mathbb{S}^2$.
The spherical coordinates of such points are given by the RVs $(\Phi, \Theta)$, where $\Phi$ denotes position angle and $\Theta$ denotes inclination angle.
As all position angles are equally likely, the distribution of $\Phi$ is uniform: $\Phi \sim \mathbb{U}[0, 2\pi)$.
In order to effect a uniform number density, the probability that a point lies within a rectangle of width $\mathrm{d}\varphi$ and height $\mathrm{d}\theta$ in the $(\varphi, \theta)$-plane equals the ratio of the solid angle of the corresponding sky patch and the sphere's total solid angle:
\begin{align}
\mathbb{P}(\varphi \leq \Phi < \varphi + \mathrm{d}\varphi, \theta \leq \Theta < \theta + \mathrm{d}\theta) = \frac{\sin\theta\ \mathrm{d}\varphi\ \mathrm{d}\theta}{4\pi}.
\end{align}
The probability that the inclination angle is found somewhere in the interval $[\theta, \theta + \mathrm{d}\theta)$, regardless of the position angle, is therefore
\begin{align}
\mathbb{P}(\theta \leq \Theta < \theta + \mathrm{d}\theta) &= \mathrm{d}F_\Theta(\theta) = f_\Theta(\theta) \mathrm{d}\theta\nonumber\\
&= \int_0^{2\pi} \frac{\sin\theta\ \mathrm{d}\theta}{4\pi}\mathrm{d}\varphi = \frac{1}{2}\sin\theta\ \mathrm{d}\theta,
\end{align}
where $F_\Theta$ is the cumulative distribution function (CDF) of $\Theta$, and $f_\Theta$ the associated probability density function (PDF).
So,
\begin{align}
f_\Theta(\theta) = \frac{1}{2} \sin\theta;\ \ F_\Theta(\theta) \coloneqq \int_0^\theta f_\Theta(\theta')\ \mathrm{d}\theta' = \frac{1 - \cos\theta}{2}.
\end{align}

\begin{figure*}
\centering
\includegraphics[width=\textwidth]{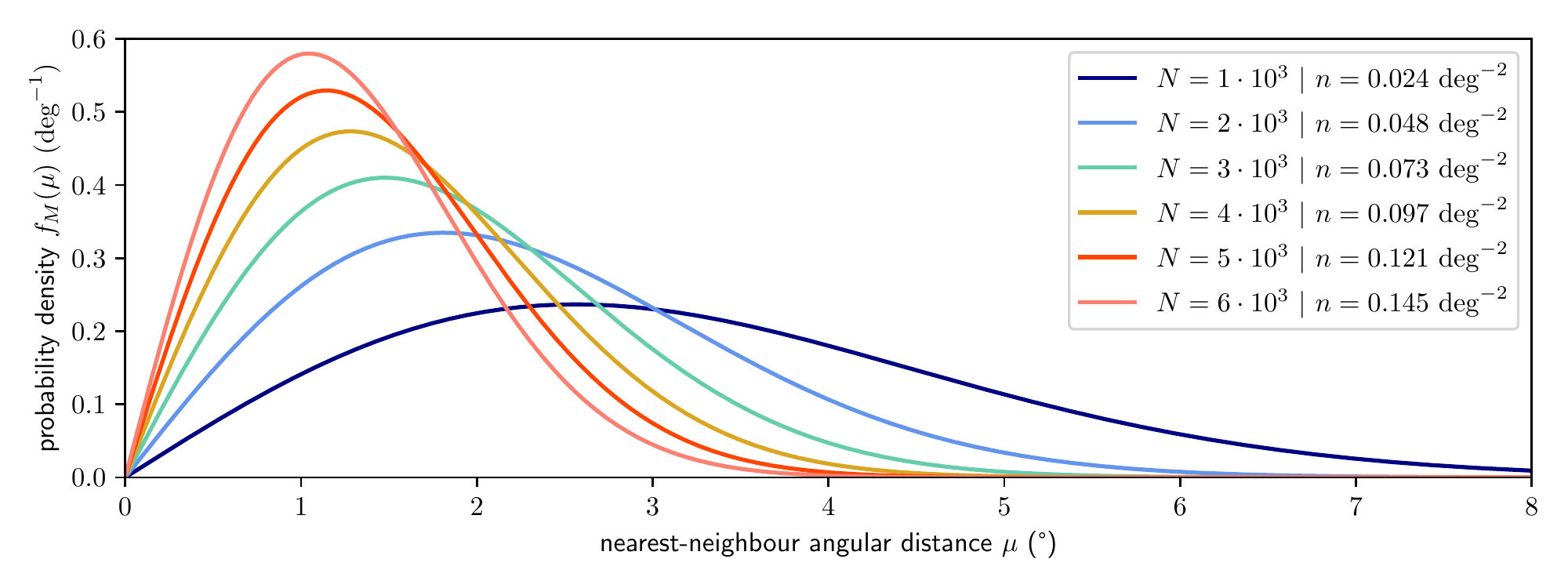}
\caption{
Probability density functions (PDFs) of the nearest-neighbour angular distance (NNAD) RV $M$ between some fixed point and $N$ other points distributed randomly over the celestial sphere.
As the sphere gets more densely packed, the probability of finding a small $M$ increases.
For each $N$, we provide the mean point number density $n$.
}
\label{fig:NNADPDF}
\end{figure*}

\paragraph{\textbf{Nearest-neighbour angular distance distribution}}
Pick a reference point and stochastically introduce $N$ other points in above fashion, which we dub its \emph{neighbours}.
We now derive the PDF of the angular distance to the nearest neighbour (NNAD).
Let $(\varphi_\mathrm{ref}, \theta_\mathrm{ref})$ be the coordinates of the reference point and let $(\varphi, \theta)$ be the coordinates of one of the neighbours.
Without loss of generality, due to spherical symmetry, we can choose to place the reference point in the direction towards the observer: $\theta_\mathrm{ref} = 0$. (Note that $\varphi_\mathrm{ref}$ is meaningless in this case.)
The angular distance between two points on $\mathbb{S}^2$ is given by the great-circle distance $\xi$.
For our choice of reference point, we immediately see that $\xi(\varphi_\mathrm{ref}, \theta_\mathrm{ref}, \varphi, \theta) = \theta$.
Because $\theta$ is a realisation of $\Theta$, $\xi$ too can be regarded as a realisation of an RV, which we call $\Xi$.
Evidently, the PDF $f_\Xi(\xi) = f_\Theta(\xi)$ and the CDF $F_\Xi(\xi) = F_\Theta(\xi)$.\\
Now consider the generation of $N$ points, whose angular distances to the reference point are the RVs $\{\Xi_i\} \coloneqq \{\Xi_1, ..., \Xi_N\}$.
The NNAD RV $M$ is the \emph{minimum} of this set: $M \coloneqq \min\{\Xi_i\}$.
What are the CDF $F_M$ and PDF $f_M$ of $M$?
\begin{align}
F_M(\mu) \coloneqq \mathbb{P}(M \leq \mu) &= \mathbb{P}(\mathrm{minimum\ of}\ \{\Xi_i\} \leq \mu)\nonumber\\
&= \mathbb{P}(\mathrm{at\ least\ one\ of\ the\ set}\ \{\Xi_i\} \leq \mu)\nonumber\\
&= 1 - \mathbb{P}(\mathrm{none\ of\ the\ set}\ \{\Xi_i\} \leq \mu)\nonumber\\
&= 1 - \mathbb{P}(\mathrm{all\ of\ the\ set}\ \{\Xi_i\} > \mu).
\end{align}
Because the $\{\Xi_i\}$ are independent and identically distributed,
\begin{align}
F_M(\mu) &= 1 - \prod_{i=1}^N \mathbb{P}\left(\Xi_i > \mu\right)\nonumber\\
&= 1 - \mathbb{P}^N(\Xi > \mu) = 1 - (1 - F_\Xi(\mu))^N.
\end{align}
By substitution, the application of a trigonometric identity and differentiation to $\mu$, we obtain the CDF and PDF of $M$:
\begin{align}
    F_M\left(\mu\right) = 1 - \cos^{2N}{\left(\frac{\mu}{2}\right)};\ f_M(\mu) = N \sin{\left(\frac{\mu}{2}\right)} \cos^{2N-1}{\left(\frac{\mu}{2}\right)}.
    \label{eq:NNADPDF}
\end{align}
In \textbf{Figure}~\ref{fig:NNADPDF}, we show this PDF for various values of $N$.\\
\noindent
The mode of $M$ (i.e. the most probable NNAD), $\mu_\mathrm{mode}$, is the solution to $\frac{\mathrm{d}f_M}{\mathrm{d}\mu}(\mu_\mathrm{mode}) = 0$.
The median of $M$, $\mu_\mathrm{median}$, is the solution to $F_M(\mu_\mathrm{median}) = \frac{1}{2}$.
Hence,
\begin{align}
\mu_\mathrm{mode} = \arccos{\left(1 - \frac{1}{N}\right)};\ \mu_\mathrm{median} = \arccos{\left(2^{1 - \frac{1}{N}}-1\right)}.
\end{align}
As common sense dictates, both equal $\frac{\pi}{2}$ for $N = 1$ and tend to $0$ as $N \to \infty$.
We find the mean of $M$ through integration by parts:
\begin{align}
\mathbb{E}\left[M\right] \coloneqq& \int_0^{\pi} \mu f_M(\mu)\ \mathrm{d}\mu = \int_0^{\pi} \mu\ \mathrm{d}F_M(\mu)\nonumber\\
&= \bigg[\mu F_M(\mu)\bigg]_0^{\pi} - \int_0^{\pi} F_M(\mu)\ \mathrm{d}\mu\nonumber\\
&= \int_0^{\pi} \cos^{2N}{\left(\frac{\mu}{2}\right)}\ \mathrm{d}\mu = 2 \int_0^{\frac{\pi}{2}} \cos^{2N}{\left(\mu\right)}\ \mathrm{d}\mu.
\end{align}
Again via integration by parts,
\begin{align}
\mathbb{E}\left[M\right] = \pi \prod_{k=1}^N \frac{2k-1}{2k} = \frac{\pi}{2^{2N}}\binom{2N}{N}.
\label{eq:NNADExpectationValueFinal}
\end{align}

\paragraph{\textbf{Maximum likelihood estimation}}
A typical application is the estimation of $N$ in the PDF $f_M(\mu\ |\ N)$ (\textbf{Equation}~\ref{eq:NNADPDF}) using data.
Let us assume we have measured $k$ NNADs, denoted by $\{\mu_1, ..., \mu_k\}$.
Let the joint PDF or \emph{likelihood} be
\begin{align}
\mathcal{L}(N) &\coloneqq \prod_{i=1}^k f_M\left(\mu_i\ \vert\ N\right)\nonumber\\
&= \left(\frac{N}{2^N}\right)^k \prod_{i = 1}^k \sin\mu_i\ (\cos\mu_i + 1)^{N-1}.
\end{align}
To find $N_\mathrm{MLE}$, we look for the value of $N$ that maximises $\mathcal{L}(N)$.
To simplify the algebra, we could however equally well maximise a $k$-th of the natural logarithm of the likelihood, or the \textit{average log-likelihood} $\hat{l} \coloneqq k^{-1}\ln\mathcal{L}(N)$, because the logarithm is a monotonically increasing function:
\begin{align}
\hat{l}(N) &\coloneqq \frac{1}{k}\ln\mathcal{L}(N) = \ln N - N \ln 2\nonumber\\
&+ \frac{1}{k} \sum_{i=1}^k \ln\sin\mu_i + (N - 1) \ln(\cos\mu_i + 1).
\end{align}
We find $N_\mathrm{MLE}$ by solving $\frac{\mathrm{d}\hat{l}}{\mathrm{d}N}(N_\mathrm{MLE}) = 0$.
This leads to
\begin{align}
N_\mathrm{MLE} = \left(\ln 2 - \frac{1}{k} \sum_{i = 1}^k \ln(\cos\mu_i + 1)\right)^{-1}.
\label{eq:NNADPDFNMLE}
\end{align}
An easy limit to evaluate is the case when $\mu_1, ..., \mu_k \to 0$. In such case, $\cos\mu_i \to 1$, and so $\frac{1}{k} \sum_{i = 1}^k \ln(\cos\mu_i + 1) \to \ln 2$. Then, $N_\mathrm{MLE} \to (0_+)^{-1} \to \infty$.
This is expected behaviour: when all measured NNADs approach 0, the number of points distributed on the sphere must be approaching infinity.

\subsubsection{Other parameters}
The other proposal parameters are each drawn from independent normal distributions centred around the current parameter values.
These proposal distributions are evidently symmetric, but have support over the full real line, so that forbidden parameter values can in principle be proposed.
As a remedy, we set the prior probability density of the proposed parameter set to 0 when the proposed opening angle is negative or exceeds $\frac{\pi}{2}\ \mathrm{rad}$, at least one of the proposed MECs is negative, or when at least one of the proposed inner truncation distances is negative or exceeds the corresponding proposed outer truncation distance.
In such cases, the posterior probability density is 0 too, as it is proportional to the prior probability density.
Consequently, the Metropolis acceptance probability vanishes and the proposal is rejected.
We do not enter forbidden regions of parameter space.
The condition of detailed balance is still respected: probability densities for transitioning \emph{towards} the forbidden region are 0, just as probability densities for \emph{being in} the forbidden region.

\subsection{Likelihood}
We assume the likelihood to be Gaussian.
To avoid dimensionality errors, we multiply the likelihood by a constant before we take the logarithm:
\begin{align}
    \ln{\left(\mathcal{L} \cdot \left(\sigma\sqrt{2\pi}\right)^{N_\mathrm{r}}\right)} = -\frac{N_\mathrm{r}}{2\sigma^2 N_\mathrm{p}}\sum_{i=1}^{N_\mathrm{p}}\left(I_{\nu,\mathrm{o}}\left[i\right] - I_{\nu,\mathrm{m}}\left[i\right]\right)^2.
\end{align}
Here, $\sigma$ is the image noise, $N_\mathrm{r} \in \mathbb{R}_{\geq 0}$ is the number of resolution elements in the image, $N_\mathrm{p} \in \mathbb{N}$ is the number of pixels in the image, and $I_{\nu,\mathrm{o}}\left[i\right]$ and $I_{\nu,\mathrm{m}}\left[i\right]$ are the $i$-th pixel values of the observed and modelled image, respectively.
For simplicity, one may multiply the likelihood by a constant factor (or, equivalently, add a constant term to the log-likelihood): the acceptance ratio will remain the same, and the MH MCMC runs correctly.

\subsection{Results for Alcyoneus}
\begin{figure}
    \centering
    \includegraphics[width=\columnwidth]{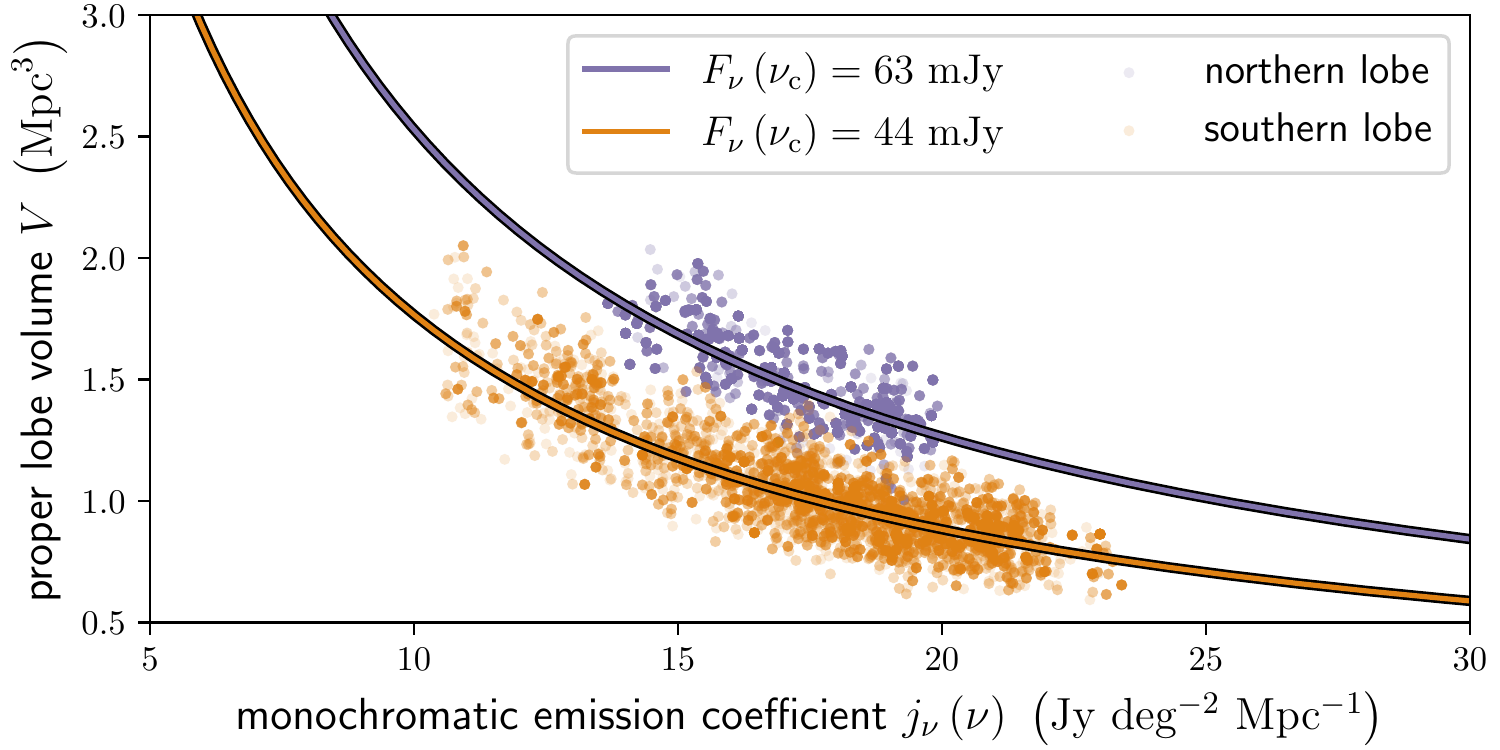}
    \caption{\textbf{
    Our Bayesian model yields strongly correlated estimates for $j_\nu\left(\nu\right)$ and $V$ that reproduce the observed lobe flux densities.
    }
    We show MECs $j_\nu\left(\nu\right)$ at $\nu = 180\ \mathrm{MHz}$ and proper volumes $V$ of Metropolis--Hastings Markov chain Monte Carlo samples for the northern lobe (purple dots) and southern lobe (orange dots).
    The curves represent all combinations $\left(j_\nu\left(\nu\right),V\right)$ that correspond to a particular flux density at the LoTSS central wavelength $\nu_\mathrm{c} = 144\ \mathrm{MHz}$.
    We show the \emph{observed} northern lobe flux density (purple curve) and the \emph{observed} southern lobe flux density (orange curve).
    }
    \label{fig:MECsVsVolumina}
\end{figure}

We apply the Bayesian model to the $90''$ LoTSS DR2 image of Alcyoneus, shown in the top panel of \textbf{Figure}~\ref{fig:doubleConeModel}.
Thus, the hyperparameters are $z = 0.24674$, $\nu_\mathrm{c} = 144\ \mathrm{MHz}$ (so that $\nu = 180\ \mathrm{MHz}$), $\theta_\mathrm{FWHM} = 90''$, $N = 750$ and $\sigma = \sqrt{2} \cdot 1.16\ \mathrm{Jy\ deg^{-2}}$.
We set the image noise to $\sqrt{2}$ times the \emph{true} image noise to account for model incompleteness.
This factor follows by assuming that the inability of the model to produce the true lobe morphology yields (Gaussian) errors comparable to the image noise.
To speed up inference, we downsample the image of 2,048 by 2,048 pixels by a factor 16 along each dimension.
We run our MH MCMC for 10,000 steps, and discard the first 1,500 steps due to burn-in.\\
\textbf{Table}~\ref{tab:MAPEstimates} lists the obtained maximum a posteriori probability (MAP) estimates and posterior mean and standard deviation (SD) of the parameters.
\begin{center}
\captionof{table}{
Maximum a posteriori probability (MAP) estimates and posterior mean and standard deviation (SD) of the parameters from the Bayesian, doubly truncated, conical radio galaxy lobe model of \textbf{Section}~\ref{sec:BayesianModel}.\\
}
\begin{tabular}{c | c | c} 
parameter & MAP estimate & posterior mean and SD\\
 [3pt] \hline
$\varphi_1$ & $307\degree$ & $307\pm 1\degree$\\
$\varphi_2$ & $140\degree$ & $139\pm 2 \degree$\\
$\vert\theta_1 - 90\degree\vert$ & $54\degree$ & $51 \pm 2\degree$\\
$\vert\theta_2 - 90\degree\vert$ & $25\degree$ & $18 \pm 7\degree$\\
$\gamma_1$ & $9\degree$ & $10 \pm 1\degree$\\
$\gamma_2$ & $24\degree$ & $26 \pm 2\degree$\\
$d_\mathrm{i,1}$ & $2.7\ \mathrm{Mpc}$ & $2.6 \pm 0.2\ \mathrm{Mpc}$\\
$d_\mathrm{o,1}$ & $4.3\ \mathrm{Mpc}$ & $4.0 \pm 0.2\ \mathrm{Mpc}$\\
$d_\mathrm{i,2}$ & $1.6\ \mathrm{Mpc}$ & $1.5 \pm 0.1\ \mathrm{Mpc}$\\
$d_\mathrm{o,2}$ & $2.0\ \mathrm{Mpc}$ & $2.0 \pm 0.1\ \mathrm{Mpc}$\\
$j_{\nu,1}\left(\nu\right)$ & $17\ \mathrm{Jy\ deg^{-2}\ Mpc^{-1}}$ & $17 \pm 2\ \mathrm{Jy\ deg^{-2}\ Mpc^{-1}}$\\
$j_{\nu,2}\left(\nu\right)$ & $22\ \mathrm{Jy\ deg^{-2}\ Mpc^{-1}}$ & $18 \pm 3\ \mathrm{Jy\ deg^{-2}\ Mpc^{-1}}$
\end{tabular}
\label{tab:MAPEstimates}
\end{center}
The proper volumes $V_1$ and $V_2$ are derived quantities:
\begin{align}
    V = \frac{\pi}{3} \tan^2{\gamma} \left(d_\mathrm{o}^3 - d_\mathrm{i}^3\right),
\label{eq:modelVolume}
\end{align}
just like the flux densities $F_{\nu,1}\left(\nu_\mathrm{c}\right)$ and $F_{\nu,2}\left(\nu_\mathrm{c}\right)$ at central frequency $\nu_\mathrm{c}$:
\begin{align}
    F_\nu\left(\nu_\mathrm{c}\right) = \frac{j_\nu\left(\nu\right) V}{\left(1+z\right)^3 r_\phi^2\left(z\right)}.
\label{eq:modelFluxDensity}
\end{align}
Together, $V$ and $F_\nu(\nu_\mathrm{c})$ imply a lobe pressure $P$ and a magnetic field strength $B$, which are additional derived quantities that we calculate through \texttt{pysynch}.\\
\textbf{Table}~\ref{tab:MAPEstimatesDerived} lists the obtained MAP estimates and posterior mean and SD of the derived quantities.
\begin{center}
\captionof{table}{
Maximum a posteriori probability (MAP) estimates and posterior mean and standard deviation (SD) of derived quantities from the Bayesian, doubly truncated, conical radio galaxy lobe model of \textbf{Section}~\ref{sec:BayesianModel}.\\
}
\begin{tabular}{c | c | c}
derived quantity & MAP estimate & posterior mean and SD\\
 [3pt] \hline
$\Delta\varphi$ & $167\degree$ & $168\pm 2\degree$\\
$V_1$ & $1.5\ \mathrm{Mpc}^3$ & $1.5 \pm 0.2\ \mathrm{Mpc^3}$\\
$V_2$ & $0.8\ \mathrm{Mpc}^3$ & $1.0 \pm 0.2\ \mathrm{Mpc^3}$\\
$F_{\nu,1}\left(\nu_\mathrm{c}\right)$ & $63\ \mathrm{mJy}$ & $63 \pm 4\ \mathrm{mJy}$\\
$F_{\nu,2}\left(\nu_\mathrm{c}\right)$ & $44\ \mathrm{mJy}$ & $45 \pm 5\ \mathrm{mJy}$\\
$P_\mathrm{min,1}$ & $4.7 \cdot 10^{-16}\ \mathrm{Pa}$ & $4.8 \pm 0.3 \cdot 10^{-16}\ \mathrm{Pa}$\\
$P_\mathrm{min,2}$ & $5.4 \cdot 10^{-16}\ \mathrm{Pa}$ & $5.0 \pm 0.6 \cdot 10^{-16}\ \mathrm{Pa}$\\
$P_\mathrm{eq,1}$ & $4.8 \cdot 10^{-16}\ \mathrm{Pa}$ & $4.9 \pm 0.3 \cdot 10^{-16}\ \mathrm{Pa}$\\
$P_\mathrm{eq,2}$ & $5.4 \cdot 10^{-16}\ \mathrm{Pa}$ & $5.0 \pm 0.6 \cdot 10^{-16}\ \mathrm{Pa}$\\
$B_\mathrm{min,1}$ & $45\ \mathrm{pT}$ & $45 \pm 1\ \mathrm{pT}$\\
$B_\mathrm{min,2}$ & $48\ \mathrm{pT}$ & $46 \pm 3\ \mathrm{pT}$\\
$B_\mathrm{eq,1}$ & $42\ \mathrm{pT}$ & $43 \pm 1\ \mathrm{pT}$\\
$B_\mathrm{eq,2}$ & $45\ \mathrm{pT}$ & $43 \pm 3\ \mathrm{pT}$\\
$E_\mathrm{min,1}$ & $6.3 \cdot 10^{52}\ \mathrm{J}$ & $6.2 \pm 0.4 \cdot 10^{52}\ \mathrm{J}$\\
$E_\mathrm{min,2}$ & $3.7 \cdot 10^{52}\ \mathrm{J}$ & $4.4 \pm 0.6 \cdot 10^{52}\ \mathrm{J}$\\
$E_\mathrm{eq,1}$ & $6.4 \cdot 10^{52}\ \mathrm{J}$ & $6.3 \pm 0.4 \cdot 10^{52}\ \mathrm{J}$\\
$E_\mathrm{eq,2}$ & $3.8 \cdot 10^{52}\ \mathrm{J}$ & $4.4 \pm 0.6 \cdot 10^{52}\ \mathrm{J}$
\end{tabular}
\label{tab:MAPEstimatesDerived}
\end{center}\noindent
The uncertainties of the parameters and derived quantities reported in \textbf{Tables}~\ref{tab:MAPEstimates} and \ref{tab:MAPEstimatesDerived} are not necessarily independent.
To demonstrate this, we present MECs and volumes from the MH MCMC samples in \textbf{Figure}~\ref{fig:MECsVsVolumina}.
\noindent
MECs and volumes do not vary independently, because their product is proportional to flux density (see \textbf{Equation}~\ref{eq:modelFluxDensity}); only realistic flux densities correspond to high-likelihood model images.\\\\
Finally, we explore a simpler variation of the model, in which we force the lobes to be coaxial.
In such a case, the true proper length $l$ and projected proper length $l_\mathrm{p}$ are additional derived quantities:
\begin{align}
    l = \frac{d_\mathrm{o,1} + d_\mathrm{o,2}}{\cos{\gamma}};\ \ l_\mathrm{p} = l \sin{\theta}.
\end{align}
For Alcyoneus, this simpler model does not provide a good fit to the data.
\end{document}